\documentclass[reprint,amsmath,amssymb,
 aps,prx]{revtex4-2}
\usepackage{graphicx}
\usepackage{dcolumn}
\usepackage{bm}
\usepackage{amssymb}
\usepackage{bbold}
\usepackage[caption=false]{subfig}
\usepackage[dvipsnames]{xcolor}
\usepackage{wrapfig}
\usepackage{comment}
\usepackage{silence}
\usepackage{scrhack}
\usepackage{hyperref}

\begin{document}

\title{Open-systems tools for non-thermalizing closed quantum systems}

\author{Unnati Akhouri}
\email{uja5020@psu.edu}

\author{Sarah Shandera} 
\email{ses47@psu.edu}
\affiliation{%
 Institute for Gravitation and the Cosmos, The Pennsylvania State University, University Park, PA 16802, USA
}%
\affiliation{Department of Physics, The Pennsylvania State University, University Park, PA 16802, USA.}

\author{Jackson Henry}
 \email{jhenry1@wpi.edu}
\affiliation{
Worcester Polytechnic Institute, 100 Institute Rd, Worcester, MA 01609, USA}%

\date{\today}

\begin{abstract}
We design several examples of constrained, symmetric quantum circuit dynamics that generate non-equilibrium steady states. The qubit networks maintain local memory of the initial conditions and display inhomogeneous subsystem dynamics over long times, clearly distinguishable from approximately thermalizing networks of the same size. Each network can be described as an ensemble of open systems, a collection of qubits evolving with phase-covariant dynamics. Constraints from the conservation law and global unitary dynamics of the entire network bound the distribution of single-qubit dynamics in the ensemble, but different steady states are distinguishable by several measures. We quantify the distance of the steady states from the homogeneous steady state and further characterize them using the complexity of their mutual information networks, the volume of state space explored, a thermodynamic utility measure using extractable work, and correlated structure in the occurrence of non-completely positive qubit propagator maps. 

\end{abstract}

\maketitle

\section{\label{sec:intro}Introduction}
A feature of thermalization in closed quantum systems is that for most initial states, most observables in most small subsystems look the same at late times, when initial state information is delocalized over many subsystems \cite{Deutsch:1991,Srednicki:1994,Rigol:2008}. In non-equilibrating systems, in contrast, we expect subsystems to display a diversity of states and dynamics over long times. Within that diversity, further structure may distinguish classes of out-of-equilibrium behavior \cite{Polkovnikov:2011}. 

To understand inhomogeneous systems requires examples of non-thermalizing dynamics. In closed quantum systems, a number of interesting examples have been developed using spins subject to kinetic constraints. In the PXP model \cite{Fendley:2004, Lesanovsky:2012}, or the quantum East model \cite{vanHorssen:2015, Pancotti:2020}, projection operators in the Hamiltonian restrict non-trivial evolution of one spin to occur only when neighboring spins are in some particular state. These Hamiltonian systems can display non-thermalizing behavior, at least for some initial states \cite{serbyn:2021,Bhore:2023,Royen:2024}. Quantum cellular automata (QCA) \cite{Farrelly:2020} expand the notion of kinetically constrained models, allowing a wide class of dynamics for individual spins constrained by the state of neighbors. QCA rules can generate a variety of dynamics, including some that lead to non-thermalizing spin networks with distinct two-spin correlation structures \cite{Hillberry:2021}. The behavior of local observables in thermalizing systems is a significant focus of many of these studies.

A different view of thermalization \cite{Davies:1974,spohn:2007}, and complementary method to explore dynamics of non-equilibrium quantum systems \cite{Prosen:2008,Prosen:2011} employs open quantum systems techniques. Here, the focus is often on defining non-unitary dynamics for an open system that drive it to a particular state. Many constructions make use of the Markovian Lindblad equation \cite{Lindblad,Vittorio:1976}, but its form requires assumptions of weak coupling, a hierarchy of time-scales, and Markovianity, all of which limit the domain of applicability \cite{Wichterich:2007}. An alternative technique, not restricted by any of those assumptions, is that of channels or dynamical maps \cite{Jagadish:2018}. In this formulation, details of the environment and the coupling of the system to the environment can enter in degenerate ways, and one may find a variety of models to generate the same dynamics. For example, collisional models \cite{Ciccarello:2022} employ discrete interactions with a refreshable bath for thermalization. More precise modeling of both energy injection (eg, Floquet driving) and dissipation generates open-system dynamics that can keep systems in non-equilibrium steady states \cite{Drummond:1980, Diehl:2011,Shirai:2014}. Maps and channels are especially suited to time-dependent dynamics, including time-dependent Hamiltonians or quantum circuits where the interactions may be adjusted at each layer. 

In this paper, we use both local observables and open-systems tools to characterize non-equilibrium dynamics in closed systems. We begin by restricting states and dynamics in such a way that all single-qubit open-system dynamics will be in a well-studied but not too restrictive class, called phase-covariant (reviewed below). Next, we introduce several examples of constrained dynamics that keep qubits in the quantum networks away from a thermal (or maximum entropy) state defined by the initial conditions. The constraints use quantum information quantities rather than just the neighboring subsystem states familiar from kinetically constrained Hamiltonians or QCA. The non-equilibrium steady states obtained with these dynamics can be straightforwardly contrasted with the state approached by random dynamics on the same network. 

We analyze each evolving network of $N$ qubits as an ensemble of $N$ open systems \cite{Prudhoe:2024yqi}. The distributions of states and dynamics within the ensemble are always constrained by the global unitary dynamics and conservation law, but there are clear distinctions between the ensembles that occur under thermalizing dynamics and those that correspond to non-equilibrium steady states on the networks. The phase-covariant dynamics used here was introduced in the context of phenomenological studies of thermalization and dephasing process beyond the Markovian approximation \cite{Holevo_PC, PhysRevLett.116.120801}; a primary result of this analysis is the role of non-Markovian sub-system dynamics in the systems that reach non-equilibrium steady states. 

This paper is organized as follows: in Section \ref{sec:networks} we define the dynamics and states used to generate out-of-equilibrium behavior. In Section \ref{sec:characterizing} we show how the resulting states compare using several measures that characterize the degree to which a network remains away from its thermal state. In Section \ref{sec:results} we show the correlations between different non-thermalizing measures, and how the prevalence of non-completely positive open system dynamics is correlated with the emergence of non-equilibrium steady states. Section \ref{sec:conc} summarizes the results.

\section{Defining the qubit networks and evolution}
\label{sec:networks}
We consider networks with $N=6,8,10,12$ or $14$ qubits with states and dynamics both subject to a symmetry constraint described in Section \ref{sec:symmetry}. The body of the paper shows results for $N=12$, while some results are shown as a function of network size in Appendix \ref{sec:appendFigures}. After initializing the system as described in Sections \ref{sec:InitializeNetworkCentral} and \ref{sec:InitializeNetworkDistributions}, the distinguishing aspect of the dynamics is the rule for choosing a decomposition of the network into 2-qubit neighborhoods to define each layer $\ell$ of the circuit. These rules are given in Section \ref{sec:EvolveNetwork}.

\subsection{Symmetry-constrained dynamics}
\label{sec:symmetry}
In order to make use of the well-studied class of phase-covariant dynamical maps \cite{Holevo:1993, Lankinen:2016, Chruscinski:2022}, we restrict the dynamics to be excitation-number preserving and restrict the initial states to be pure states of definite excitation number or classical mixtures of such states. This restriction also enables a relatively simple interpretation of thermodynamic quantities \cite{Roie:2021}. 

To ensure phase-covariance, we define a common excitation basis for the qubits and a corresponding free Hamiltonian using the Pauli matrix $\sigma_z$ (shifted so that the total energy is simply identified with excitation number):
\begin{equation}
    \hat{H}_0=\sum_{i=1}^N\frac{1}{2}(\mathbb{1}-\hat{\sigma}_z^{(i)})\,,
\end{equation}
and only consider unitary gates $\hat{U}$ that satisfy
\begin{equation}
\label{eq:PCcond}
   [\hat{H}_0,\hat{U}]=0\,.
\end{equation}
The initial states of the $N$-qubit system are tensor products of single-qubit states, each diagonal in the excitation-number basis:
\begin{equation}
    \label{eq:initialState}
    \rho^{(N)}(0)=\rho_1(0)\otimes\rho_2(0)\dots\otimes\rho_N(0)\,,
\end{equation}
where
\begin{equation}
\label{eq:qdiag}
\rho_q(0) = \begin{bmatrix}
1-p_q(0) & 0 \\
0 & p_q(0) 
\end{bmatrix}\,.
\end{equation}
or, $\rho_q(0)=\frac{1}{2}(\mathbb{1}+(1-2p_q(0)) \sigma_z)\equiv \frac{1}{2}(\mathbb{1}+z_q(0)\sigma_z)$. The initial state of the full system is characterized by the set of initial populations of the excited state $\{p_q(0)\}$. 

These conditions on the initial state and evolution have two important consequences. First, for single qubits, any evolution in this class is Gibbs-preserving. That is, the state of each qubit remains diagonal (in the same form as Eq.(\ref{eq:qdiag})) at all times. After $\ell$ layers of the circuit, the set of single-qubit states is fully described by $\{p_q(\ell)\}$. Since single-qubit states are always Gibbs states, we may use $p_q(\ell)$ to define a temperature for each state, at each time. 

In addition, Eq.(\ref{eq:PCcond}) and Equations (\ref{eq:initialState}), (\ref{eq:qdiag}) result in single-qubit dynamics that is always described by a phase-covariant dynamical map \cite{Holevo:1993, Lankinen:2016, Chruscinski:2022}. That is, the state of the $q^{\rm th}$ qubit after layer $\ell$ of the circuit is related to its initial state by 
\begin{equation}
\label{eq:Lambdarho}
\rho_q(\ell)=\Lambda_q(\ell,0)\circ\rho_q(0)\,,
\end{equation}
where, writing $\rho_q$ as a vector in the Pauli basis,
\begin{align}
 \label{eqn::PhaseCovMap}
\Lambda_{q}&(\ell,0)= \\
&\begin{pmatrix}
        1&0&0&0\\
        0&\lambda_q (\ell) \cos{\phi_q (\ell)}&-\lambda_q (\ell) \sin{\phi_q (\ell)}&0\\
          0&\lambda_q (\ell) \sin{\phi_q (\ell})&\lambda_q (\ell) \cos{\phi_q (\ell)}&0\\
          \tau_{z,q} (\ell)&0&0&\lambda_{z,q} (\ell)
    \end{pmatrix}\,.\nonumber
\end{align}
The parameter $\lambda_q$ describes the isotropic dilation of the $x$ and $y$ components of the state, and $\lambda_{z,q}$ and $\tau_{z,q}$ are the dilation and shift, respectively, of the $z$ component. The rotation angle $\phi_q$ is determined by the unitary gate and is the part of the map that does not change the mixedness of any state. Qubits undergoing a combination of pure dephasing with generalized amplitude damping have dynamics described by phase-covariant maps. 

Each channel defined by $\Lambda_{q}(\ell,0)$ has an associated single-qubit invariant state,
\begin{equation}
\label{eq:mapInvariantState}
\rho_{q,\ell}^{*} =  \frac{1}{2}\left(\mathbb{1}+ \left(\frac{\tau_{z,q}(\ell)}{1-\lambda_{z,q}(\ell)}\right)\hat{\sigma}_z\right)\,.
\end{equation}

Since the initial state is a tensor product, the maps $\Lambda_{q}(\ell,0)$ will be completely positive, and so satisfy the following conditions at all $\ell$:
\begin{align}
    \label{eqn::CPcond}
    |\lambda_{z,q} | + |\tau_{z,q}|\leq 1 \\
    4 \lambda_q^2 + \tau_{z,q} ^2\leq (1+\lambda_{z,q})^2\,.
\end{align}
However, the propagator ``maps" between later time steps, $\Lambda_{q}(\ell,\ell_*\neq0)$, will be of the form shown in Eq.(\ref{eqn::PhaseCovMap}) but need not be (completely) positive if sufficient quantum correlations are generated between qubit $q$ and the rest of the network. That is, the evolution of the individual qubits in the network may be quantum non-Markovian \cite{Filippov:2019, Siudzinska:2023a}. We may also consider the quantity $\frac{\tau_{z,q}}{1-\lambda_{z,q}}$ from these propagator maps, although the interpretation as the z-component of a stationary state as in Eq.(\ref{eq:mapInvariantState}) is not valid if $\left|{\frac{\tau_{z,q}}{1-\lambda_{z,q}}}\right|>1$. 

For simplicity, we consider circuits that differ only in the arrangement of neighborhoods (the interaction graph) at each layer of the circuit, with the type of interaction restricted to a single 2-qubit gate $U_{*}$ consistent with Eq.(\ref{eq:PCcond}). We choose $U_{*}$ to act in energy (or excitation) subspace $E=1$. The most general gate satisfying this restriction is
\begin{equation}
\label{eq:2Qunitary}
U_{*}= 
\begin{pmatrix} 
1 & 0 & 0 & 0 \\
0 & e^{-i \frac{\phi + \omega}{2}} \cos\left(\theta\right) & -e^{i \frac{\phi - \omega}{2}} \sin\left(\theta\right) & 0 \\ 
0 & e^{-i \frac{\phi - \omega}{2}} \sin\left(\theta\right) & e^{i \frac{\phi + \omega}{2}} \cos\left(\theta\right) & 0 \\
0 & 0 & 0 & 1
\end{pmatrix}\,.
\end{equation}
We choose $\phi=\omega=0$ for simplicity and $\theta=\frac{\pi}{15}$. The restricted dynamics means that our circuits do not explore the full space of symmetric dynamics on $N$ qubits \cite{Marvian:2020}, but the simplicity of $U=U_*$ allows us to isolate key features in the open-system evolution of individual qubits, as well as the relationship between the distribution of resources in the initial state and the evolution rule \footnote{We previously considered a more varied set of gates with only random circuits in \cite{Akhouri_2023}.}. Specifically, this choice fixes the value of $\lambda_{z,q}(\ell)$, so all variation between dynamical maps occurs in $\tau_{z,q}(\ell)$. Since there is only a single parameter in the maps, any dynamics between circuit layers that is not completely positive is not even positive (see Section \ref{sec:dynamMaps}). 

\subsection{\label{sec:InitializeNetworkCentral}Initializing the networks: central states}
The dynamics restricted by the choice of a single, fixed, two-local gate is convenient since many properties of the out-of-equilibrium behavior can be understood in terms of just a few functions of two-qubit states and single-qubit time evolution. To complement this simplicity of dynamics, we use ensembles of nearby initial states to obtain a statistical sample of each class of (finite-size) thermalizing or non-thermalizing behavior. 

The initial states we consider will consist of ensembles of states around one of four central state, {\bf CSP, CS1, CS2}, or {\bf CS3}. The central states are each tensor products of $N$ single-qubit states, defined as follows:
\begin{itemize}
    \item {\bf CSP (Pure)}: Each qubit is in a pure state, with $N-1$ in the ground state, $|0\rangle$, and one in the excited state, $|1\rangle$.
    \item {\bf CS1 (Thermal Resource)}: Each qubit is in a mixed state. $N-1$ qubits are in the state associated to a ``cold" temperature given by excited-state population $p_{c,{\rm therm}}$. The remaining qubit is in the state associated to a ``hot" temperature, with population $p_{h,{\rm therm}}>p_{c,{\rm therm}}$. The total energy is $E = (N-1)p_{c,{\rm therm}}+ p_{h,{\rm therm}}$. The difference in temperature creates an energy gradient in the system that can be exploited by the dynamics. The total entropy of this state is simply given by the sum of the entropy of the individual qubits. We show results for $p_{c,{\rm therm}}=0.1$, $p_{h,{\rm therm}}=0.4$
    \item {\bf CS2, CS3 (Inhomogeneous Thermal)}: Each qubit is in a mixed state, with generically different populations $p_{q}$, but constrained such that the total energy and the sum of single-qubit entropies are the same as for the thermal resource central state, {\bf CS1}. That is, 
    \begin{equation}
        \sum_q ^N p_q = (N-1)p_{c,{\rm therm}} + p_{h,{\rm therm}} 
    \end{equation} 
    and 
    \begin{align}
        \sum_q ^N S(p_q) =& (N-1)S(p_{c,{\rm therm}}) + S(p_{h,{\rm therm}})\,. \nonumber\\
    \end{align}
    
    These conditions do not uniquely specify a set $p_{i,{\rm inh}}$, so there are many inhomogeneous central states associated with a single central state of thermal resources. We show results below for two different cases that differ in the populations of the first eight qubits, where for central state {\bf CS2} $\{p_q\}= \{$0.02352335, 0.08, 0.28, 0.12, 0.12, 0.28, 0.08, 0.11647665$\}$, while for {\bf CS3} $\{p\}= \{$0.04340705, 0.12, 0.09, 0.15, 0.3, 0.14, 0.23, 0.02659295$\}$. For both cases, additional qubits added to the network all have a $p_q= 0.1$.
\end{itemize}

For any of the $N$-qubit central states defined above, there is a related equilibrium state that will be useful in this analysis. It is the state that maximizes the sum of single-qubit entropies, subject to the constraint imposed by the conserved charge, $E=\sum_{q=1}^N z_q=\sum_{q=1}^N (1-2p_q)$. The reference state is
\begin{equation}
\label{eq:rhoNequil}
    \bar{\rho}^{(N)}(E)\equiv\bar{\rho}\otimes\dots\otimes\bar{\rho}\,,
\end{equation}
where the tensor product is over $N$ copies of a single-qubit state
\begin{equation}
\label{eq:rhobar}
\bar{\rho}(E) = \frac{1}{2}\left(\mathbb{1} + \frac{E}{N}\hat{\sigma}_z\right)\,.
\end{equation}  
In other words, this is the state where any information about the initial differences in $\langle \hat{\sigma}_{z,q} \rangle$ has been moved entirely to the correlations. The excited state population of any qubit in $\bar{\rho}$ is $\bar{p}=\frac{1}{2}(1-\frac{E}{N})$.
    
We may also identify a family of single-qubit channels, $\{\bar{\Lambda}\}$, whose fixed point, $\rho^{*}$ as given in Eq.(\ref{eq:mapInvariantState}), is $\bar{\rho}(E)$. This family consists of all channels with parameters satisfying
\begin{equation}
\label{eq:thermalizingTau}
\frac{\tau_{z}}{1-\lambda_{z}}= \frac{E}{N}\,.
\end{equation}
The parameter controlling the $xy$-deformation, $\lambda$, is unconstrained by this requirement. Since the simple choice of $U_*$ has fixed $\lambda_z$, there is a unique $\tau_z$ associated to the fixed point $\rho^{*}=\bar{\rho}(E)$. Repeated application of any of this channel will take any qubit state and relax it to $\bar{\rho}(E)$. That is,
\begin{align} 
\label{eq:Lambdabar}
\bar{\Lambda}[\rho^*=\bar{\rho}] &= \bar{\rho}\,,\\
\bar{\Lambda}\circ \bar{\Lambda} \circ  \cdots\bar{\Lambda}[\rho] &\rightarrow  \bar{\rho}\,, \quad \forall \rho\,.
\end{align}
The action of a thermalizing random quantum circuit will relax the systems it acts on to configurations where the one-point measures of the qubits look identical, up to fluctuations. This is the $N$-approximation of the state $\bar{\rho}^{(\infty)}$. The dynamics of each qubit is then very close to that given by the channel $\bar{\Lambda}$, which is, of course, completely positive.

The state $\bar{\rho}$ provides several ways to quantify differences between the central resource states. For each qubit, one can compute the relative entropy between it and $\bar{\rho}$, and then sum over all qubits:
\begin{align}
\label{eq:curlyD}
 \mathcal{D}(\{\rho_q\},\bar{\rho})&=\sum_{q=1}^N D(\rho_q||\bar{\rho})\,,
\end{align}
where $D(\rho_q||\bar{\rho})={\rm Tr}[\rho_q(\log\rho_q-\log\bar{\rho})]$. Since the qubits begin in a product state, 
\begin{equation}
    \mathcal{D}(\{\rho_q(0)\},\bar{\rho})=D(\rho_1 (0)\otimes \ldots \rho_N(0)||\bar{\rho}^{(N)}).
\end{equation}
All central states involving mixed states ({\bf CS1}, {\bf CS2}, {\bf CS3}), have the same value of $\mathcal{D}$. The value is also independent of the number of qubits at $N=8$ and higher. 

The central states do differ in their trace distance from the $\bar{\rho}^{(N)}$, 
\begin{equation}
\label{eq:TrDistance}
    {\rm Tr}(\rho^{(N)},\bar{\rho}^{(N)})=\frac{1}{2}\sum_{q=1}^N(p_q-\bar{p})\,.
\end{equation} 
The central states are most similar at $N=8$ but diverge from each other as the number of qubits in the network increases.

\subsection{\label{sec:InitializeNetworkDistributions}Initializing the networks: initial state ensembles} 
For each of the central states above, we generate an ensemble of nearby states that will be used as initial conditions for the networks. To generate these we define 100 random circuits of depth 10 for each $N$, assuming full connectivity of the network. Each step of the circuit randomly assigns each qubit to a two-member neighborhood and applies an identical two-local gate, Eq.(\ref{eq:2Qunitary}), on each neighborhood. The family of initial states is the result of running each central state through this ensemble of random circuits. These ensembles provide the statistics to characterize fluctuations of the random (thermalized) systems at finite size, and to contrast the typical behavior of the systems with constrained dynamics. In the rest of the paper, quantities that are averaged over all initial states in a given central state ensemble are indicated by $\langle\cdot\rangle_{\rm ens}$. We include these 10 steps in the layer count, but keep in mind that the differing dynamics begins at layer 11.

The state space defined by this procedure can be partially visualized using Principal Component Analysis (PCA) on the set of single-qubit populations. PCA is a dimensionality reduction technique that projects a high-dimensional data set onto a lower-dimensional subspace constructed from the directions, or the principal components, $\mathbf{X}_{\text{PCA}}$, that capture the most variance in the data. Appendix \ref{sec:AppPCA} contains a more detailed description. 

\begin{figure*}
\includegraphics[width=\textwidth]{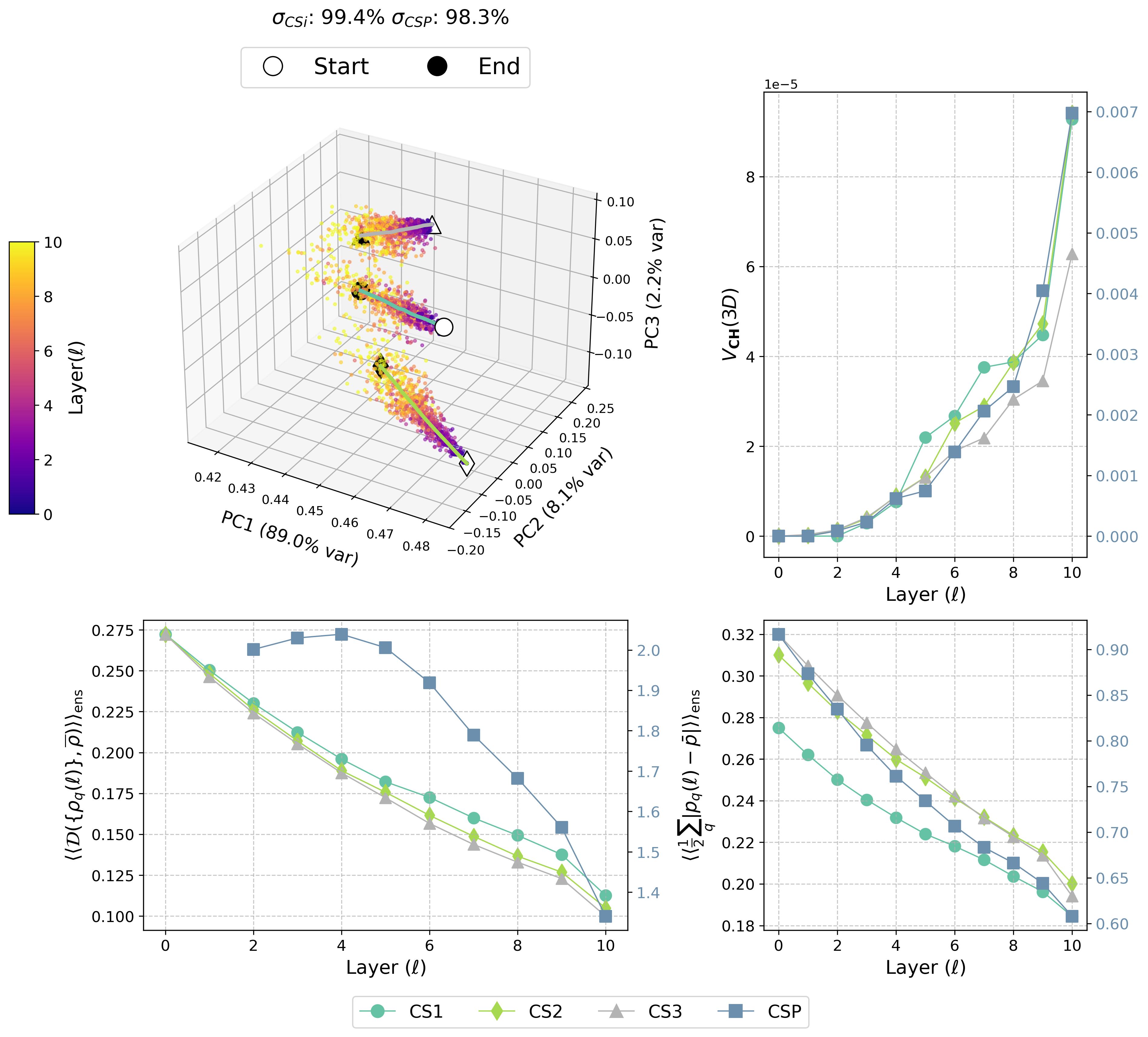}%
\caption{\label{fig:DistributionsPCA} Characteristics of the initial state ensembles generated about central states. The distribution about the pure central state, {\bf CSP}, is quite distinct from those about {\bf CS1}, {\bf CS2}, {\bf CS3}. {\bf CSP} cannot be accurately captured with the same principal components, so it is omitted from the top left panel. The numerical values needed in the top right and both bottom panels of a different magnitude, so {\bf CSP} uses the axes on the right side of each panel. The other central states use the left axes. {\bf Top left panel:} Visualization of the 12-qubit states in each ensemble, using the three most descriptive principle component axes. The points after 10 layers of random, fully connected circuit evolution using $U_*$ (yellow, clustered around the large filled symbol for the CS) are the set of initial states used in the rest of the paper. {\bf Top right:} The volume of the convex hull in the three-dimensional PCA, of the set of all $p_q$ points reached by a particular ensemble as a function of number of random circuit layers applied. {\bf Bottom left:} Evolution $\langle\mathcal{D}\rangle_{\rm ens}$ (a measure of relative entropy from the $\bar{\rho}^{(N)}$, Eq.(\ref{eq:curlyD})), averaged over the 100 states in each ensemble. {\bf Bottom right:} Evolution of the total trace distance from $\bar{\rho}^{(N)}$, Eq.(\ref{eq:TrDistance}), averaged over the 100 states in each ensemble.}
\label{fig:first_10_steps_IC}
\end{figure*}

To visualize the state space spanned by the ensembles of initial states about each central state, we first determine the principal components of the $N\times10\times100\times3$-dimensional data consisting of the excited state populations, $p_q$, of each of the $N$ qubits, over 10 circuit layers of random evolution, for 100 choices of random circuit, for each of the three central states containing mixed states ({\bf CS1}, {\bf CS2}, {\bf CS1}). Three axes capture more than 99\% of the variation in these data, so a three-dimensional visualization provides nearly all information about the spread of the states. The top left panel in Figure \ref{fig:DistributionsPCA} uses these three components to show how the ensembles about central states {\bf CS1}, {\bf CS2}, {\bf CS1} develop over the 10 random layers, for a network of 12 qubits. The ensemble about the pure central state is not well captured by the same principal components, so we perform independent PCA on that. 

PCA also provides an approximate measure of the state space spanned by each ensemble, through the convex hull volume, $V_{\rm CH}(\mathbf{X}_{\text{PCA}})$, the volume of the smallest convex set that contains all points. This volume is computed using the axes (and number of axes) appropriate to each data set. The top right panel of Figure \ref{fig:DistributionsPCA} shows how the volume of the convex hull evolves for 100 initializations of the central states for $N=12$ qubits, as a function of the number of random circuit layers applied. Finally, the two bottom panels show the evolution of the average trace distance and average relative entropy of the ensemble from the central state ($\ell=0$), to the ensemble used as initial states ($\ell= 10$). The steps of random evolution, on average, carry the states closer to the average thermal state. However, as the PCA trajectories in the top left panel show, the ensembles remain sufficiently separated that they can be treated as distinct families of initial conditions. The convex hull volumes of the initial ensembles differ by at most about 5\%, so the volume of states eventually reached by each family, given a particular non-random dynamics, can be sensibly compared to the volume reached by the other central state ensembles.

\subsection{Evolving the network}
\label{sec:EvolveNetwork}
To generate networks that stay away from the average thermal state, $\bar{\rho}^{(N)}$, we introduce several dynamically constructed circuits, where the arrangement of gates in the $\ell+1$ layer is determined by the extremum of a scalar function, $\mathcal{M}$, of the thermodynamic properties of the qubit network after layer $\ell$. No measurements are made, but the implementation of the rules requires an additional device beyond the network with exact knowledge of the initial state, and which maintains a record of the circuit. This device performs a calculation to extremize $\mathcal{M}$ and then applies the next set of gates based on the result. In this sense, while the dynamics of the full system is unitary, its dynamics are not self-contained, or closed: the effective Hamiltonian depends on time, and on some properties of the state. Since we have restricted to 2-local gates, the two-qubit density matrices after each layer of the circuit contain sufficient information to perform the maximization. However, it is crucial that no measurement is performed, since correlations in the full density matrix may still affect the overall state and evolution of the network.

Each network has physical connectivity $C$ labeling the number of nearest neighbors that each qubit has. This is the coupling graph in quantum computational literature - only nearest neighbors may be jointly evolved by a (unitary) gate.  For each circuit layer $\ell$, the applied gates define the interaction graph, $\mathcal{I}_{\ell}$, \cite{Bandic_2023}. We label the set of all possible interaction graphs, or neighborhood decompositions, of the $N$-qubit network as $\{\mathcal{I}^{(N)}_k\}$, and an interaction graph actually implemented at layer $\ell$ of the circuit as $\mathcal{I}^{(N)}_{\ell}$. Since each interaction graph is built of $N/2$ disjoint subgraphs, or neighborhoods, with two nodes each, we can write $\mathcal{I}^{(N)}_k=\sum_{n=1}^{N/2}\mathcal{I}^{(2)}_n$. The set of all possible two-node subgraphs that qubit $q$ can be a part of is denoted $\{\mathcal{I}^{(2)}_{j|q}\}$.

Each layer of the circuit is determined based on some scalar function of the state of the entire network, extremized over the possible ways of applying $N/2$ copies of $U_{*}$. That is, the extremization sweeps the set of neighborhoods, or interaction graphs, $\mathcal{I} = \{\mathcal{I}_1,\mathcal{I}_2,\ldots \}$, permitted by the physical spatial connectivity, $C$, of qubits. Each qubit is in exactly one neighborhood for every possible $\mathcal{I}_k$. 

Several of the update rules we consider reference the thermal state, $\bar{\rho}$, given by Eq.(\ref{eq:rhoNequil}) and Eq.(\ref{eq:rhobar}), and determined by the central state. We also investigate rules that use a more thermodynamic quantity, the extractable work. This compares the non-equilibrium free energy of the system in state $\rho$ to that of a reference equilibrium state at temperature $T$:
\begin{equation}
\label{eq:Wextractable}
    W^{\rm ex} = (E_\rho - T S_\rho) - ( E_{\sigma_{\rm th}} - T S_{\sigma_{\rm th}}) = T D(\rho || \sigma_{\rm th})\,.
\end{equation}
Here $E={\rm Tr}[\rho\hat{H}]$ is the energy of the system. $W^{\rm ex}$ captures the maximum amount of work that can be done if the system is brought into contact with a reservoir at temperature $T$, interacts with it, and then is removed \cite{Jarzynski:1997, Parrondo:2015,Kolchinsky:2025}. It is also the the amount of work required to bring the system to the thermal at temperature $T$.

For $W^{\rm ex}$ to quantify the utility of a given qubit state compared to its neighbors, we take $\sigma_{\rm th}$ to be a state constructed from course-grained information, the population or effective temperature of each single-qubit state, at a given time. That is, for qubit $j$, after layer $\ell$ of the circuit, the effective environmental temperature is
\begin{equation}
  \label{eq:landscapeTref}
  T_{ref|q_{j}}(\ell) = \frac{1}{N-1}\sum_{i\neq j}^N \frac{1}{\log\left[\frac{1-p_i(\ell)}{p_i(\ell)}\right]}\,.
\end{equation} 
The choice of temperature for each qubit is unambiguous given the conservation law and the restriction to single-qubit density matrices to define the temperature. However, there is some ambiguity in which network qubits should be used to define the temperature (i.e., only nearest neighbors). Extending this notion to larger subsystems is additionally complicated by correlations \cite{Kliesch2018,Mehboudi:2019,Alipour:2021,Allahverdyan:2022}. At the level of single-qubit $W^{\rm ex}$, we expect the conclusions we draw below to hold at a qualitative level in spite of these ambiguities.

Notice that any change in extractable work, $\Delta W^{\rm ex}$, will be determined by both the change in reference temperature and the change in relative entropy. For any classical process simultaneously evolving the system and reference thermal state, or quantum evolution that is either unitary or represented by a positive channel, the relative entropy can only decrease. An increase in extractable work then occurs only when the reference temperature decreases \cite{Kolchinsky:2025}. In the qubit networks considered here, different processes evolve the single-qubit state compared to the reference state, and quantum correlations that build up can lead to the single-qubit ``channels" describing evolution between two time steps that are not positive. So, positive $\Delta W^{\rm ex}$ for qubits in the network can arise for several reasons \cite{Akhouri_2023}.

Building on the states and thermodynamic measures defined above, we consider the following rules for assigning interaction neighborhoods (building the circuit) at each step:
\begin{itemize}
\item{\bf R1: Random:} At each layer, a decomposition of the network into 2-qubit neighborhoods is chosen at random from the neighborhoods allowed by the connectivity of the network.

\item {\bf R2: Subsystems avoid the global thermal state}\\
This rule maximizes the sum of distances of each qubit from the equilibrium state determined by the appropriate central state (see Eq.(\ref{eq:rhobar})). That is, for each trial interaction graph $\mathcal{I}_k$, $\text{Tr}$ from Eq.(\ref{eq:TrDistance}) is the scalar function to be extremized:
\begin{align}
\mathcal{M}_{\rm R2}(\mathcal{I}^{(N)}_k|\rho^{(N)},\bar{\rho})=\text{Tr}(\{\rho_{q|\mathcal{I}^{(N)}_k} (\ell+1)\},\bar{\rho})\nonumber\,,\\
\end{align}
where $\{\rho_{q|\mathcal{I}^{(N)}_k} (\ell+1)\}$ corresponds to the set of single-qubit states after the network evolves the full state $\rho^{(N)}$ via unitary $U^*$ applied using interaction graph $\mathcal{I}^{(N)}_k$, the $\ell+1$ layer of the circuit. The circuit that maximizes this function determines the interaction network actually applied in the $\ell +1$ layer of the circuit:
\begin{equation}
  \mathcal{I}^{(N)}_{\ell+1,{\rm R2}} = \arg\max_{\mathcal{I}^{(N)}_k} \mathcal{M}_{\rm R2}(\mathcal{I}^{(N)}_k|\rho^{(N)},\bar{\rho})\,.
\end{equation}

\item{\bf R3: Subsystems collectively maximize $\Delta W^{\rm ex}$.}\\ This rule maximizes the sum of the change in extractable work for each qubit. That is, for each trial interaction graph $\mathcal{I}_k$, the scalar function to be extremized is
\begin{align}
    \mathcal{M}_{\rm R3}(\mathcal{I}^{(N)}_k|\rho^{(N)})=& \sum_{q=1} ^{N} \Delta W_{q|\mathcal{I}^{(N)}_k} ^{\rm ex}\\
    &\equiv \Delta W^{\rm ex}_{N,\mathcal{I}^{(N)}_k}
\end{align}
where
\begin{equation}
\label{eq:DeltaWevolve}
  \Delta W_{q|\mathcal{I}^{(n)}_k}^{\rm ex}=W_{q|\mathcal{I}^{(n)}_k} ^{\rm ex}(\ell +1) - W_q ^{\rm ex}(\ell)\,,
\end{equation}
corresponds to the change in extractable work of qubit $q$ after the network evolves the full state $\rho^{(N)}$ via unitary $U^*$ applied using interaction graph $\mathcal{I}_k$.

The circuit that maximizes this function determines the interactions in the $\ell +1$ layer of the circuit:
\begin{equation}
  \mathcal{I}^{(N)}_{\ell+1,{\rm R3}} = \arg\max_{\mathcal{I}^{(N)}_k} \mathcal{M}_{\rm R3}(\mathcal{I}^{(N)}_k|\rho^{(N)})\,.
\end{equation}
   
\item {\bf R4: Subsystems with the most resources maximize their $\Delta W^{\rm ex}$ first.}\\ Rather than a global extremization of the sum of $\Delta W^{\rm ex}$, this rule lets the qubits with largest $\Delta W^{\rm ex}$ from previous step determine the interaction neighborhood that will give generate the largest $\Delta W^{\rm ex}$ at the next step. Maximizing a single qubit's neighborhood in this way determines a part of the circuit, or one subgraph $\mathcal{I}^{(2)}$. 

Consider the list of qubits, ordered by the values of their $\Delta W^{\rm ex}_q$ just after the $\ell$th layer, with $1$ labeling the qubit with smallest (likely negative) $\Delta W^{\rm ex}$ and $N$ labeling the qubit with highest $\Delta W^{\rm ex}$. Then  
\begin{align}
    &\mathcal{I}^{\rm R4}_{\ell+1}=\arg\max_{\mathcal{I}_{j,N}^{(2)}} \Delta W^{\rm ex}_{N|\mathcal{I}_{j,N}^{(2)}}(\mathcal{I}^{(2)}_{j|N},\rho^{(N)})\nonumber\\
    &+\arg\max_{\mathcal{I}^{(2)}_{j|N-1}\backslash \mathcal{I}^{(2)*}_N} \Delta W^{\rm ex}_{N-1|\mathcal{I}_{j,N-1}^{(2)}}(\mathcal{I}^{(2)}_{k|N-1},\rho^{(N)})\nonumber\\
    &+\arg\max_{\mathcal{I}^{(2)}_{j|N-2} \backslash \mathcal{I}^{(2)*}_N\cup\mathcal{I}^{(2)*}_{N-1}}\nonumber\\
    &\hspace{1.5cm}\left[\Delta W^{\rm ex}_{N-2|\mathcal{I}_{j,N-2}^{(2)}}(\mathcal{I}^{(2)}_{k|N-2},\rho^{(N)})\right]\nonumber\\
    &+\dots
\end{align}
where $\Delta W_{q|\mathcal{I}_k}^{\rm ex}$ is again given by Eq.(\ref{eq:DeltaWevolve}). Since each qubit is a member of only one neighborhood, the sum above has only $N/2$ non-zero terms.

\item {\bf R5: Strategy Mimic.}\\ As in the previous rule, the circuit is built up sequentially. For a fixed qubit $Q_i$ the algorithm sweeps all the possible neighbors allowed by the spatial connectivity $C$ and finds the neighbor who achieved the highest $\Delta W^{\rm ex}$ across the previous circuit layer ($\ell-1)$. The excited-state population difference between that successful qubit and its partner, $(\Delta p)_{{\rm target},i}$, becomes the target to determine $Q_i$'s partner in the next layer. The algorithm again sweeps the neighbors of qubit $Q_i$ qubit, seeking the resource qubit $Q_r$ with excited state population closest to
\begin{equation}
    p^{{(\rm target)},i}=p_{Q_i}-(\Delta p)_{,i}\,.
\end{equation}
The qubit that most closely matches this criteria is assigned to the interaction neighborhood of qubit $Q_i$. Formally, the resulting graph consists of two vertices ($v_1$, $v_2$) and one edge, so it is a complete graph with two vertices, $K_2(Q_i,Q_{r_i})$.

To choose the order of qubit pairings that will build up the next layer of the circuit, the algorithm orders qubits by the value of their achieved $\Delta W^{\rm ex}$ after layer $\ell$. The qubit with the {\it smallest} $\Delta W^{\rm ex}$, $Q_1$, goes first and the algorithm continues until all qubits are paired. That is,
\begin{align}
    \mathcal{I}_{\ell+1,R5}=K_2&(Q_1,Q_{r_1})\\
    &\cup K_2(Q_{\min \{2,3\}\backslash r_1},Q_{r_2})\cup \dots\nonumber
\end{align}
Since each qubit is a member of only one neighborhood, there are $N/2$ sub-graphs. This rule models the effects of imperfect local knowledge determining a strategy, partnering qubits with a neighbor that looks like a resource in terms of temperature, ignorant of the correlations that may be present.    
\end{itemize}
Notice that for the smallest connectivity, C2, the first step of any rule completely constrains the interaction structure for all the remaining qubits. 

Since the network is closed, each thermodynamic rule must balance the benefit of evolving some qubits far from $\bar{\rho}$, or to high $\Delta W_{\rm ex}$, with the cost of some qubits approaching $\bar{\rho}$ or having $\Delta W_{\rm ex}<0$. The room for these trade-offs can only decrease when more layers of the circuit result in more correlated qubits. This is especially clear for $R2$, since the 
monotonicity of trace distance and invariance of $\bar{\rho}$ under any evolution satisfying Eq.(\ref{eq:PCcond}) imply
\begin{align}
\label{eq:MonoT}
  \text{Tr}(\rho^{(N)}(t),\bar{\rho}^{\otimes N})&\leq \text{Tr}(\rho_1 (0)\otimes \ldots \rho_N(0),\bar{\rho}^{\otimes N})\,.\nonumber\\
  &
\end{align}
Similarly, the relative entropy also satisfies, 
\begin{align}
\label{eq:MonoRelS}
  D(\rho^{(N)}(t),\bar{\rho}^{\otimes N})&\leq D(\rho_1 (0)\otimes \ldots \rho_N(0),\bar{\rho}^{\otimes N})\,.\nonumber\\
  &
\end{align}
We may use the super-additivity property for bipartite systems to write an inequality between the left-hand side of Eq.(\ref{eq:MonoRelS}) and $\mathcal{D}$. First,
\begin{align}
  D(\rho^{(N-1)}||\bar{\rho}^{\otimes N-1}) + &D(\rho_1 ||\bar{\rho})\nonumber\\
  &\leq D(\rho^{(N)}||\bar{\rho}^{\otimes N})\,
\end{align}
Repeated partitioning off of single qubits from $\rho^{(N-1)}$ gives
\begin{equation}
\label{eq:partitionedRelS}
  \sum_q D(\rho_q (t)||\bar{\rho})\leq D(\rho^{(N)}(t)||\bar{\rho}^{\otimes N})\,.
\end{equation}
Finally, combining Eq.(\ref{eq:partitionedRelS}) and Eq.(\ref{eq:MonoRelS}) gives
\begin{align}
  \sum_q D(\rho_q (t)||\bar{\rho})&\leq  \sum_i D(\rho_q (0)||\bar{\rho})\nonumber\\
  \mathcal{D}(\{\rho_q (t)\}||\bar{\rho})&\leq\mathcal{D}(\{\rho_q (0)\}||\bar{\rho})\,.
\end{align}
The constraints imposed by the closed nature of the system lead to the development of locked interaction neighborhoods in many cases, as illustrated in the next section.

Even though the maximization for $R3$, $R4$ occurs at the level of extractable work, which depends only on the set of single-qubit populations $\{p_q\}$ (via Eq.(\ref{eq:Wextractable}) and Eq.(\ref{eq:landscapeTref})), the underlying correlation structure of the full system still matters. In Section \ref{sec:correlations}, we show explicitly how two-qubit correlations in the two-qubit density matrices affect the dynamics.

\section{Characterizing dynamics on co-evolving networks}
\label{sec:characterizing} 
In this Section we characterize the dynamics induced by the update rules in several ways. First (Section \ref{sec:EmergentGraph}) we define an emergent connectivity graph based on the frequency of each interaction neighborhood. This provides a notion of a Markovian limit of each dynamics. Next (Section \ref{sec:dynamMaps}), we consider each network as an ensemble of single-qubit states and single-qubit (open system) dynamics. We contrast the inhomogeneous subsystem behavior in non-thermalizing systems to the homogeneous, thermalizing case. Much of the behavior of the single-qubit dynamics is driven by two-qubit correlations that develop. In Section \ref{sec:correlations}, we look at properties of the magnitude and distribution of those correlations, demonstrating the characteristics that support the single-qubit ensemble behavior. Finally, in Section \ref{sec:Utility}, we define several measures of the possible relative utility of different non-thermalizing dynamics, including the accessible state space, mutual information, and the occurrence and persistence of positive changes in extractable work. 

\subsection{Emergent connectivity graph from update rules and initial state}
\label{sec:EmergentGraph}
\begin{figure*}[htb]
\includegraphics[width=\textwidth]{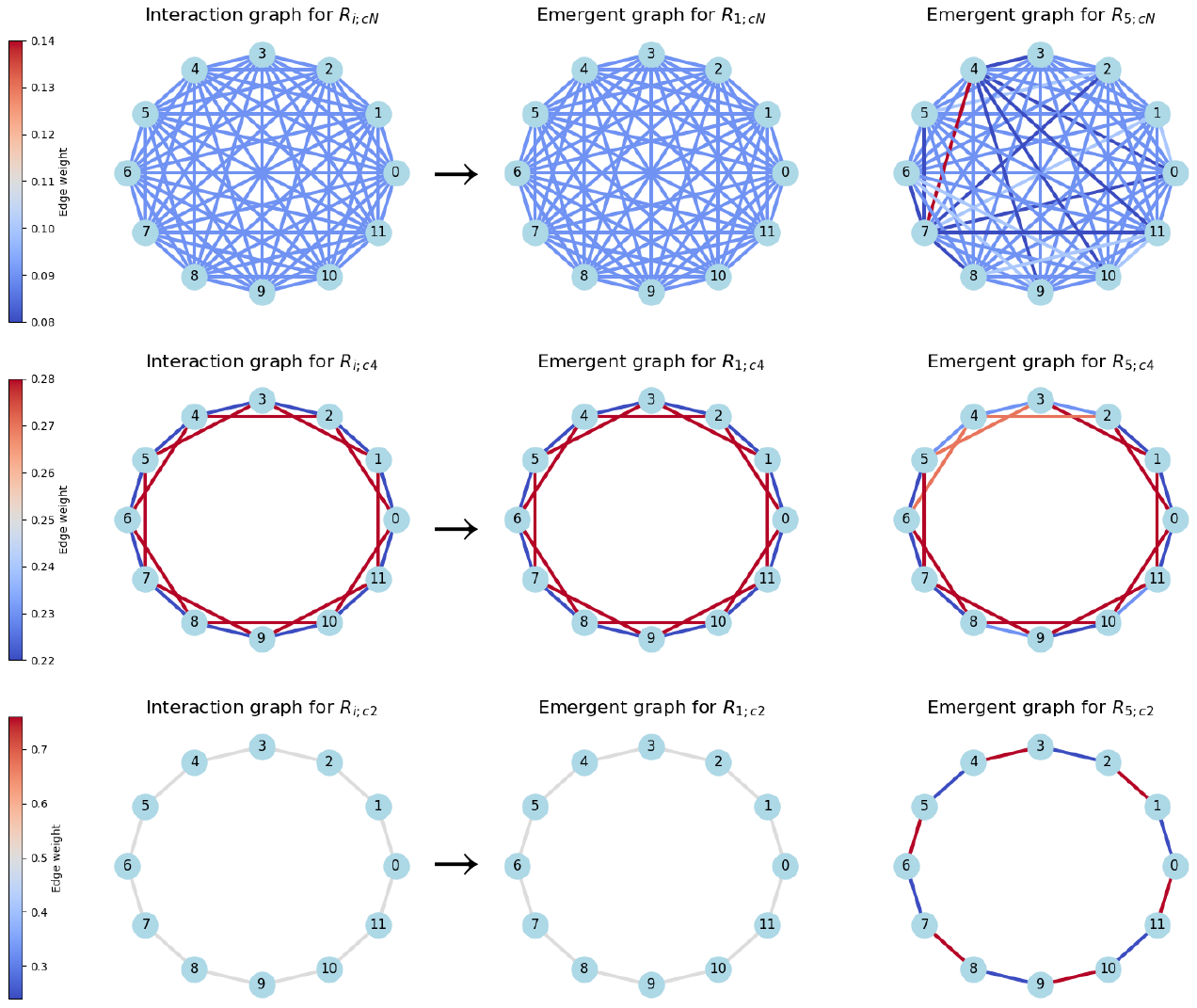}
\caption{\label{fig::emergent_network} Average initial interaction graphs (left column) and emergent networks for 12 qubits evolved with $R1$ (random, middle column) and $R5$ (strategy mimic, right column). The frequency of pairings chosen by each update rule, over 990 steps of evolution for each member of the {\rm CS3} ensemble, determines the emergent network. }
\end{figure*}

The update rules above, except the random rule, $R1$, generate preferential domains of interaction that depend on the rule and on the initial state. One way to visualize this is to use the frequency of each two-qubit interaction to construct a weighted graph defining an effective interaction network for each system. For example, suppose a network of $N$ qubits is fully connected. That is, the coupling graph (or physical connectivity) is $C=N-1$. Under the random evolution rule, $R1$, interactions are equally likely to be applied to each two-qubit pair (and every qubit participates in an interaction at every step). The emergent connectivity graph for a very high-depth random circuit would have equal weights of $1/(N-1)$ along each edge, making it indistinguishable from the coupling graph. The top row of Figure~\ref{fig::emergent_network} shows this coupling graph for 12 qubits (left), as well as the actual weights obtained after 990 steps of random evolution, $R1$, averaged over the 100 initial states in the ensemble about the inhomogeneous central state {\bf CS3} (middle panel). 

In contrast, the right panel in the top row shows the effective network that emerges after 990 steps of $R5$ (the strategy mimic rule), again averaged over the {\bf CS3} ensemble. Although the initial states are not identical, they do maintain quite a bit of the inhomgeneity of the central state (see the PCA analysis in the previous section. The inhomogeneous edge weights show that this evolution, in contrast to random, is affected by the initial state even after many circuit layers. A few nodes have effectively lower connectivity (around 6), but most qubits have some interactions with every other qubit on the network.

The interplay of initial state and dynamics becomes more intricate when the coupling is restricted. Consider 12 qubits with coupling graph $C=4$. In the set of all possible interaction graphs, restricted to only those graphs where each qubit participates in one and only one two-qubit neighborhood, some interactions occur more frequently than others. The first column in the second row of Figure~\ref{fig::emergent_network} illustrates this, showing the coupling graph for $C=4$, with an edge weight that indicates how often a particular edge occurs in the full set of interaction graphs. The middle panel in that row shows that 990 steps of random evolution, $R1$, averaged over the 100 initial states in the ensemble about an inhomogeneous central state nearly recovers the original statistical weights of the coupling graph. The right panel in the middle row again shows the unequal weight distribution that results from $R5$ evolution. 

Finally, the last row of Figure~\ref{fig::emergent_network} shows the emergent network for a $C=2$ coupling graph. In both the interaction graph and the network that emerges under random interaction, each qubit has equal probability to interact with either neighbor. But, under $R5$ evolution, the inhomogeneity of the initial state results in a strongly preferred interaction partner for each qubit. 

One helpful use of this visualization is to characterize the role of multi-partite correlations compared to the domain size for local interactions. That is, if the emergent network from some dynamics on $N$ fully connected qubits has an effective connectivity less than $N-\eta$ (which need not be an integer), then any difference in thermodynamic measures between networks with connectivity $N-\eta<C\leq N-1$ must be due to correlations that are long-range compared to the domain of the unitary. On the other hand, for the same initial central state and update rule, networks with connectivity $C<N-\eta$ will likely have thermodynamic properties that differ from each other primarily due to the limited transport of energy and temperature (excitation number) across the network. 

In the main body of the paper, we present results for $C=2$ networks. We find that as the connectivity increases, the $C=4$, then the update rules and 2-local interactions are not as successful at maintaining out-of-equilibrium dynamics. Appendix \ref{sec:C4} presents several results for $C=4$.

The emergent networks can also be used to understand the role of non-Markovianity of the dynamics, which affects how the information about the inhomogeneity of the central state persists over time. To extract this information, consider a class of Markovian circuits where the frequency of each interaction graph is fixed to match the relative weightings of an emergent network generated by some constrained dynamics and initial state, but the interaction graphs are randomly chosen at each layer. For example, consider 12 qubits with $C=2$ (the third row in Figure~\ref{fig::emergent_network}). There are just two possible interaction graphs, distinguished by whether qubit 0 is coupled to qubit 1 or qubit 11. Labeling these $\mathcal{I}^{(12)}_{0,1}$ and $\mathcal{I}^{(12)}_{0,11}$, they can be written formally as
\begin{align}
    I^{(12)}_{0,1}=&K_2(Q_0,Q_1)\cup K_2(Q_2,Q_3)\cup\dots\cup K_2(Q_{10},Q_{11})\nonumber\\
    I^{(12)}_{0,11}=&K_2(Q_0,Q_{11})\cup K_2(Q_{10},Q_9)\cup\dots\cup K_2(Q_1,Q_2)\,.
\end{align}
Then, if the set of unitaries (gates) corresponding to a layer built from one of these interaction graphs are labeled $U^{(12)}_{0,1}$ and $U^{(12)}_{0,11}$, the emergent network defines a class of Markovian circuits built by applying either unitary with some probability at each layer $\ell$ of the circuit:
\begin{align}
\label{eq:BernoulliCircuit}
    U(\ell)=\prod_{\ell} \left(b_{\ell} U^{(12)}_{0,1}+(1-b_{\ell})U^{(12)}_{0,11}\right)\,.
\end{align}
The $\{b_{\ell}\}$ form a Bernoulli process, with probability $p_b$ that $b_{\ell}=1$ determined from the emergent network. From the emergent networks shown in the last row of Figure~\ref{fig::emergent_network}, the random rule can be compared to a Bernoulli circuit with $p_b=0.5$, while the mimic rule and inhomogeneous initial condition can be compared to a Bernoulli circuit with $p_b=0.24$. For higher connectivity emergent graphs, the circuit structure will generalize such that the distributions for the $\mathcal{N}$ possible circuits $\mathcal{I}^{(N)}$ form a Bernoulli scheme. Clearly, the weights for the Bernoulli process depend on the initial state. So, this is a Markovian process, but one where the dynamics at all times is biased by the initial state.

Each realization of the circuit in Eq.(\ref{eq:BernoulliCircuit}) can be understood as stochastic evolution in the space of all possible trajectories (sequences) of elements in $\{\mathcal{I}\}$. These circuits resemble a $U(1)$ symmetric version \cite{Liu:2024vea} of the Brownian circuits used to study scrambling time in \cite{Brown:2012,Lashkari_2013,Onorati:2016met,Zhou_2019,Bentsen_2021, Jian:2022}. It would be interesting to clarify further the distinction between the Gaussian distributed couplings used there and the Bernoulli distribution that more naturally follows from the symmetries and constraints here. Section \ref{sec:results} will compare the Bernoulli circuit dynamics with those that arise using the update rules from Section \ref{sec:BernouliResults}.

\subsection{Ensembles of single qubit states and dynamics}
\label{sec:dynamMaps}
The emergent networks above already indicate that under at least some evolutions $R2-R5$, information about the initial state is still locally accessible at late times, and so at least some parts of the system are not thermalized. The ensemble of the states and dynamics of the smallest subsystems, individual qubits, characterize the non-thermalizing nature of these networks in greater detail.

\begin{figure*}[ht]  
  \centering
  \begin{minipage}{0.24\textwidth}
    \includegraphics[width=\textwidth]{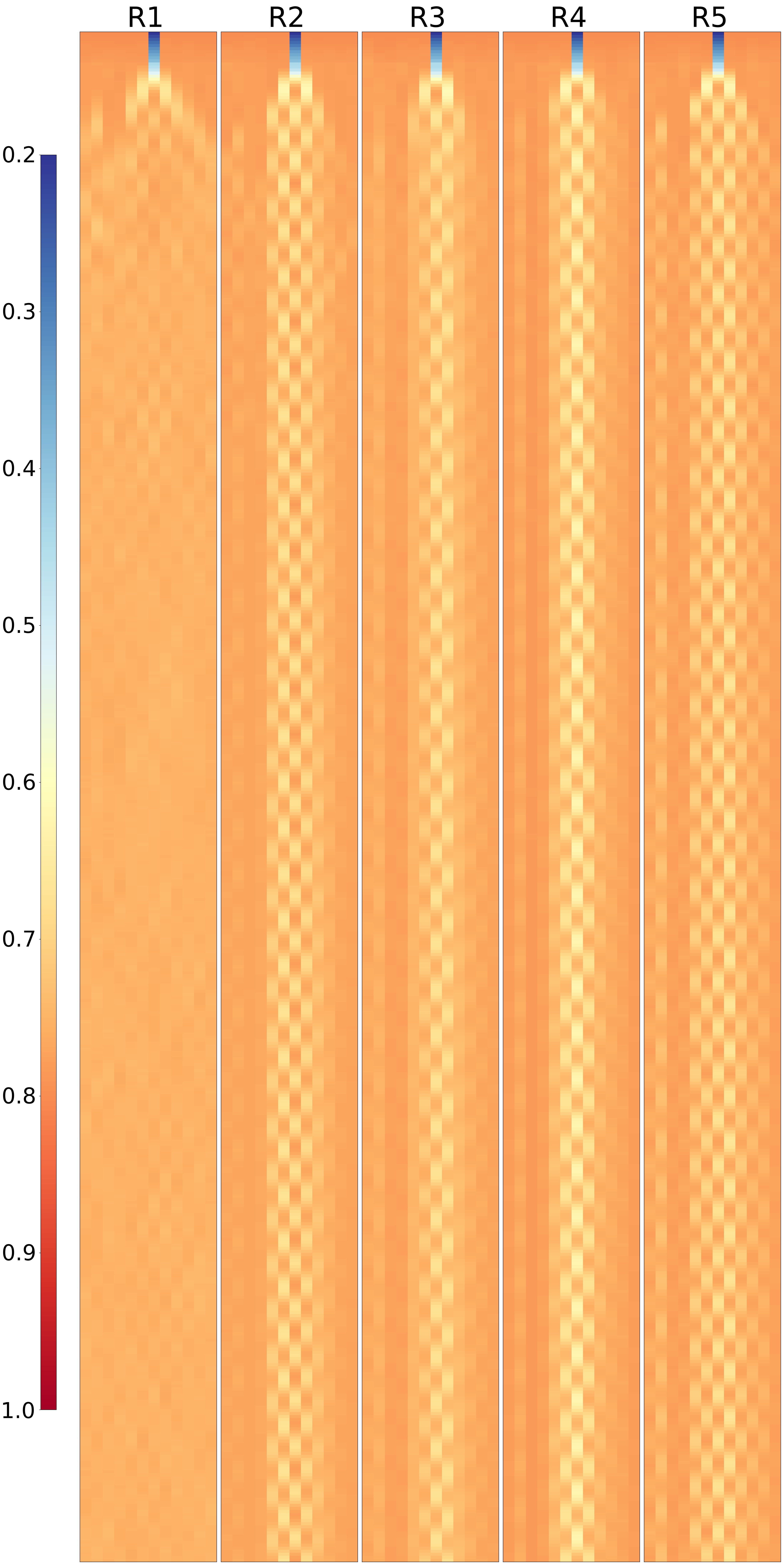}
    \caption*{CS1}
    \label{fig:CS1_heatmap}
  \end{minipage}
  \hfill
  \begin{minipage}{0.215\textwidth}
\includegraphics[width=\textwidth]{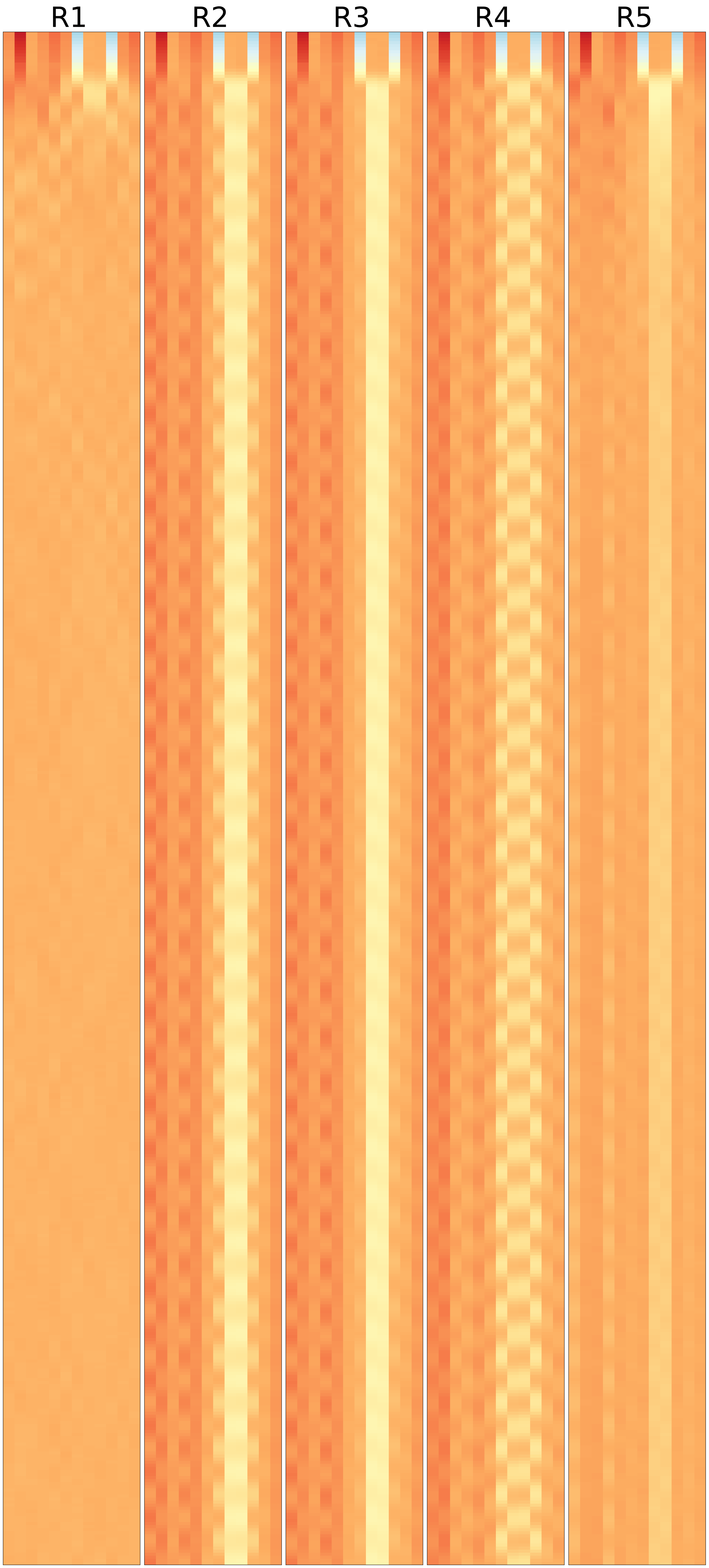}
    \caption*{CS2}
    \label{fig:CS2_heatmpa}
  \end{minipage}
  \hfill
  \begin{minipage}{0.215\textwidth}
    \includegraphics[width=\textwidth]{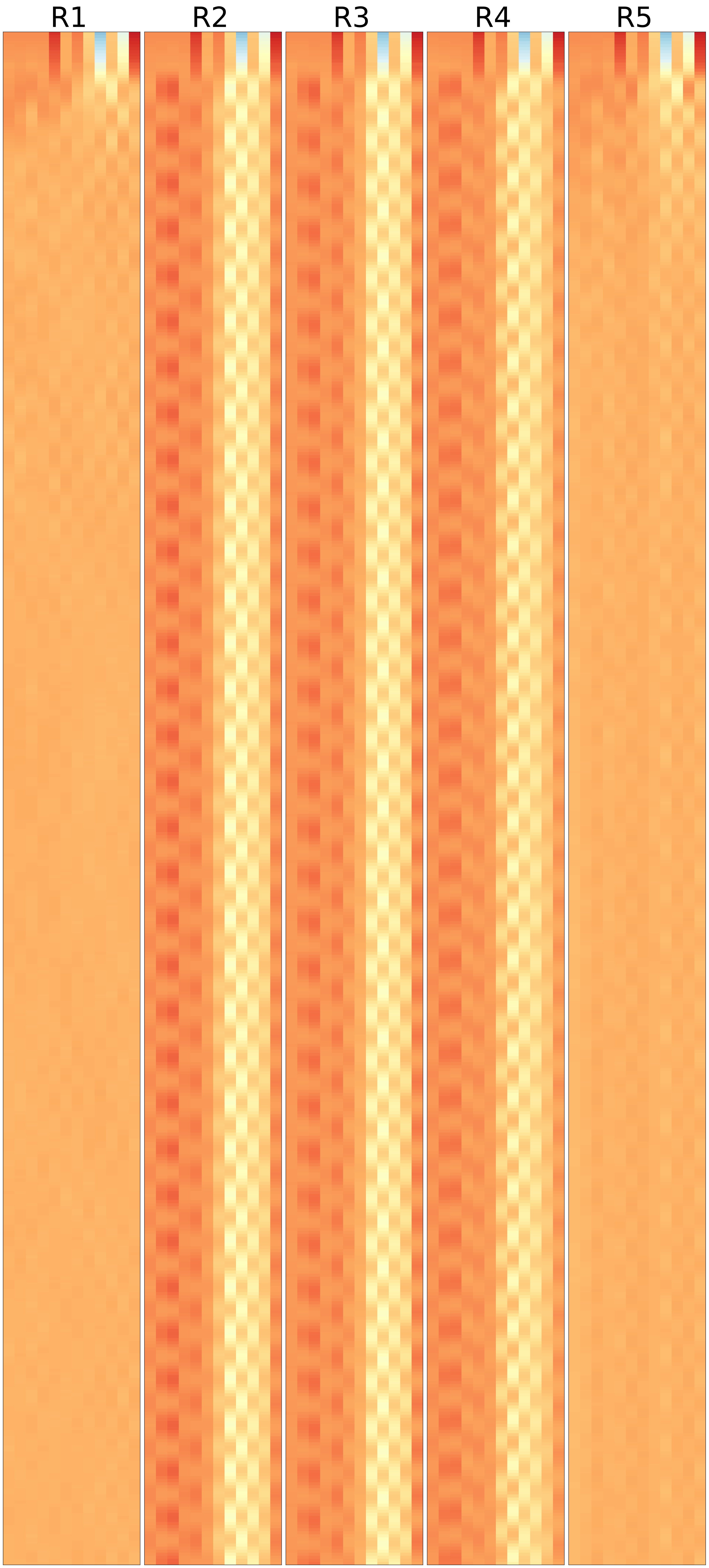}
    \caption*{CS3}
    \label{fig:CS3_heatmap}
  \end{minipage}
  \hfill
  \begin{minipage}{0.247\textwidth}
    \includegraphics[width=\textwidth]{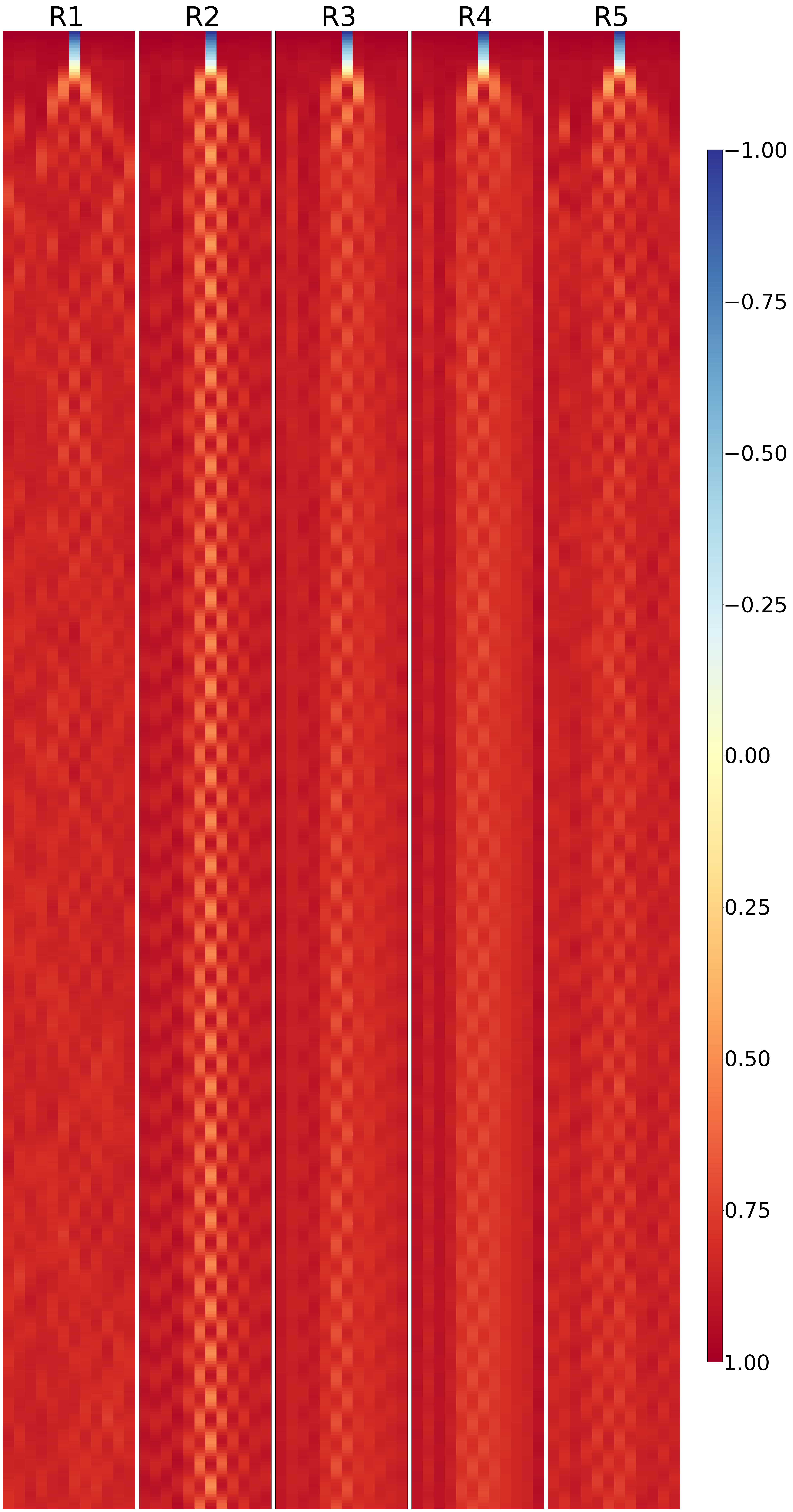}
    \caption*{CSP}
    \label{fig:CSP_heatmap}
  \end{minipage}
  \caption{Heatmap of $\langle z_{q}(\ell)\rangle_{\rm ens}$, for the four central states and all update rules, as a function of circuit layer. Within each sub-panel, labeled by a central state and an update rule, pixels represent the individual qubits at each layer of the circuit. The initial state is shown at the top, with increasing circuit layer running down the page. These are 12-qubit networks with connectivity $C=2$. Magnitudes for central states {\bf CS1}, {\bf CS2}, and {\bf CS3} use the left-hand fixed color bar, while {\bf CSP} qubits have a different range of $\langle z_{q}(\ell)\rangle_{\rm ens}$, shown in the color bar on the right. \label{fig:temperatureHeatMaps}} 
\end{figure*}

Figure \ref{fig:temperatureHeatMaps} shows the evolution of $\langle\langle\sigma^z_{q}(\ell)\rangle\rangle_{\rm ens}$, the individual qubit $\langle\sigma^z(\ell)\rangle$ at a given circuit layer, for all rules, averaged over the ensemble associated to each central state. The excited state populations, $p_q$, and temperatures, $T_q$, are closely related to $\langle\sigma^q\rangle$ (see Eq.(\ref{eq:qdiag})) and show similar behavior. The figure shows that under random evolution, the single-qubit states become more homogeneous. Update rules $R2-R4$ maintain inhomogeneous states even after many layers of evolution, frequently with locked neighborhoods forming when qubits have a preferred interaction partner. The approximate optimization rule, $R5$, displays behavior that sometimes looks like $R1$, and sometimes displays locked states or neighborhoods, depending on the central state.

More detail in the different dynamics of the networks can be seen in the ensembles of propagator maps describing single-qubit evolution between circuit layers. Given the restricted class of dynamics described in Section \ref{sec:symmetry}, we can derive several useful equations describing single-qubit dynamics that hold for all rules $R1-R5$. 

Across layer $\ell$, the unitary evolution of a two-qubit neighborhood containing qubits $Q_{a}$ and $Q_b$ is
\begin{equation}
    \rho_{ab,\ell}=U_*\rho_{ab}(\ell-1)U_*^{\dagger}\,.
\end{equation}
The conditions in Eq.(\ref{eq:PCcond}), Eq.(\ref{eq:qdiag}) imply that the density matrix for $Q_a$ and $Q_b$ is always of the form
\begin{align}
    \label{eq:twoQrho}
    \rho_{ab}(\ell)=\frac{1}{4}&\left(\mathbb{1}^{(4)}+z_{a,\ell}\sigma_z\otimes\mathbb{1}^{(2)}+z_{b,\ell}\mathbb{1}^{(2)}\otimes\sigma_z\right.\nonumber\\
    &\left.+C^{(zz)}_{ab,\ell}\sigma_z\otimes\sigma_z+{C^{xx}_{ab,\ell}}(\sigma_x\otimes\sigma_x+\sigma_y\otimes\sigma_y)\right)\,.
\end{align}
Then the open-system evolution of just one of the qubits, e.g. $Q_{a}$, across the layer is given by $\Lambda_{q}(\ell,\ell-1)$, where
\begin{equation}
\rho_a (\ell)=\Lambda_{a}(\ell,\ell-1)\circ\rho_a(\ell-1)\,.
\end{equation}
As in Section~\ref{sec:networks}, $\Lambda_{q}$ is a four-by-four matrix acting on the Bloch vector corresponding to state $\rho_a$. Given the symmetries imposed on the states and evolution, only the $z$-component of the Bloch vector, $z_a$ changes. Its evolution can be written in terms of the map components as
\begin{equation}
\label{eq:zEvolve}
    z_{a,\ell}= \lambda_{z,a}(\ell) z_{a,\ell-1}+ \tau_{z,a}(\ell)\,.
 \end{equation}
Furthermore, the map components only depend on the $z$-component of the Bloch vector for $Q_b$ and the correlation between the qubits before the unitary is applied, $z_{b,\ell-1}$ and $C_{ab,\ell-1}$, respectively, as well as the unitary rotation angle applied by the gate at layer $\ell$. In other words, with the restriction to a single, one-parameter unitary, Eq.(\ref{eq:2Qunitary}), the map parameters are
\begin{align}
\label{eq:lambdaTau}
    \lambda_{z,a}(\ell)& = \cos^2 (\theta)\nonumber\\
    \tau_{z,a}(\ell) &= z_{b,\ell-1} \sin^2 (\theta) + {C^{xx}_{ab,\ell-1}} \sin(2\theta)\,.
\end{align}
Since $\theta$ is fixed, all variation in the propagator maps occurs in the $\tau_{z,q}(\ell)$. The global $U(1)$ constraint requires
 \begin{equation}
     \sum_q z_{q}(0) = \sum_q (\lambda_{z,q}(\ell)z_{q,\ell-1} + \tau_{z,q}(\ell) ) = E \,.
 \end{equation} 
 
In addition, there is a fixed value $\tau_z=\bar{\tau}_z$, determined by the central state, that corresponds to the thermalizing map given by Eq.(\ref{eq:thermalizingTau}). For all mixed state central states ({\bf CS1}, {\bf CS2}, {\bf CS3}), $\bar{\tau}^{\rm mixed}_z=0.0324$. For {\bf CSP}, $\bar{\tau}^{\rm pure}_z=0.0360$. 

The simple form of the unitary reduces the conditions for complete positivity and positivity (the two lines of Eq.(\ref{eqn::CPcond})) to the same constraint:
\begin{equation}
\label{eq:oneParamCP}
    (z_{b,\ell-1} \sin^2 (\theta) + {C^{xx}_{ab,\ell-1}} \sin(2\theta))^2\leq \sin^4\theta\,.
\end{equation}
Any propagator maps that break (complete) positivity do so because of sufficiently large correlations $C^{xx}_{ab,\ell-1}$ between the two qubits. For a fixed value of $\theta$, and Bloch vectors components $z_q$ bounded by the largest $z_q$ of the central state, the value of $C^{xx}$ required to break (C)P can be computed. Figure \ref{fig:nonCPconditions} in the Appendix shows more detail.

Figure \ref{fig:tz_time_avg} shows an example of the evolution of $\tau_{z,q}(\ell)$ each of 12 qubits, for one initial state in the {\bf CS1} ensemble evolved with each update rule. After layer $\ell=250$, the trajectories have been smoothed over a window of 30 circuit layers to reduce noise. Under random evolution, $R1$, the dynamics of the qubits are indistinguishable after a few circuit layers, with the late time dynamics of each fluctuating in a noisy way $\bar{\tau}$. For all other rules, the qubit maps are generically distinguishable from each other, with a regular period of oscillation inherited from the $\theta=\pi/15$ angle in the unitary gate. In addition, by smoothing the late time points over a moving $\Delta\ell=30$ window, it is clear that the qubit dynamics oscillate around values stabilized on one side or the other of the fixed-point channel.
\begin{figure*}[hbt]
    \centering
    \includegraphics[width=\textwidth]{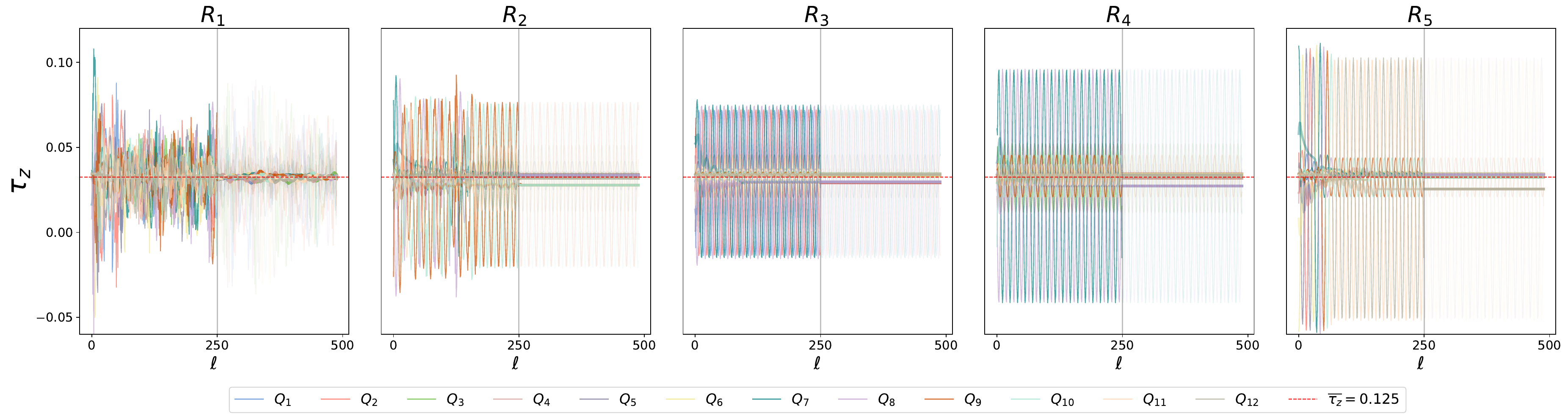}
    \caption{The evolution of $\tau_{z,q}$ of a single initial state from the {\bf CS1} ensemble. After layer $\ell=250$, the full dynamics is shown in lighter weight in the background, with dynamics smoothed with a moving time-averaging window of 30 circuit layers over-plotted in heavier-weight lines. The horizontal red-dashed line shows the value of $\tau_z$ that corresponds to the thermalizing map, Eq.(\ref{eq:thermalizingTau}), for this central state.  \label{fig:tz_time_avg}}
\end{figure*}

For any single circuit, across any layer, consider the dynamics averaged over the set of qubits $Q$ in the network $\langle\Lambda(\ell+1,\ell)\rangle_Q$. The dynamics of any single qubit can be written
\begin{equation}
    \Lambda_{q}(\ell,\ell-1) =\langle\Lambda(\ell,\ell-1)\rangle_Q + \delta \Lambda_{q}(\ell,\ell-1)\,.
\end{equation}
If the map parameters for all the qubits converge as the depth of the circuit, $\mathcal{L}$, increases, then the effective behavior of that update rule is equilibrating and
\begin{equation}
    \lim_{\mathcal{L}\rightarrow \infty} \langle|\Lambda_{q_a}(\mathcal{L},\mathcal{L}-1) - \Lambda_{q_b}(\mathcal{L},\mathcal{L}-1)|\rangle = \epsilon(N)\,,
\end{equation}
where the fluctuations $\epsilon(N)$ depend on system size and, for the dynamics in this paper, are entirely in the $\tau_z$ component. Similarly, the populations (and effective temperatures and $\langle\sigma^z\rangle$) of each of the qubits will all converge to the average population defined by the initial state:
\begin{equation}
    z_q = \frac{\langle\tau_z\rangle \pm\delta \tau_{z,q}}{1-\lambda_{z,q}} =  \bar{z} \pm \delta z_q=\frac{E}{N}\pm \delta z_q\,.
\end{equation} 
 In other words, in equilibrating systems, the state will approach $\bar{\rho}^{(N)}$ (Eq.~(\ref{eq:rhoNequil})) and the maps will approach $\bar\Lambda$, Eq.~(\ref{eq:Lambdabar}). If, however, the standard deviation in the map parameters stays large after many circuit layers, the individual dynamical maps, and therefore the single-qubit states, will not converge.
 
\begin{figure*}[hbt]
\includegraphics[width=\textwidth]{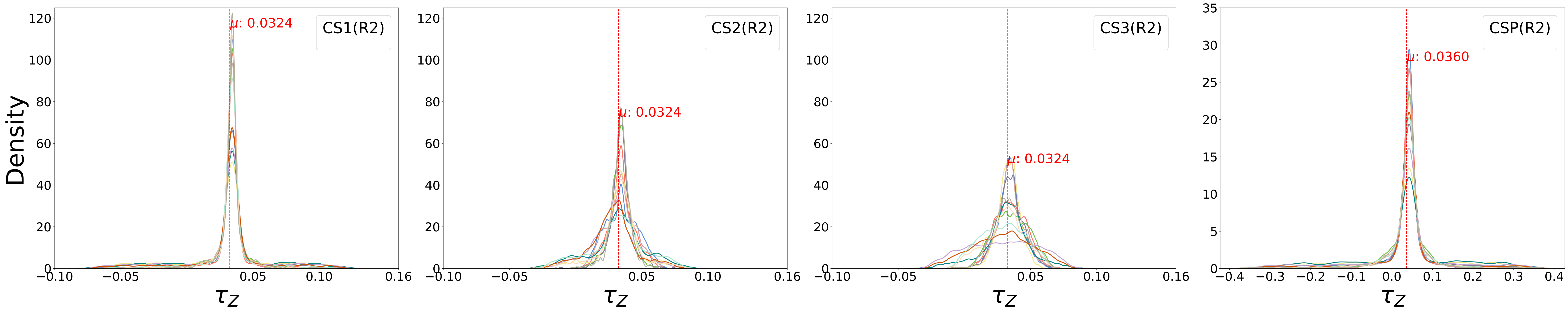}
    \caption{Normalized distribution of $\tau_{q,z}$, for each individual qubit, including all members of the ensemble of circuit evolutions for the central state. The vertical red-dotted line shows the calculated mean, $\mu$, which agrees with the value from the thermalizing map. All networks had connectivity $C=2$.  \label{fig:TauzDistributions}}
\end{figure*}

 Figure \ref{fig:TauzDistributions} shows the distribution of the propagator map parameter, $\tau_{z,q}$, for each qubit in the networks evolving with $R2$, a non-random evolution. The distributions for each central state contain $\tau_{z,q}(\ell)$ for all initial conditions associated with the central state, across all layers of each circuit. Several features of Figure \ref{fig:tz_time_avg} are reflected in Figure \ref{fig:TauzDistributions}. For example, most qubits in the non-random rules underwent dynamics with $\tau_z$ taking oscillating values on both sides of $\bar{\tau}_z$. However, they fluctuated about average values different from the thermalizing map, $\bar{\tau}_z$. This is apparent in the figure, where each distribution has mean equal to $\bar{\tau}_z$, but the distributions are skewed such that the peak does not align with the mean. 

\begin{figure*}[htb]
    \includegraphics[width=\textwidth]{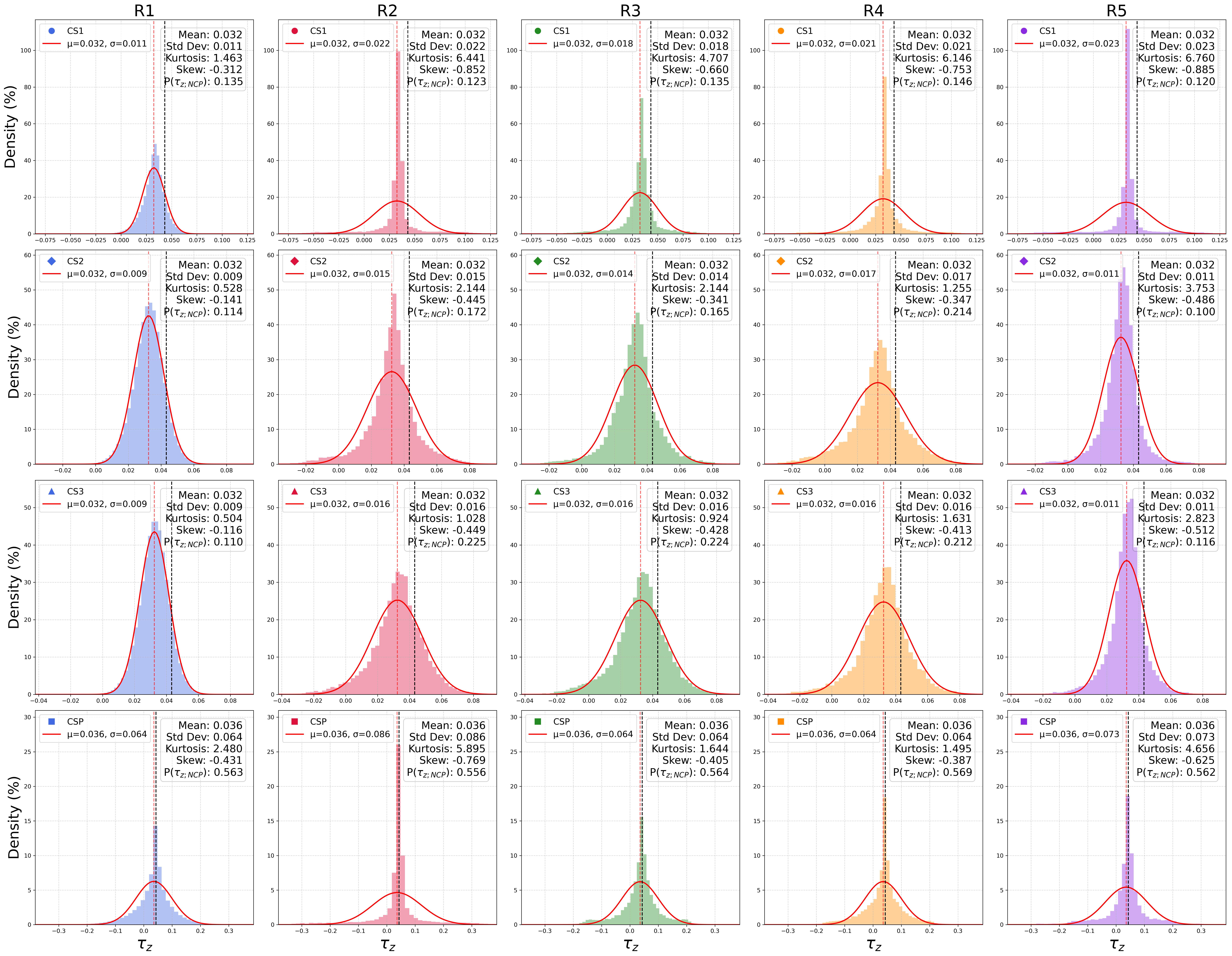}
    \caption{The distribution of shifts, $\tau_z$, in the propagator maps for 12-qubit networks with connectivity $C_2$. For a given central state, the distribution contains $\tau_z$ for each qubit, across circuit layers 300-500, for all initial states in the central state ensemble. From left to right in each row (fixed central state), the panels show the distribution generated by update rules $R_1 - R_5$. The red dotted line shows the value of $\tau_z$ associated with the (central state-dependent) thermalizing map, Eq.(\ref{eq:thermalizingTau}). The black vertical dashed line shows the minimum positive value of the $\tau_z$ that would break complete positivity of the dynamical/propagator map. Since the condition is $|\tau_z|>1-|\lambda_z|$, there is a corresponding negative value of $\tau_z$ that can also break the condition.  The solid red line is a normal distribution with the same mean and variance as the data.\label{fig:tzEachIcEachRule}}
\end{figure*}

Finally, Figure \ref{fig:tzEachIcEachRule} shows the distribution of map parameters over all initial members of each central state ensemble, collected from circuit layers 300-500 to capture the non-equilibrium steady state distribution for each update rule and each central state. Each row shows results for a particular central state and each column shows a particular update rule. As expected, the means of all distributions agree with the values of $\bar{\tau}_z$ calculated for each central state. For the three mixed central states ({\bf CS1}, {\bf CS2}, {\bf CS3}), random evolution ($R1$) shown in the first column has a distribution closest to Gaussian. All non-random rules have distributions with a higher standard deviation and higher kurtosis. Since for each circuit large individual $\tau_z$ can only occur at the price of many small $\tau_z$, the non-equilibrium steady state distributions with less homogeneity must also have a large kurtosis. The pure central state (bottom row) shows a different behavior, with even $R1$ evolution generating a distribution away from Gaussian. This is an indication that the very constrained dynamics used here is not sufficient to thermalize the pure central state ensemble. We will see additional evidence of this below, and for this reason we do not use the $R1$ ensemble to calibrate the thermalizing limit for {\bf CSP}.

The black dashed line in each histogram shows the magnitude of $\tau_z$ above which a propagator map is not completely positive, via Eq.(\ref{eqn::CPcond}). The total fraction of propagators with $|\tau_z|$ above this value is shown in the inset of each figure. The implications of these dynamics is discussed further in Section \ref{sec:BernouliResults}.

\subsection{Ensemble of two-qubit correlations}
\label{sec:correlations}
As Eq.(\ref{eq:lambdaTau}) showed, the behavior of the single-qubit dynamics is driven by the distribution of correlations as well as by the states of the interaction partners. Examining the distribution of the size of correlations in the different network dynamics of $R1-R5$ provides additional information about how the network supports inhomogeneous single-qubit dynamics. 

The previous section demonstrated that for the non-random rules the distribution of single-qubit dynamics stabilized away from the thermalizing channel, with different qubits experiencing different dynamics. The expression for $\tau_z$, Eq.(\ref{eq:thermalizingTau}), suggests that this result for dynamics of the smallest subsystems must be accompanied by a related stabilization in the correlations between subsystems. To check this connection, consider the sum of correlation magnitudes in all two-qubit density matrices, Eq.(\ref{eq:twoQrho}), at a fixed circuit layer, averaged over the initial states of a central state ensemble:
\begin{equation}
 \label{eq:CorrTotal}
   C_{\rm total}(\ell)\equiv \left\langle\frac{1}{4}\sum_{(a,b)}|C_{ab}^{xx}(\ell)|\right\rangle_{\rm ens}\,.
\end{equation}
Since the networks spend most of the time in their steady states (See Appendix \ref{sec:appendFigures}, Figure \ref{fig:Corr_growth}), statistics of the correlations will largely reflect properties of the non-equilibrium steady states of each class of networks.

\begin{figure*}
    \centering
\includegraphics[width=\textwidth]{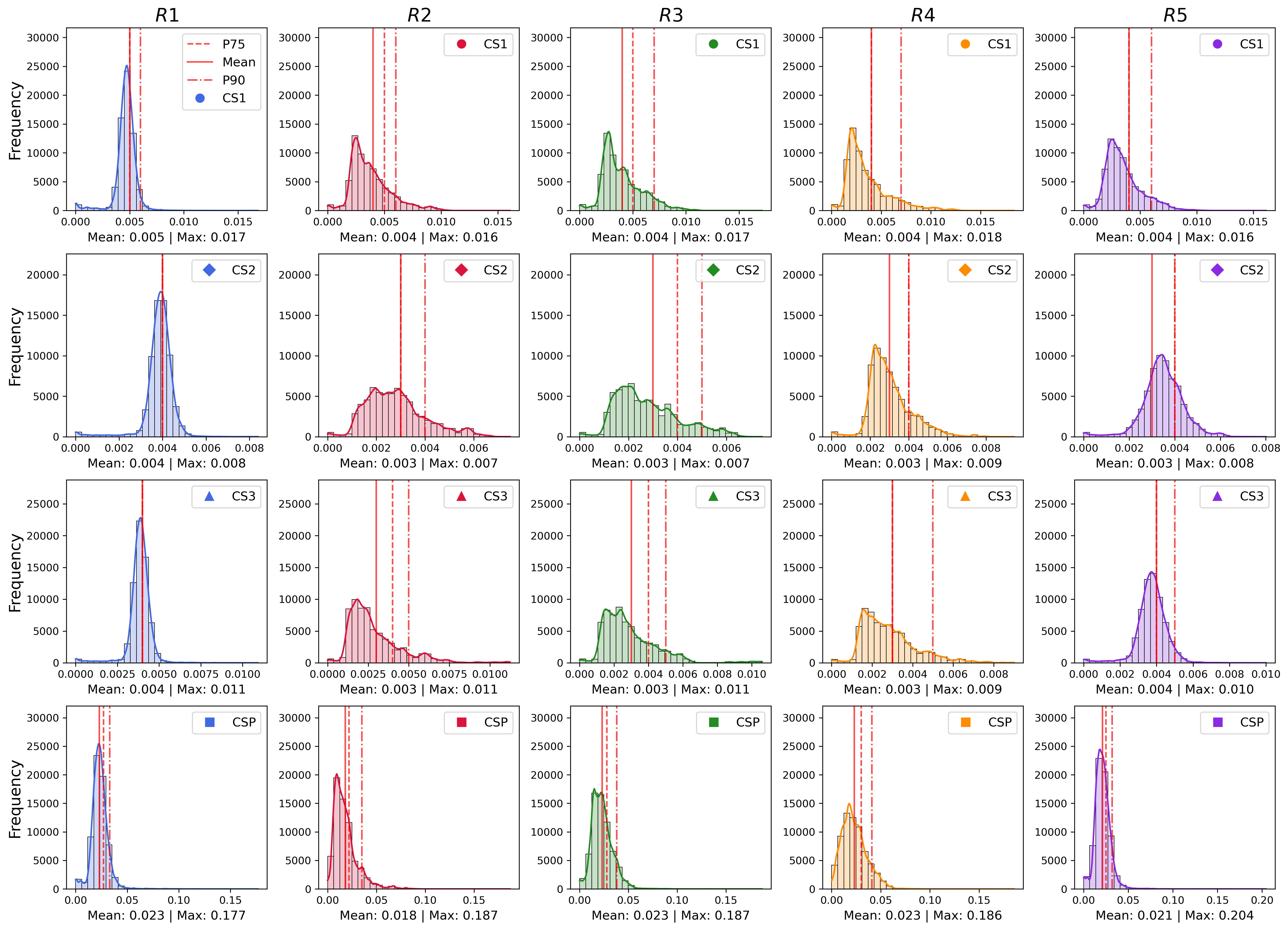}
    \caption{The distribution of the correlation magnitudes, $\frac{1}{4}|C^{xx}_{ab}|$, between all possible pairs of qubits ($Q_a$, $Q_b$) on 12-qubit networks with $C=2$ connectivity. The distributions include correlations at all circuit layers and for all members of the central state ensembles. The mean and maximum amplitudes are noted under each plot, and the solid line shows the mean. The dashed and dot-dashed vertical lines show the values below which $75\%$, and $90\%$ of values occur, respectively. For the random rule, $R1$, all the distributions are closest to Gaussian, with central-state dependent mean and the variance. For the non-random rules, the distributions depart significantly from Gaussian. \label{fig:Corr_12_Q}}
\end{figure*}

Figure \ref{fig:Corr_12_Q} shows the distributions of $C_{\rm total}(\ell)$, confirming that the single-qubit dynamics seen in Section \ref{sec:dynamMaps} is indeed supported by similar behavior in the correlations. The ensembles with more inhomogeneous propagator maps are generated by dynamics with more inhomogeneous (less Gaussian) distributions of correlations between qubits. The distributions for the non-random rules indicate that most qubits are more weakly correlated than in the thermal equilibrium states, while a few qubits share order-of-magnitude stronger correlations. The next section will use a graphical representation of mutual information to elaborate on that result. 

Of course, the diagonal and off-diagonal components of the density matrix do not evolve independently. Some additional insight into the structure of correlations can be obtained from examining a constraint between the components. The unitary rotations generated by the gates must preserve the trace of the density matrix and preserve purity, $\gamma = Tr(\rho^2)$. Given the $U(1)$ symmetry imposed, this also means that the trace and purity within each energy (excitation) subspace is preserved. 

To see what this implies, recall that each network begins from a diagonal parent central state. Writing a diagonal density matrix within a subspace of excitation number $n$ as $\rho_{(n)}= \text{diag}(l_1,l_2,\ldots,l_d)$, where $d = \binom{N}{n}$ is the dimension of the subspace, the initial purity of subspace $n$ is 
\begin{equation}
\gamma_{(n)}(0) = \sum_{m=1} ^d l_m^2\,.
\end{equation}
Evolution will generate off-diagonal terms within the subspace, so that the subspace purity after $\ell$ layers of the circuit is
\begin{align}
\gamma_{(n)}(\ell) &= Tr[\rho_{(n)}(\ell)^2 ] \nonumber\\
&= \sum_{m=1} ^d |\rho_{(n),mm}(\ell)|^2 + \sum_{m\neq k} |\rho_{(n)mk}(\ell)|^2\,.
\end{align}
Since this purity is conserved, the magnitude of off-diagonal terms is constrained to satisfy 
\begin{equation}
    \label{eq:purityconstraint}
 \sum_{m\neq k} |\rho_{(n),mk}(\ell)|^2  = \sum_{m=1} ^d l_m^2-\sum_{m=1} ^d |\rho_{(n)mm}(\ell)|^2
\end{equation}
For all networks that began in a particular central state, $\sum_{m=1} ^d l_m^2$ is fixed, and so is $\sum_{m=1} ^d \rho_{(n)mm}(\ell)$. But, there is room for networks to evolve different distributions of correlations, including differing overall magnitudes, as long as Eq.~(\ref{eq:purityconstraint}) is satisfied. The difference in $\sum_{m=1} ^d |\rho_{(n)mm}(\ell)|^2$ between networks determines how different the correlations can be.

\subsection{The utility of staying away from equilibrium}
\label{sec:Utility}
The measures above can diagnose and quantify the degree to which the qubit networks retain local information about the initial state at late times, how quickly a steady state is reached (if at all) and how that steady state compares to the equilibrium, maximum entropy (thermal) state defined by the initial conditions. However, within the very broad class of non-thermalizing systems we would like to know which dynamics is most thermodynamically interesting. In this section, we introduce several measures that will help determine the relative thermodynamic utility of dynamics that remains inhomogeneous. 

\subsubsection{State space explored}
\label{sec:StateSpace}
The state space explored by quantum systems determines all thermodynamic and information theoretic measures of the dynamics. While the evolution of the full state is not easy to visualize, the constrained nature of the dynamics used here (as demonstrated in Eq~(\ref{eq:purityconstraint}), for example) means that even the ensemble of single-qubit states is quite informative. The principal component analysis, PCA, introduced in Section \ref{sec:InitializeNetworkCentral} to characterize the ensembles of initial states can therefore illuminate several features of the full dynamics. Since the mutual information and extractable work considered in later subsections are also functions of the one- or two-qubit reduced systems, we expect features of the single-qubit state space analysis to be correlated with those additional measures.

We construct three different datasets to use for PCA. The principal components will differ depending on the data set used, so each combination illustrates a particular feature of the data. However, plots illustrating PCA analysis for different datasets should be compared with caution. All data sets are aggregates of some single-qubit $\langle\sigma_{z,q}(\ell)\rangle$.
\begin{itemize}
    \item {\bf Update rule aggregates over all time, mixed CSs only:} We combine $\langle\sigma_{z,q}(\ell)\rangle$ for all initial conditions in the three mixed CS ensembles, at all layers from circuits with a fixed update rule. Then, trajectories for all the {\bf CS1}, {\bf CS2}, {\bf CS3} ensembles (but fixed update rule) may be shown on the same axes. Trajectories are the projections onto the PCs of datasets containing $\langle\sigma_{z,q}(\ell)\rangle$ for all qubits in all members of the CS ensemble at each $\ell$, evolved with a particular update rule.
    \item {\bf Central state aggregates over all time:} For a fixed central state, we combine $\langle\sigma_{z,q}(\ell)\rangle$ for all initial conditions in a single CS ensemble, at all layers from circuits generated by all update rules. Then, trajectories for any update rule but fixed CS, may be shown on the same axes. Trajectories are the projections onto the PCs of datasets containing $\langle\sigma_{z,q}(\ell)\rangle$ for all qubits, evolved with in all members in the fixed CS ensemble at each $\ell$.
    \item {\bf Central state aggregates at late time:} For a fixed central state, we combine $\langle\sigma_{z,q}(\ell)\rangle$ for all initial conditions in a single CS ensemble, at circuit layers $\ell>$. This isolates the late time, non-equilibrium stead state behavior. Late-time trajectories for any update rule, but fixed CS, may be shown on the same axes.
\end{itemize}
In addition, for this analysis we remove the noise associated with the members of each central state ensemble as described in Appendix \ref{sec:AppPCA}.

Figure \ref{fig:PCA_across_rules} shows the result of PCA on the first dataset, aggregated by rule. The first panel shows that under thermalizing dynamics, $R_1$, the three central states converge in the PCA space. For $R2-R5$, the central states move away from their starting points, but remain in distinct regions of the PCA space.
\begin{figure*}[htb]
    \centering
    \includegraphics[width=\textwidth]{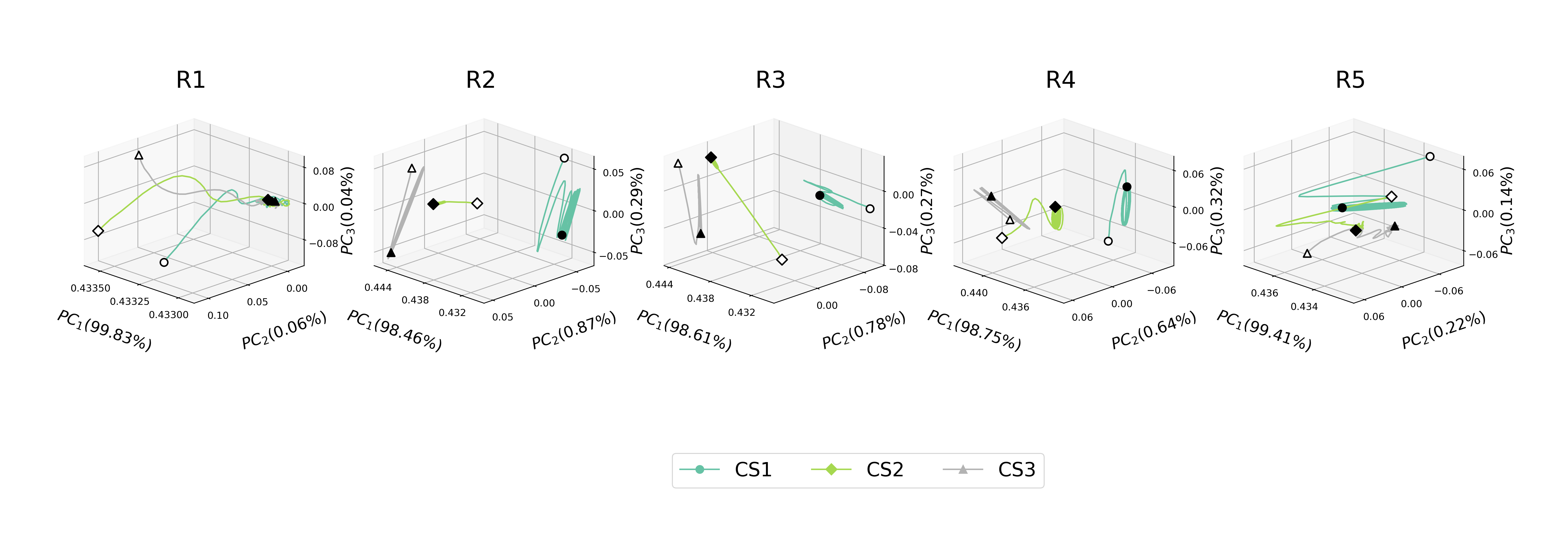}
    \caption{The 3-dimensional PCA of $\langle\sigma_{z,q}(\ell)\rangle$ aggregated by update rule. For each central state and rule, the line traces the projection of the ensemble-averaged state vector as the network evolves. Open symbols are circuit layer zero, filled symbols are layer 500.}
    \label{fig:PCA_across_rules}
\end{figure*}

\begin{figure*}[htb]
    \centering
    \includegraphics[width=\textwidth]{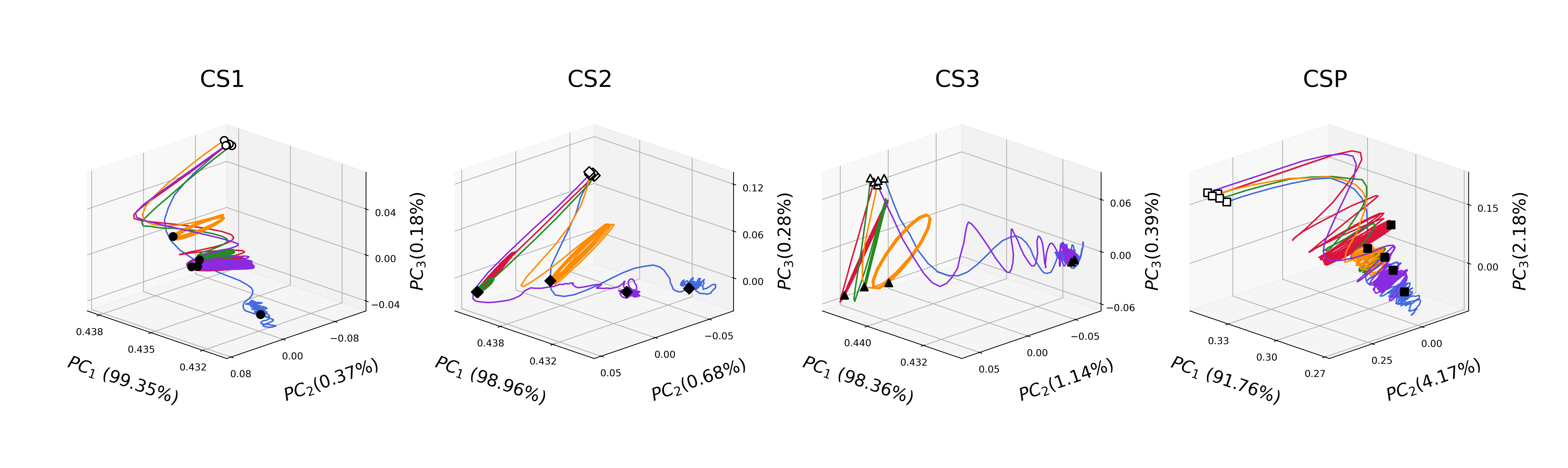}
    \vspace{0.1cm} 
    \includegraphics[width=\textwidth]{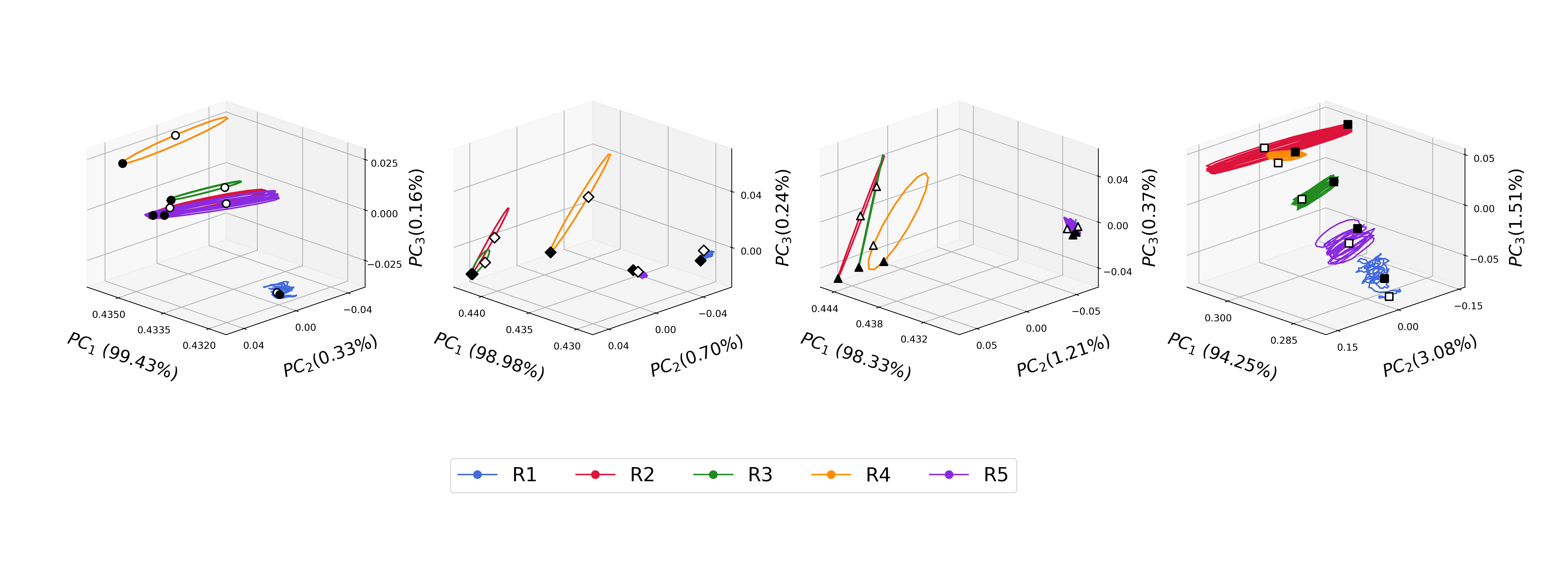}
    \label{fig:PCA_across_ICs_late_time}
    \caption{A visualization of the states explored by the dynamical rules. Each figure shows 3-dimensional PCA of $\langle\sigma_{z,q}(\ell)\rangle$, aggregated by central state. {\bf Top panel:} PCA and trajectories including all circuit layers. Open symbols are circuit layer zero, filled symbols are layer 500.  {\bf Bottom panel:} PCA and trajectories including all only circuit layers $\ell>300$. Open symbols are circuit layer 300, filled symbols are layer 500.  \label{fig:PCA_across_ICs}}
\end{figure*}

Figure \ref{fig:PCA_across_ICs} shows the results for PCA aggregated by central states, both for across all layers (top panels) and isolating the dynamics after $l=300$ to $500$ layers (bottom panels). In both top and bottom panels, there is a clear distinction between the state space trajectories of $R1$ compared to $R2$, $R3$, and $R4$. The top panels show that rules $R2-R4$ have a phase of exploration before settling into cycles. (Figure \ref{fig:Corr_growth} in the Appendix shows time evolution for correlations, which gives an indication of how many layers are required before the steady state is reached.) The approximate maximization rule, $R5$ displays intermediate behavior, and trajectories that differ more significantly depending on the central state. The bottom panels, where only late time behavior is used, show that $R2$, $R3$, and $R4$ can generate late-time trajectories are exactly cyclic, indicating very stable non-equilibrium steady states, at least by this measure. 

The PCA evolution for the pure central state (fourth column) shows all the trajectories drifting in the same direction. The cyclical patterns are also less defined. This may indicate that for $CSP$ and evolution by $U_{*}$, 500 layers may not be sufficient to completely characterize the (quasi-)static states. 

\subsubsection{Mutual information graphs: Clustering and disparity}
\label{sec:complexity}
Mutual information is a measure of information shared between subsystems in the network, and provides additional detail beyond the correlation magnitudes presented in the previous section. Some work has suggested that the graph complexity of mutual information may be an indication of interesting physics \cite{Cardillo:2013,valdez:2017,Kuwahara:2020,Hillberry:2021}, including phase transitions and far-from equilibrium behavior, although the general implications are not completely established.

For two qubits, $Q_i$ and $Q_j$, the mutual information is 
 \begin{equation}
    M_{ij}= \frac{1}{2}\left[(S(\rho_i) + S(\rho_j) - S(\rho_{ij})\right]  
 \end{equation}
where we have chosen the same normalization as in \cite{Hillberry:2021,Kuwahara:2020} for easy comparison to those works. All the mutual information shared by $N$ qubits in the network can be visualized as a graph, where vertices representing $Q_i$ and $Q_j$ are connected with an edge whose weight scales with their mutual information $M_{ij}$. This value evolves as a function of circuit and define a global quantity $\mathcal{M} = \sum_{i,j = 1}^{N}M_{ij}$ which gives the total mutual information on the network. In the results section, we discuss the time late time value of the ensemble average total mutual information on the network. 

We will consider two measures of graph complexity, the clustering coefficient and the disparity, for the mutual information graphs. The (global) clustering coefficient of a graph quantifies how likely it is for vertex $Q_i$ to be connected to
vertex $Q_j$ if both $Q_i$ and $Q_j$ are connected to $Q_k$. For a full, unweighted graph, the clustering is computed by computing the number of closed triangles divided by the number of possible closed triangles. Here, the value of mutual information weights the graph, and one can generalize the definition of clustering \cite{Ahnert:2007} to
\begin{equation}
\label{eq:clustering}
    \mathcal{C} := \frac{Tr[M^3]}{\sum_{j,k = 0; j\neq k} ^{N-1} [M^2]_{jk}}\,.
\end{equation}
``Complex" networks tend to have larger clustering coefficient \cite{Watts:1998}.

The disparity of a graph measures the non-uniformity in the strength, $s_i=\sum_j M_{ij}$, of the vertices in the graph. For vertex $i$, a measure of its strength compared to other vertices is
\begin{equation}
    \mathcal{Y}_i = \frac{\sum_{j=1} ^{N} (M_{ij} ^2)}{(\sum_{j=1} ^{N} M_{ij})^2}\,.
\end{equation}
A vertex connected by $L$ edges, with equal correlation $\gamma$ along each, has disparity $\mathcal{Y}_i = \frac{L \gamma^2}{L^2 \gamma^2} = \frac{1}{L}$. A vertex with one edge with weight $\Gamma\gg \gamma_i$, much greater than the weight along any other edge, has disparity $Y_i = \frac{ (\gamma_1 ^2 + \gamma_2 ^2 + \Gamma + \gamma_3 ^2 + \cdots)}{(\gamma_1 + \gamma_2 + \Gamma + \gamma_3 + \cdots)^2} = \frac{\Gamma^2}{\Gamma^2} =1$. 
The graph disparity is the average disparity over vertices,
\begin{equation}
\label{eq:disparity}
    \mathcal{Y} = \frac{\sum_{i=1} ^{N} Y_i}{N}\,.
\end{equation}
A graph with low disparity will have vertices of nearly uniform strength, while a high disparity graph will have some ``hub", or central, vertices connected to many others, but most vertices only coupled to the few, central vertices. 

\begin{figure*}[htb]
    \centering
    \includegraphics[width=\textwidth]{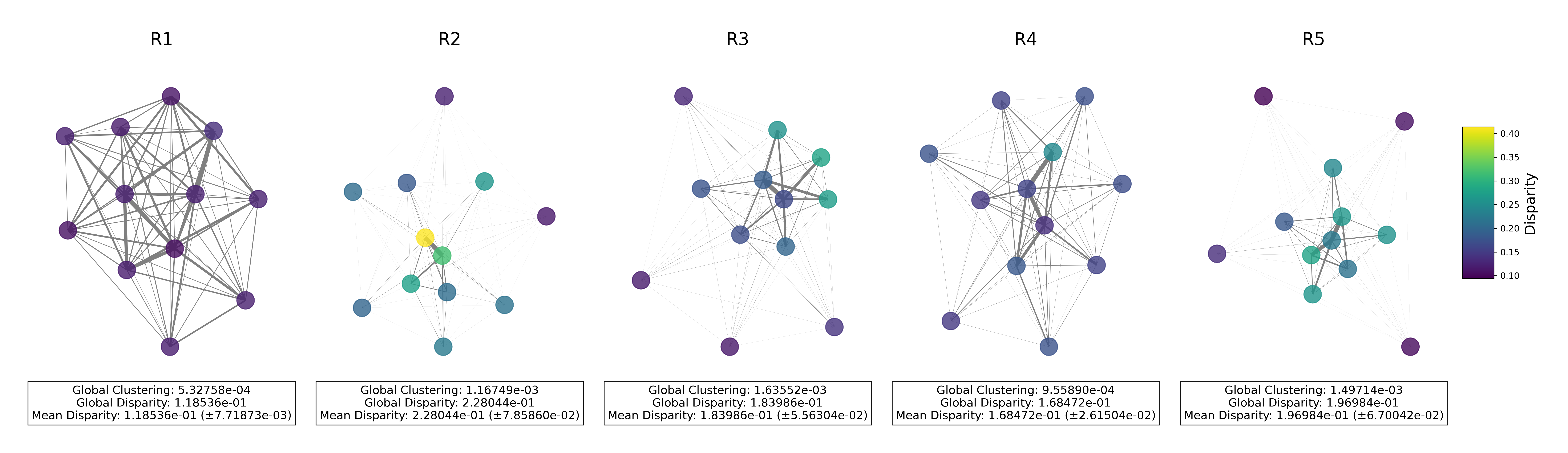}
    \caption{Mutual information network for one member of each {\bf CS1} ensemble, evolved with each update rule and connectivity $C=2$. Well-connected nodes are positioned more closely together, and nodes are colored according to their disparity. The thickness of lines connecting nodes is scaled with the value of mutual information shared by the pair. \label{fig:MI_network_late_time_avg}}
\end{figure*}

Figure \ref{fig:MI_network_late_time_avg} shows the graphs associated with the mutual information at late times ($\ell = 300-499 $) developed in the evolution a single member of the {\bf CS1} central state ensemble under each update rule. Below each graph the values of the clustering and disparity are reported. As expected, random evolution produces an MI graph that is most homogeneous, with the smallest clustering and disparity. Section \ref{sec:ResultsCorrelations} will show the relationship between mutual information graph complexity measures and other statistics for each rule and central state.

\subsubsection{Extractable work: Total and persistence}
\label{sec:Counting Delta W}
When the qubits in the network start with different populations, or temperatures, the process of thermalization will lead to instances where the extractable work achievable by some qubits increases \cite{Akhouri_2023}. However, if some subsystems stay usefully out of equilibrium, then they may continue to have positive $\Delta W^{\rm ex}$ even at late times. In addition, rather than fluctuations of single steps where a subsystem experiences positive $\Delta W^{\rm ex}$, some subsystems will experience repeated, sequential increases in extractable work, over many layers of the circuit. 

\begin{figure*}[htb]
    \centering
    \includegraphics[width=\textwidth]{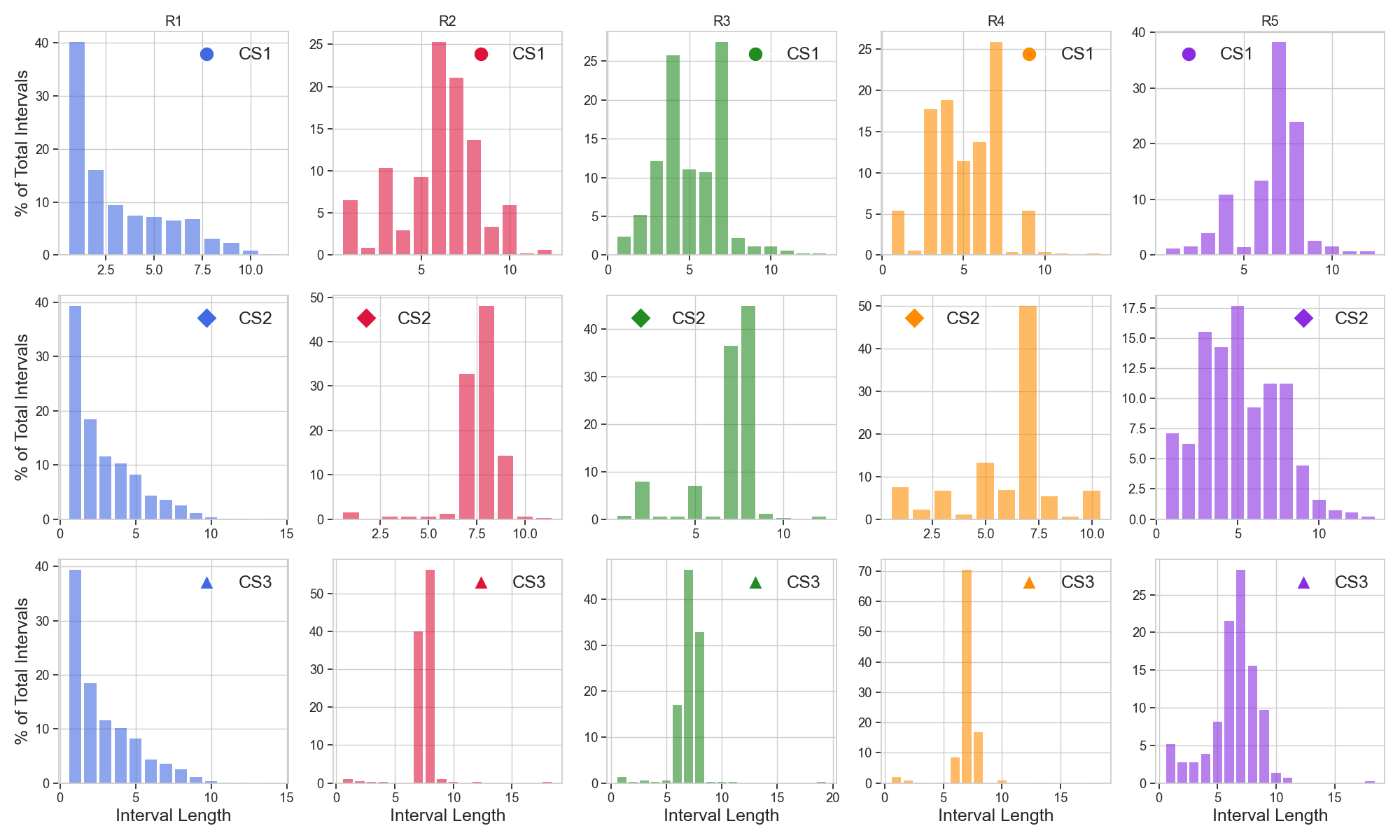}
    \caption{Histograms of the interval lengths, $\Delta\ell$, with consecutive positive change in extractable work. For each central state, all members of the ensemble are included. \label{fig:ExWhistograms}}
\end{figure*}

To quantify this, we record the length of all intervals of consecutive positive changes in $\Delta W^{\rm ex}$, for each qubit $q$. We then combine the data for all qubits in the network and find the frequency of each interval length $L$ in the list, $f(L)$. Figure \ref{fig:ExWhistograms} shows the resulting histograms for each rule and frequency lists that combine all members of the central state ensemble. As expected, the random update rule shows an exponential decay in frequency with increasing lengths. For the non-random update rules, this is no longer the case. Instead, these networks show sustained positive change in extractable work over intervals an order of magnitude longer than in $R1$. 

\begin{figure*}[htb]
    \centering
    \includegraphics[width=\textwidth]{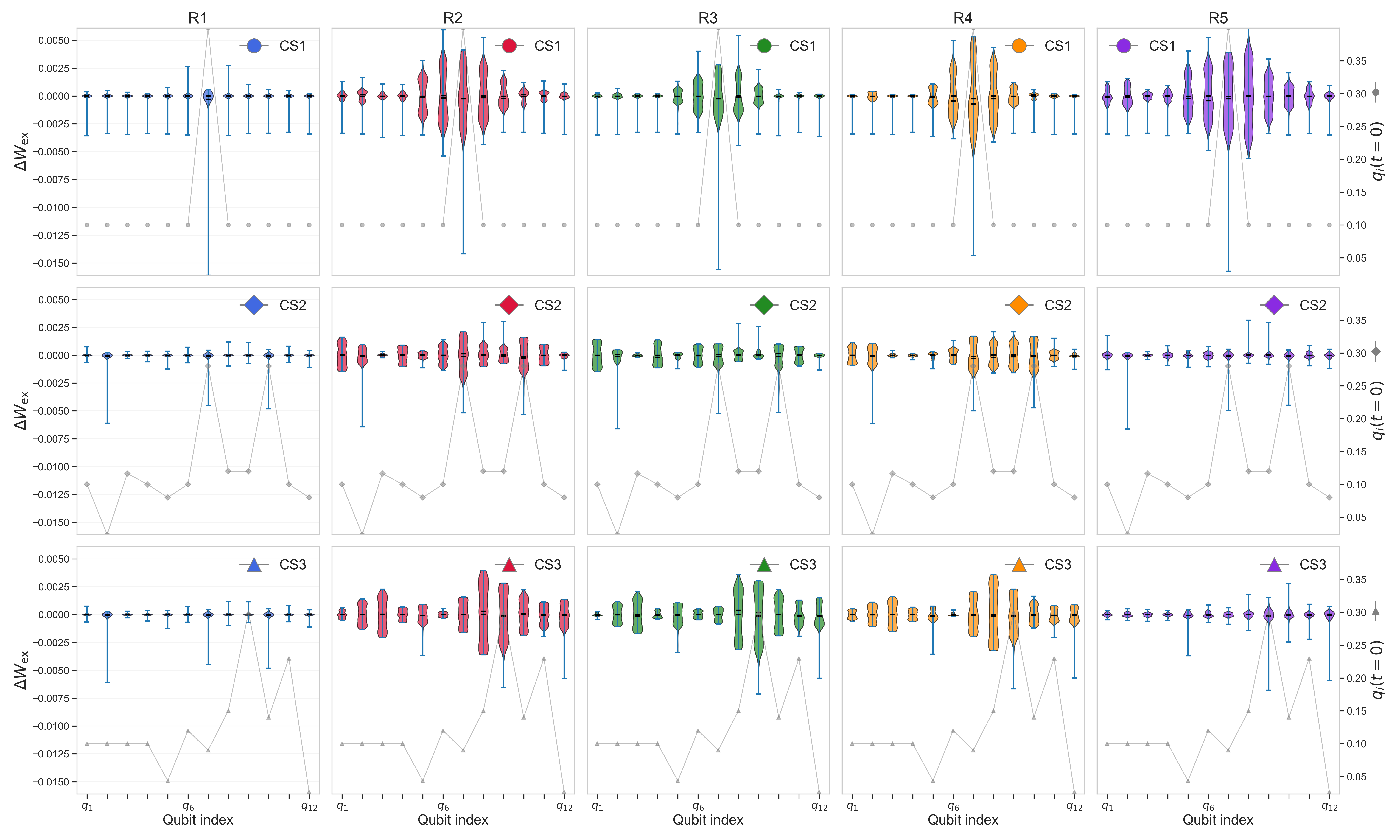}
    \caption{Violin plots for the change in extractable work for $N=12$ qubit networks for the three mixed central states, {\bf CS1}, {\bf CS2}, and {\bf CS3}, averaged over all members of the central state ensemble, and over all layers of the circuit. The population distribution of the central state is shown by the gray points and line, according to the right-hand axes labels. }
    \label{fig:violin}
\end{figure*}

Figure \ref{fig:violin} presents violin plots that show, for each qubit, the distribution of $\Delta W^{\rm ex}$ experienced across single layers of the circuit. All circuits for a given central state ensemble are included. Overall the magnitude and variation in $\Delta W^{\rm ex}$ is much lower for the thermalizing rule, $R_1$, compared to the other rules, for any of the central state. Within a central state, the qubit with the highest variation is generally the one that started with the highest initial-state population. This is especially prominent in {\bf CS1}. Rules $R2$-$R5$ generically generate a higher probability for larger magnitudes of $\pm \Delta W^{\rm ex}$ than $R1$ does. For {\bf CS1}, distributions with significant $|\Delta W^{\rm ex}|$ are centered around the resource qubit. Central states {\bf CS2} and {\bf CS3}, on the other hand, start with a more inhomogeneous population distribution and end up with more qubits experiencing significant $|\Delta W^{\rm ex}|$. The plot also shows that rules perform differently across the central states, most prominently for $R_5$ which shows the greatest variety of behavior. As other measures have also shown, with {\bf CS1}, $R5$ is far from the thermalizing while with {\bf CS2}, it more closely resembles $R1$, random, evolution. 

To more simply compare between central states and rules, we define a single statistic for sequential increases in extractable work for a single circuit evolution:
\begin{equation}
\label{eq:persistance}
L_{\Delta W^{\rm ex}} = L_{\rm mode} \times f(L_{\rm mode})
\end{equation}
where $L_{\rm mode} = \text{argmax} f_{\chi}(L)$ is the length of the most frequent length in the list and $f(L_{\rm mode})$ is the frequency. The relationship of this measure to the others will be presented in the next section.

\section{Results \label{sec:results}}
This Section presents the relationship between different measures of the non-equilibrium steady states. 

Section \ref{sec:ResultsCorrelations} first presents the strongest correlations between various measures. Since all networks reach a steady state by circuit layer 300, we use late-time, average quantities. For any measure $X$, $\langle X \rangle_l$ is the average over circuit layers 300-500. In additions, we take the average over all members of the central state ensemble, so 
\begin{equation}
    \langle \langle X\rangle \rangle_{\rm ens} = \frac{1}{N}\sum_{i=1}^{100} \langle X_i \rangle_l
\end{equation}
where $\langle X_i \rangle_l$ is the late-time mean value of the measure for the $i^{\rm th}$ member of the ensemble. Where appropriate, we show the percent difference from results for the same central state but random ($R1$) dynamics, 
\begin{equation}
\label{eq:Frac_random}
    \Delta  \langle \langle X\rangle \rangle_{\rm ens} = \frac{ \langle \langle X\rangle^{} \rangle_{\rm ens}- \langle \langle X\rangle^{R1} \rangle_{\rm ens}}{ \langle \langle X\rangle^{R1} \rangle_{\rm ens}}\,.
\end{equation}
Since $\langle \langle \tau_z\rangle \rangle_{\rm ens}$ is a constant, independent of dynamics, we instead report $\langle \sigma(\tau_{z})\rangle_{\rm ens}$. 

Then, Section \ref{sec:BernouliResults} discusses two different types of non-Markovianity that are present in the dynamics, and the correlation with other properties of the steady states.

\subsection{Correlations between non-equilibrium measures}
\label{sec:ResultsCorrelations}
\begin{figure}
\includegraphics[width=\columnwidth]{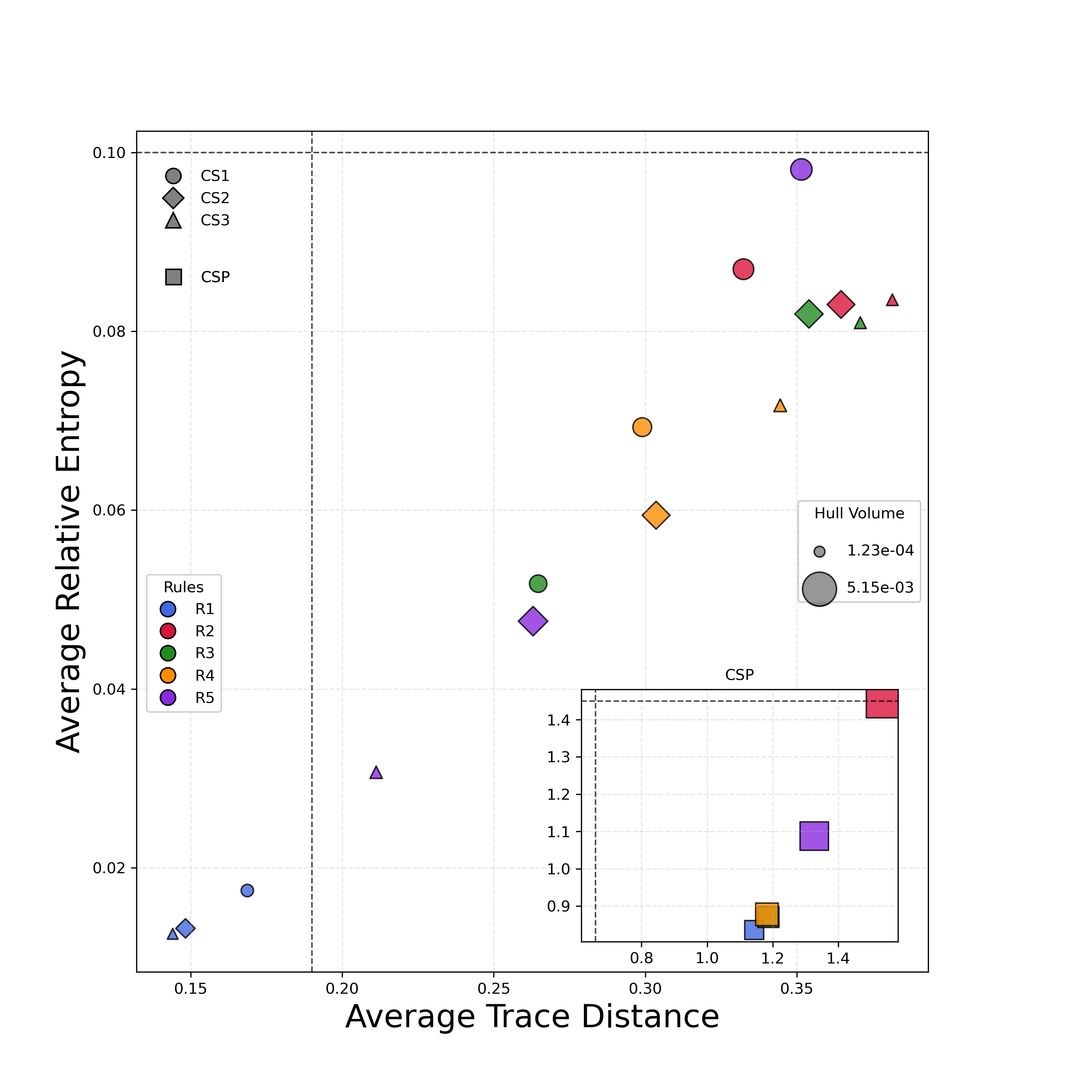}
\vspace{0.1cm}
\includegraphics[width=\columnwidth]{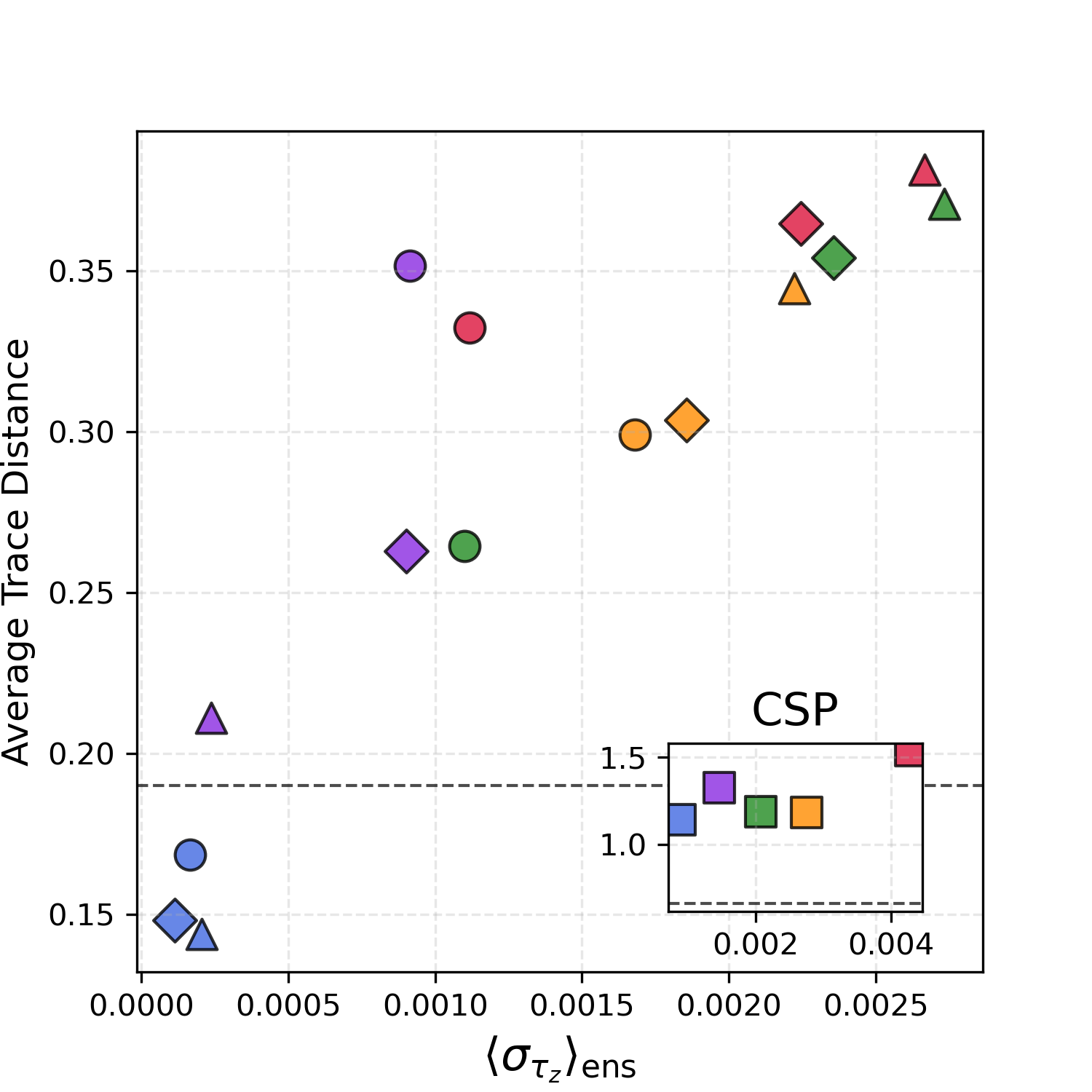}
    \caption{\textbf{Top panel:} Ensemble averages of the late-time total relative entropy $\langle \langle \mathcal{D}(\{\rho_q\},\bar{\rho})\rangle\rangle_{\rm ens}$, total trace distance, $\langle \langle {Tr(\rho^{(N)},\bar{\rho}^{(N)}}\rangle\rangle_{\rm ens}$, and volume of the convex hull $V_{CH}$. Symbol size scales with the convex hull volume, with the relationship shown for two examples in the inset legend. The vertical dashed line is the average trace distance of the initial state distributions about the three mixed central states. The horizontal dashed line is the mean relative entropy of the three central states after the 10 layers. \textbf{Bottom panel:} The average trace distance vs $\langle \langle\sigma(\tau_z)\rangle\rangle_{\rm ens}$. \label{fig:KL_vol_TD}}
\end{figure}

Figure \ref{fig:KL_vol_TD} shows how the late-time characteristics of the {\bf CS1}, {\bf CS2}, and {\bf CS3} network states compare on several measures presented in previous sections: average relative entropy (Eq.(\ref{eq:curlyD})), average trace distance from the reference thermal state (Eq.(\ref{eq:TrDistance})), and the PCA convex hull volume (Appendix \ref{sec:AppPCA}). The PCA uses axes determined from only the distributions that share both a central state and an update rule, and that capture greater than 85$\%$ of the variance in the $\langle \sigma_z\rangle$. For the three mixed central states, this requires only a two-dimensional PCA space, while the pure central state requires 11 dimensions. 

The top panel of the Figure \ref{fig:KL_vol_TD} highlights the effectiveness of the non-random rules in achieving non-thermalizing behavior. The dotted lines show the initial average relative entropy and the average trace distance from the thermal state (see the lower right panel of Figure \ref{fig:DistributionsPCA}, at $\ell=10$). Under ergodic or randomizing processes, the relative entropy and trace distance must decay, so a constant offset in the relative entropy or the absence of decay in trace distance can be indicative of a non-ergodic evolution. The random rule, $R1$, always takes the system to states with lower trace distance and relative entropy. However, the other rules are able to prevent the relative entropy from decreasing by much, and can increase the average trace distance from the thermal reference state. 

In addition, the non-random rules achieve different results depending on the central state. Perhaps the most significant contrast is between $R2$ and $R5$: $R2$ uses an exact global optimization and achieves roughly comparable results for all central states. The global constraint is likely most sensitive to the similarity in thermodynamic measures shared by the central states. On the other hand, $R5$ is an approximate optimization, one qubit at a time. This update rule shows the greatest variation in performance between central states. For {\bf CS1} it nearly maintains the initial average relative entropy compared to the thermal state. This result is reminiscent of the superior performance of similar dynamics in other complex optimization problems \cite{Gomez:2019,Lazer:2007,Olson_2022}, where an element of error, or approximate information, in optimization prevents the dynamics from becoming stuck in local minima.

The convex hull volume of the steady state also distinguishes the update rules. The volume is represented by the symbol size in the Figure. The smallest symbols show comparable volumes to that of the initial ensembles (see the top right panel of Figure \ref{fig:DistributionsPCA}), while the largest symbols indicate dynamics whose late-time convex hull increases by about an order of magnitude. For $R1$, the convex-hull volume does not increase much, for any central state. The other update rules increase the volume at least somewhat, with all rules achieving an approximately similar increase for a given central state. 

The bottom panel of Figure \ref{fig:KL_vol_TD} shows the relationship between a state-space measure, the average trace distance of the qubits from the reference thermal state, and the inhomogeneity of subsystem dynamics ($\sigma(
\tau_z)$). The random rule, $R1$, evolves networks to states with low average total trace distance and low standard deviation in the map parameters, consistent with thermalizing behavior. All the non-random rules generate late-time states with the higher deviation in map parameters, which is correlated with increasing inhomogeneity. Notice that $R2$, $R3$ and $R4$, the global optimization rules, all follow the same linear relation, while the approximate optimization rule, $R5$, falls on a line of steeper slope. In addition, the global optimization rule restores the initial ordering of central states by trace distance, $\textbf{CS1}<\textbf{CS2}<\textbf{CS3}$ (see bottom right of Figure \ref{fig:DistributionsPCA} at $\ell=0$ vs $\ell=10$). However, the approximate rule, $R5$ inverts that order. 

\begin{figure*}[htb]
\includegraphics[width=\textwidth]{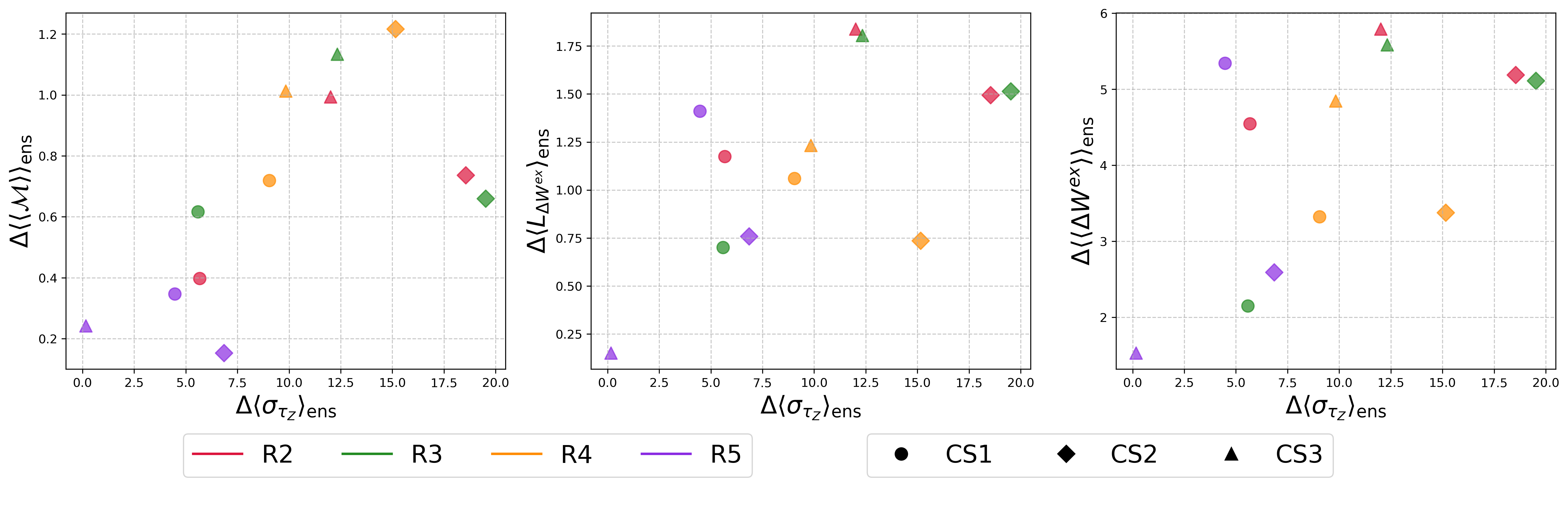}
    \caption{The relationship between late-time, ensemble-average of the inhomogeneity of the propagator maps, $\langle\sigma_{\tau_z}\rangle_{\rm ens}$ and (left) the persistence measure for consecutive increases in extractable work (Eq.(\ref{eq:persistance})), and (right) the total positive $\Delta W^{\rm ex}$ for all qubits, over all layers. All quantities are presented compared to random, see Eq.(\ref{eq:Frac_random}).
    \label{fig:scatter_plot_tauz} }
\end{figure*}

\begin{figure*}[htb]
\includegraphics[width=\textwidth]{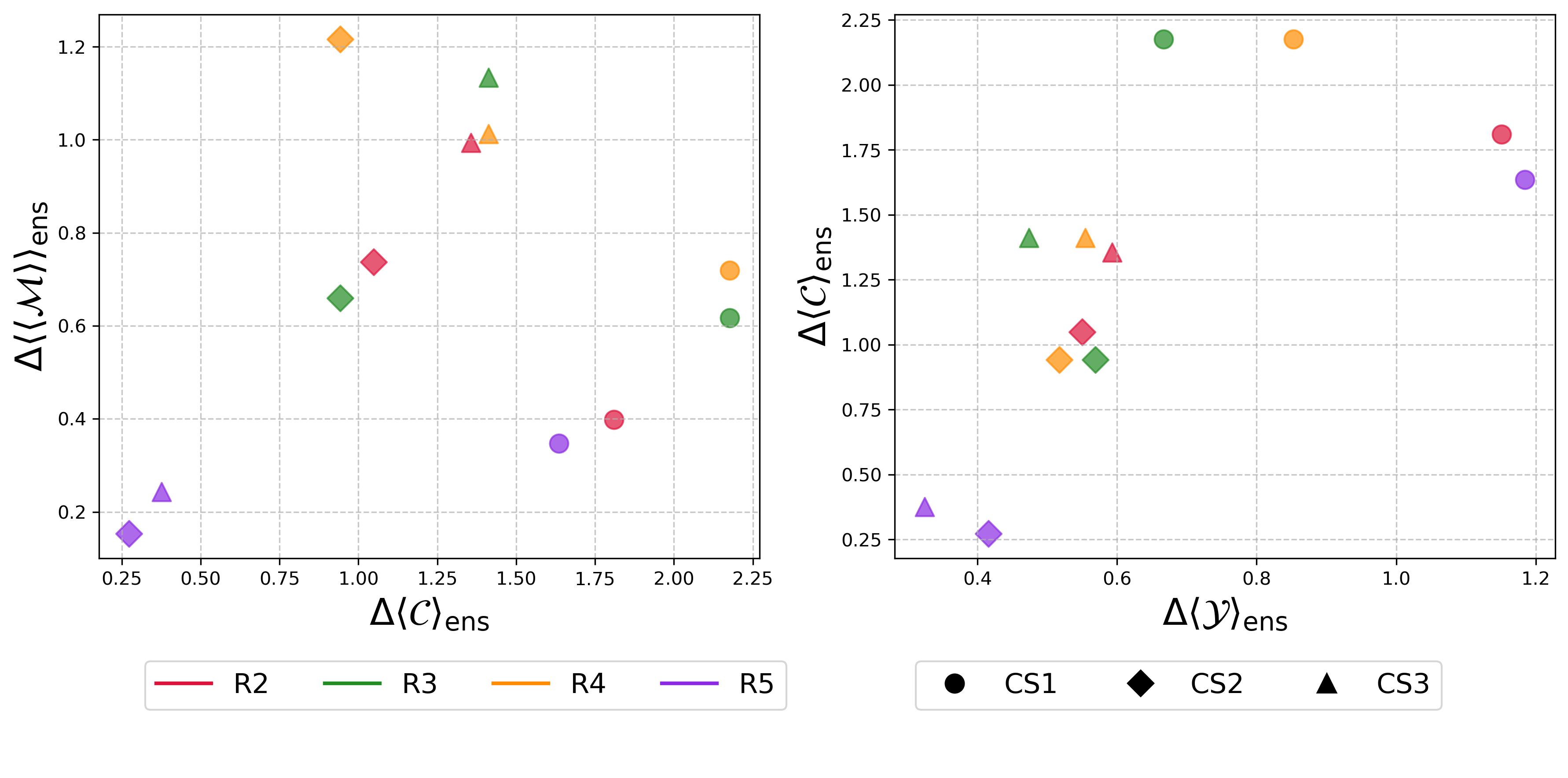}
    \caption{The relationships between mutual information ($\langle\mathcal{M}\rangle$), clustering ($\mathcal{C}$, Eq.(\ref{eq:clustering})) and disparity ($\mathcal{Y}$, Eq.(\ref{eq:disparity}), all presented as the percent different from random dynamics, across qubit networks of size 12. \label{fig:scatter_result_corr}}
\end{figure*}

Next, Figure \ref{fig:scatter_plot_tauz} presents the relationship between several measures of the utility of away-from-equilibrium states and the degree of inhomogeneity in the single-qubit evolution, $\langle\langle\sigma_{\tau_z}\rangle\rangle_{\rm ens}$. Figure \ref{fig:scatter_result_corr} shows the relationship between different measures of out-of-equilibrium utility. The left panel shows that the mean value of the mutual information at late time has a positive correlation with the mean value of clustering coefficient, although again the {\bf CS1} central states fall on a different slope from {\bf CS2}, {\bf CS3}. The right panel shows that the two graph complexity measures for mutual information are also approximately correlated, but there is increasing dispersion as the complexity measures grow. This plot also again shows the strikingly different results achieved by $R5$ on {\bf CS1}.

Taken together, these figures indicate that dynamics with same general degree of inhomogeneity in single-qubit propagator maps can perform quite differently in terms of complexity of mutual information and extractable work. The strongest correlation is between total mutual information and $\sigma(\tau_z)$, but even there the relation depends on the central state. We also find that the MI graph complexity measures are not strongly correlated with each other. 

To summarize this subsection: the inhomogeneity of single-qubit dynamics (propagator maps) is well-correlated with single-qubit measures of distance from the thermal state (trace distance, and relative entropy). However, the single-qubit inhomogeneity measures cannot be used to predict performance on utility measures. In other words, and unsurprisingly, there are a variety of out-of-equilibrium steady states. 

\subsection{The role of non-Markovianity: (C)P-divisibility and Bernoulli circuits}
\label{sec:BernouliResults}
In Section \ref{sec:correlations}, we saw that networks that displayed less homogeneity in the single-qubit dynamics ($\tau_z$) also displayed a less symmetric distribution of correlations (Figure \ref{fig:Corr_12_Q}). The mutual information graphs gave one measure of how the distribution of correlations across qubits is important in the non-equilibrium steady state. However, a second consequence of large correlations is the presence of propagator maps that are not (completely) positive. That is, those propagator maps do not give a physical evolution for all states within the Bloch sphere. Figure \ref{fig:tzEachIcEachRule} reported the fraction of propagator maps that are not (C)P for each central state and rule. Figure \ref{fig:NCP_prob} in the Appendix shows more detail on the circuit layers where transitions are most likely to be non-(C)P. Here, we study the importance of spatial structure in the non-(C)P maps.

As Figure \ref{fig:tzEachIcEachRule} reported, even networks undergoing random evolution have a considerable fraction of non-(C)P propagator maps. This is an effect of the small size of the network: although correlations spread out nearly uniformly as the network approaches its equilibrium steady state, typical fluctuations may still be large enough to violate the condition in Eq.(\ref{eqn::CPcond}). Appendix \ref{sec:NCPscaling} elaborates on this point.  

In order to display the prevalence of non-(C)P propagator maps with the small-system thermal fluctuations removed, we construct noise-reduced propagator maps, $\Lambda_{NR}(\ell,\ell-1)$, from the central-state ensembles. This is defined as the propagator map that takes the ensemble-averaged state $\rho_{NR}(\ell-1)=\langle\rho(\ell-1)\rangle$ (the element-by-element average) before layer $\ell$ to the ensemble-averaged state after $\ell$. In practice, for a qubit, $Q_a$, we only need to compute the two-qubit reduced states containing $Q_a$ to find this map. That is, we compute the average of the correlations ($C^{xx}_{ab,\ell}$) and partner Bloch vectors ($z_{b,\ell}$) in every two-qubit density matrix involving $Q_a$ before a given layer $\ell$ and across all members of the central state ensemble (see Eq.(\ref{eq:twoQrho})). From the average of these quantities, we compute a corresponding single-qubit propagator map across each layer. 

\begin{figure*}[htb]
    \centering
    \includegraphics[width=\textwidth]{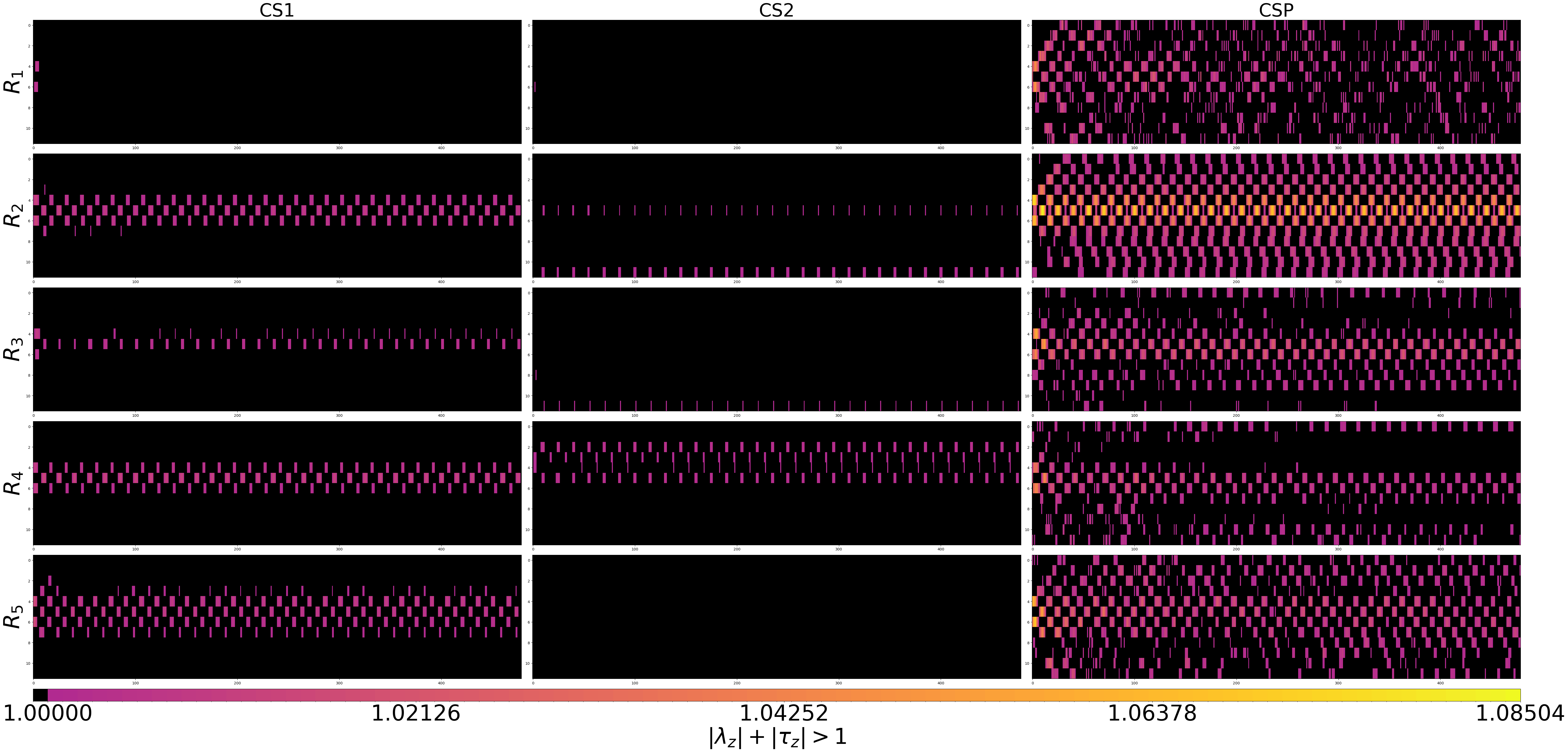}
    \caption{Breaking of the positivity condition in the noise-reduced propagator maps on the 12-qubit networks. The breaking of the positivity condition comes from the presence of strong correlations between the qubits. The maps rapidly become CP whereas for the pure states, even after ensemble averaging, the map maintains non-CP behavior at late times.}
    \label{fig:NoiseCancelledCP}
\end{figure*}

Figure \ref{fig:NoiseCancelledCP} shows the result. The noise-reduced maps for random evolution, $R1$, are largely (C)P-divisible. In contrast, for non-thermalizing update rules the noise-reduces maps are frequently non-(C)P, with a clear pattern related to the pattern in $\langle\sigma_z\rangle$, Figure \ref{fig:temperatureHeatMaps}, and the correlations that build up due to preferred interaction partners. This in turn depended on the distribution of resources in the initial (central) state.

\begin{figure}
\includegraphics[width=\columnwidth]{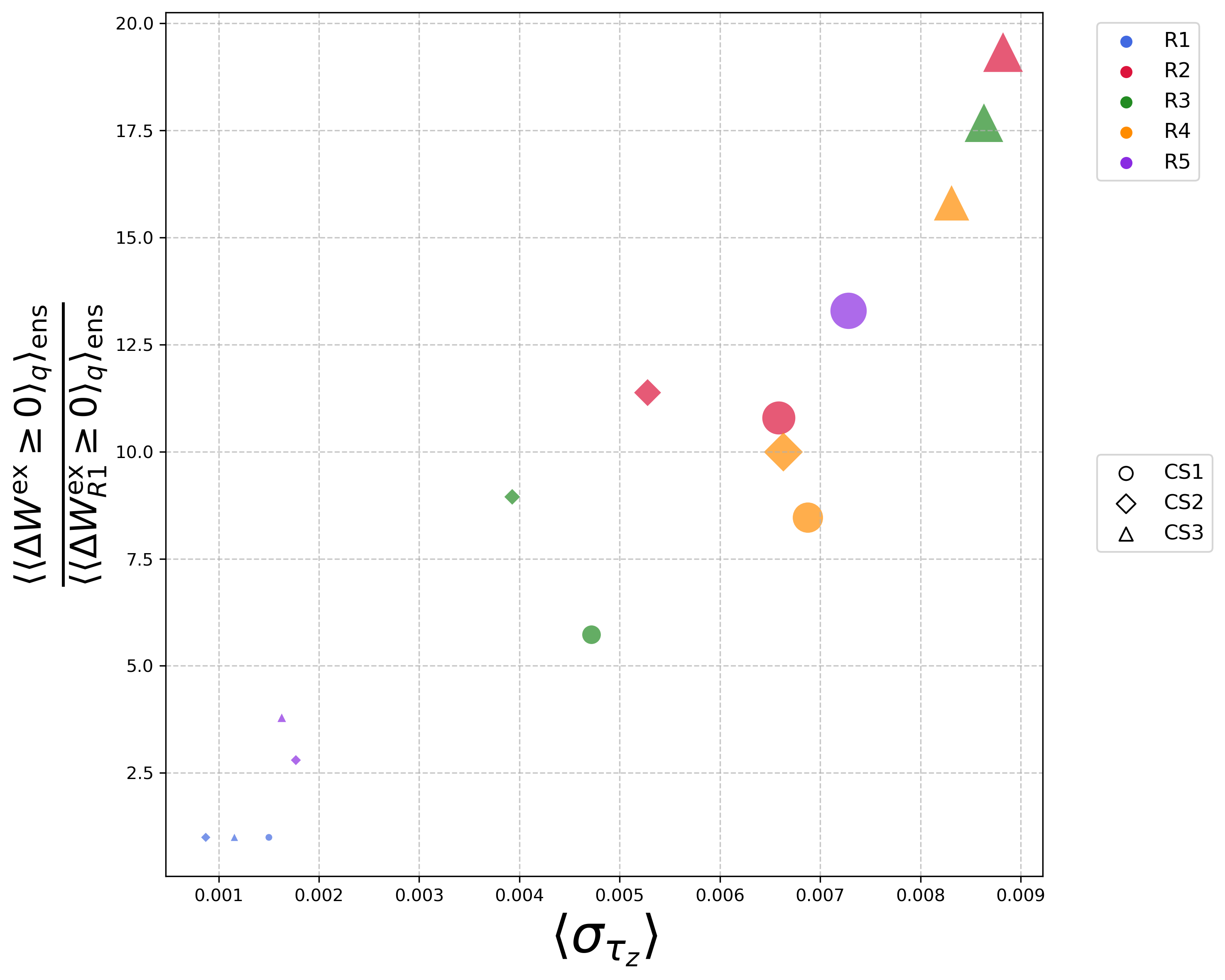}
\caption{The ratio in average positive change in extractable work normalized by the change in extractable work for the random rule versus the late time standard deviation in the shift parameter of the. The marker size scales with the fraction of non-(C)P noise-reduced maps, which are defined on the ensemble-averaged state. \label{fig:ncpCorrelations}}
\end{figure}

Among the possible utility measures, the non-(C)P propagator maps on the noise-reduced systems have the highest correlation the spatial clustering of high values of extractable work, as seen in the violin plots of Figure \ref{fig:violin}. Figure \ref{fig:ncpCorrelations} shows the relationship between the basic out-of-equilibrium measure, $\sigma(\tau_z)$, the total magnitude of positive change in extractable work, and the occurrence of non-(C)P propagator maps on the noise-reduced systems. There is a clear correlation between the presence of non-(C)P dynamics and out-of-equilibrium behavior as diagnosed by the variance in $\tau_z$. This correlation can likely be traced to the fact that both extractable work and breaking of the (C)P condition depend on the relationship between the single-qubit populations and two-qubit correlations.

The presence of non-(C)P propagator maps in continuous-time systems is one of the proposed definitions of quantum non-Markovian evolution \cite{Denisov:1989,Wolf:2008,Davalos:2019,Filippov:2019}. More precisely, if a CPTP map between times $\ell_0$ and $\ell_2$, $\Lambda_{\ell_2,\ell_0}$, is (C)P-divisible, then for every intermediate time $\ell_1$ it satisfies
\begin{equation}
\Lambda_{\ell_2,\ell_0} = \Phi_{\ell_2,\ell_1}\circ \Lambda_{\ell_1,\ell_0}\,,
\end{equation}
where $\Phi$, the propagator map, is (completely) positive. CP-divisibility is a necessary, but not sufficient, condition for a quantum process to be Markovian \cite{Milz:2019}.

This definition is tailored to apply to continuous-time evolution, where a CP map $\Lambda(t,0)$ is only (C)P divisible if the propagator map at every intermediate time $t>t_1>0$ is (C)P. For the circuit dynamics considered here\footnote{Notice that even for the random rule, for which the noise-reduced evolution does not show non-(C)P dynamics, the evolution is not time homogeneous ($\Lambda_{t+s,0}\neq \Lambda_{s}\cdot \Lambda_t)$.}, the circuit layers determine a finite set of natural points of divisibility, in the spirit of inhomogeneous collision models \cite{Ciccarello:2022}. Although the full connection to non-Markovianity remains to be explored, it is interesting to examine the presence of non-(C)P propagator maps in the context of the non-equilibrium steady states.  

The non-(C)P propagator maps occur because of spatial and temporal structure in the correlations among the qubits. Those correlations are also used by the update rules to perform the maximization. However, the net effect of the update rules is to generate a sequence of gate configurations. The differing sequences, combined with the initial state, directly determine the non-equilibrium steady state achieved. We return to the comparison between the full gate sequences and the simpler dynamics of the Bernoulli circuits defined in Section \ref{sec:EmergentGraph}.

On networks with connectivity $C=2$, there are only two circuit configurations. These can be visualized as the two ways of laying bricks in the traditional brickwork circuit. In this case, the Bernoulli circuits use the central state-dependent emergent networks to assign probabilities ($p_\alpha$, $1-p_\alpha$) to each of the two types of layers. The resulting trajectories in the space of circuit configurations display a biased random evolution, with each step independent from the others. We can compare the results achieved via the the Bernoulli circuits with those that correspond to brickwork pattern with the same frequency of each type of layer, but in arranged in a non-random sequence. 

\begin{figure*}[htb]
    \centering
    \includegraphics[width=\textwidth]{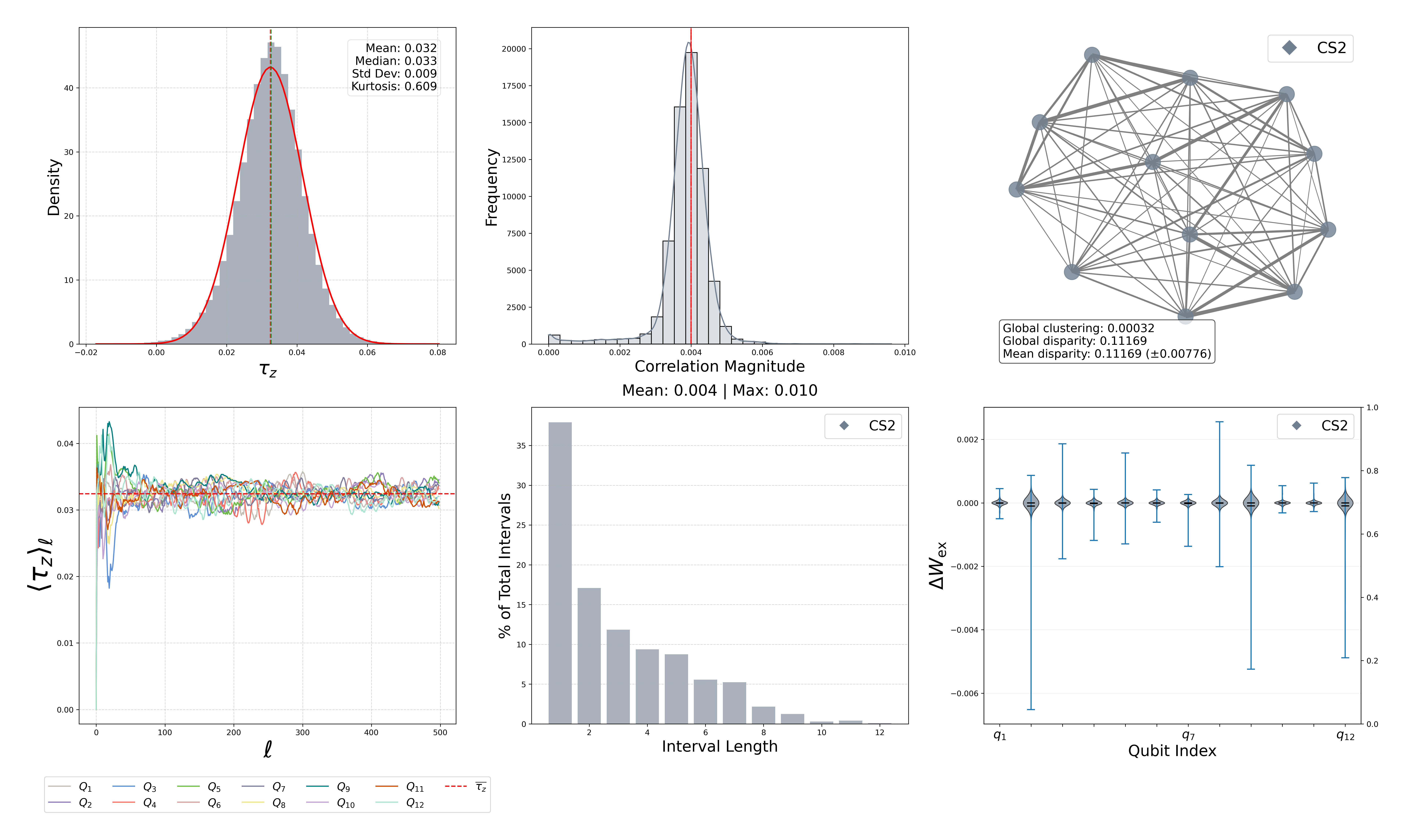}
    \vspace{0.25cm}
\includegraphics[width=\columnwidth]{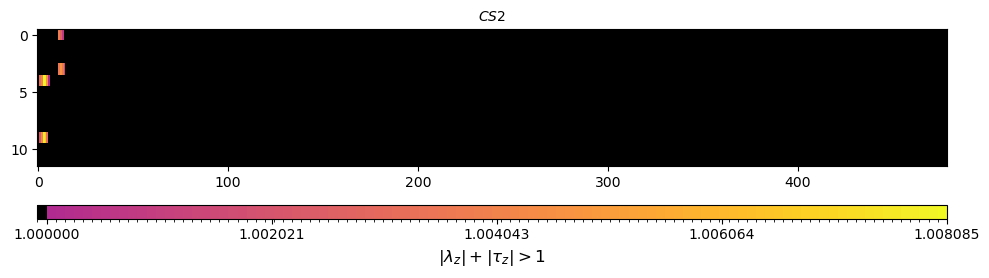}
    \caption{The 12-qubit, {\bf CS2}, network statistics under Bernoulli circuit evolution derived from the emergent network corresponding to $R2$ evolution.\label{fig:bernoulli_r2}}   
\end{figure*}

Figure \ref{fig:bernoulli_r2}, shows the outcome of the Bernoulli circuit derived from $R2$ on networks of 12 qubits and the {\bf CS2} ensemble. We chose $R2$ because it developed significantly unequal probabilities for the two mixed central states, (0.27,0.73) for {\bf CS2}. This Bernoulli circuit tends to behave like the $R1$ circuit on all measures. In other words, the bias based on initial conditions is not sufficient to maintain local information for very long against the randomizing effects of the circuit layer application.

The random rule and the Bernoulli circuit both have a larger magnitude of total correlation and higher frequency of correlations at the mean value compared to the other rules. Both also have non-(C)P propagator maps with a noise-only structure. The bottom panel of Fig\ref{fig:bernoulli_r2} shows that the ensemble-averaged circuit has no non-(C)P maps beyond the first few layers. 
These results illustrate the role of the interaction trajectories of non-random update rules that are layer- and state- dependent, as opposed to a biased random update which scramble information.  Specifically, the Bernoulli circuit fails to maintain the magnitude and persistence of correlations between the qubits. This indicates that adaptive circuit dynamics, that take both local and two qubit correlations into account, are needed for reaching and maintaining out-of-equilibrium steady states. 

\section{Conclusions}
\label{sec:conc}
In closed quantum systems that thermalize, the late-time dynamics of subsystems are quite homogeneous, allowing only for small fluctuations about the thermal state. In the examples studied here, we find several distinct non-equilibrium steady states whose subsystem dynamics are significantly inhomogeneous. The use of symmetry-restricted states and dynamics allowed us to characterize the ensemble of single-qubit open system dynamics in terms of a single distribution, for the shift $\tau_z$ in phase-covariant dynamical maps. The non-equilibrium steady states that correspond to inhomogeneous dynamics are simple - well described by a single principal component for the set of reduced, single-qubit density matrices. The state space available is large enough to contain many such states that may have different characteristics. 

We used several measures to distinguish the steady states, which also distinguish combinations of initial states and dynamics. The late-time complexity of the mutual information network, and persistence of increasing extractable work measures differed significantly between the different non-random rules, even when they were operating on central states that shared a common value of the conserved charge and an identical relative entropy measure, $\mathcal{D}$. The degree of inhomogeneity of the single-qubit dynamics does not contain sufficient information to predict the variations in the MI complexity or extractable work statistics of the steady state. 

In the different non-equilibrium steady states found here, the subsystem behavior is akin to that of driven, dissipative systems, where the rest of the closed system provides both driving and dissipation. This contrasts with thermalizing closed systems, where the complement of the system considered plays the role of a thermal bath. On each network, the non-equilibrium steady state will be slightly different, with the ensemble of steady states constrained by both dynamics (the Hamiltonian family) and the global initial state. This construction may provide a new method to generate non-equilibrium steady states in the lab \cite{Zhang:2022, Chen:2024}, since it has a simple circuit implementation. It would be interesting to see if the networks presented here may be analyzed by the techniques of \cite{Strachan:2024}, who considered infinite baths, but non-Markovian dynamics and dynamical maps. In addition, the utility measures using extractable work may be related to entropy production constraints for open systems \cite{Spohn:1978, Lacerda:2024}.

\begin{acknowledgments}
 The work of U.A. was supported by the Department of Energy grant number DE-SC0020360. The work of Jackson Henry was supported by NSF grant PHY-2206591, S.S. thanks the U. Maryland for hospitality for an extended visit and discussions with T. Jacobson, B.L. Hu, N. Yunger-Halpern. We thank Andrew Belmonte for stimulating discussions on strategies in the rock-paper-scissors game as motivation for the strategy mimic update rule. The authors are grateful for computational resources supported the NSF through OAC-2201445.
\end{acknowledgments}

\widetext
\appendix

\section{Supplemental Figures}
\label{sec:appendFigures}
This section contains several additional results and figures. Section \ref{sec:C4} shows how changing the connectivity of the interaction network changes the results. Section \ref{sec:timeEvol} shows the time-evolution of a number of quantities, illustrating the approach to the steady state and stabilization at high circuit depth. Section \ref{sec:NCPscaling} supports the interpretation of the noise-reduced propagator maps by showing how several results scale as a function of the number of qubits on the network. 

\subsection{Results for connectivity \texorpdfstring{$C=4$}{C=4}}
\label{sec:C4}
In the body of the paper, we presented results for networks whose interaction graph has connectivity $C=2$. In this section, we show several of the same analyses for networks with $C=4$. These interaction graphs are shown in left column of Figure \ref{fig::emergent_network}, bottom and middle panels. In all other respects the procedure, the procedure is the same: we use the same $U_*$, the same central state ensembles given by the $\ell=10$ (yellow) points in Figure \ref{fig:DistributionsPCA} and the same update rules. 

 The higher connectivity allows next-to-next-to nearest neighbor interactions. For a 12-qubit network, there are 36 configurations of two-qubit neighborhoods and so 36 different arrangement of gates possible in each layer. The non-random rules explore this space of configurations to optimize the cost functions, $\mathcal{M}$ defined in Section \ref{sec:EvolveNetwork}.
 
\begin{figure}
    \centering
      \begin{minipage}[t]{0.24\textwidth}
        \vspace{0pt}
        \centering
        \includegraphics[width=\textwidth, height=0.32\textheight]{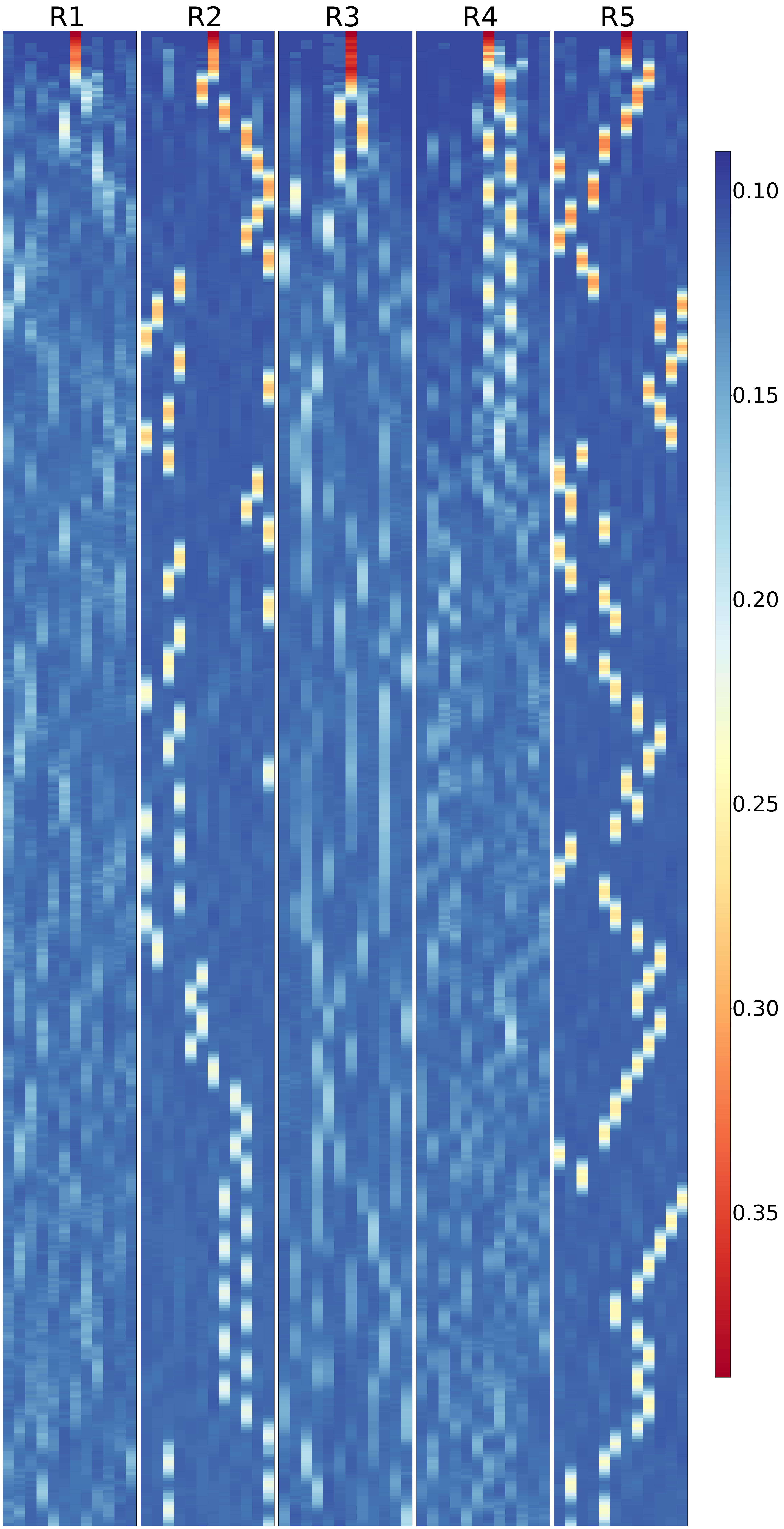}
    \end{minipage}
        \hfill
    \begin{minipage}[t]{0.75\textwidth}
        \vspace{0pt}
        \centering
        \includegraphics[width=\textwidth, height=0.16\textheight]{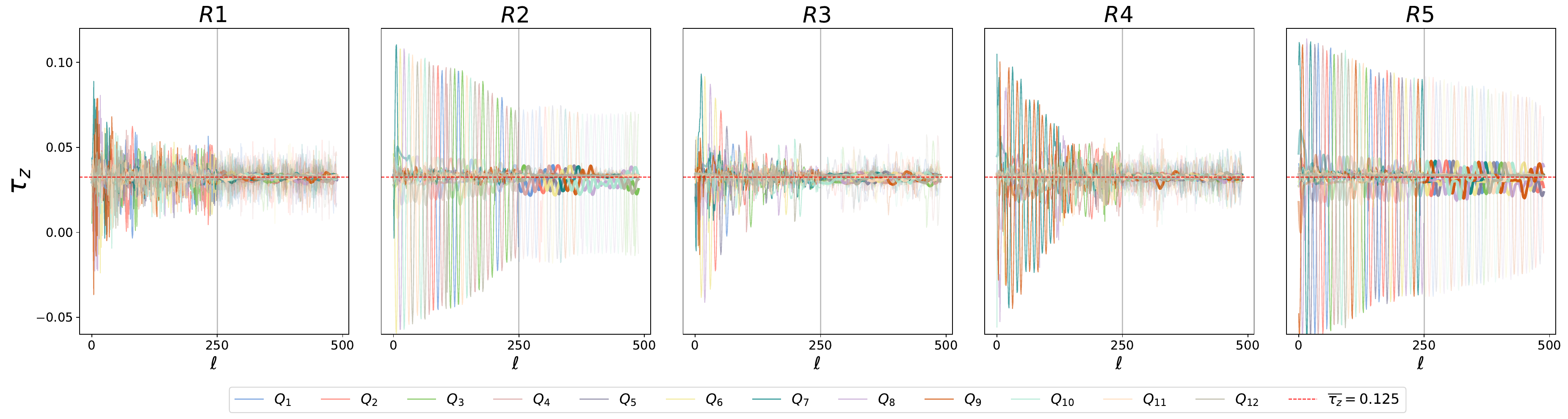}
        \vspace{0cm} 
        \includegraphics[width=\textwidth, height=0.16\textheight]{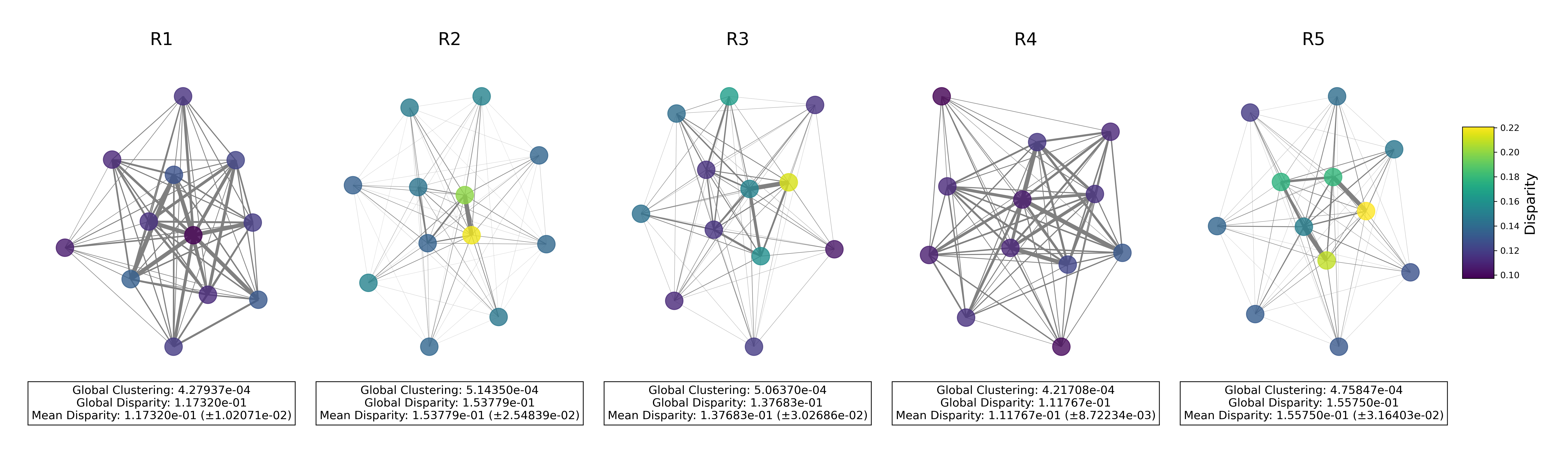}
    \end{minipage}%
    \caption{{\bf Left:} Heatmap of $\langle\langle\sigma^z_{q}(\ell)\rangle\rangle_{\rm ens}$, for the {\bf CS1} central state and all update rules, as a function of circuit layer on 12-qubit networks with connectivity $C=4$. Pixels represent the individual qubits at each layer of the circuit. The initial state is shown at the top, with increasing circuit layer running down the page. {\bf Upper Right:} The evolution of $\tau_{z,q}$ of a single initial state from the {\bf CS1} ensemble. After layer $\ell=250$, the full dynamics is shown in lighter weight in the background, with dynamics smoothed with a moving time-averaging window of 30 circuit layers over-plotted in heavier-weight lines. The horizontal red-dashed line shows the value of $\tau_z$ that corresponds to the thermalizing map, Eq.(\ref{eq:thermalizingTau}), for this central state. {\bf Lower Right:} Mutual information network for one member of each {\bf CS1} ensemble, evolved with each update rule and $C=4$. Well-connected nodes are positioned more closely together, and nodes are colored according to their disparity. The thickness of lines connecting nodes is scaled with the value of mutual information shared by the pair.  \label{fig:C4_seed_45}}
\end{figure}

 The increased connectivity, given the two-local gate, generally leads to increased homogenization. However, for cases that do reach a non-equilibrium steady state, the structure is less locked into place among the qubits. These points are evident in the left panel of Figure \ref{fig:C4_seed_45}, which shows heat maps of $\langle\sigma_z\rangle$ for one member of the {\bf CS1} ensemble. For comparison at $C=2$, see Figure \ref{fig:temperatureHeatMaps}
 
 The top right panel of Figure \ref{fig:C4_seed_45} shows the evolution of the shift parameter of the dynamical map for a member of the {\bf CS1} ensemble for all rules. Unlike the non-equilibrium fixed points found in $C=2$ connectivity, the windowed time-average after $\ell=250$ shows that propagator maps for each qubit do not stabilize to fluctuate about fixed value. (Compare Figure \ref{fig:tz_time_avg}.) Even in the cases where the standard deviation is highest, $R2$ and $R5$, individual qubits do not settle down. This is consistent with trends shown in the heat map. Finally, the bottom right panel of Figure \ref{fig:C4_seed_45} shows the mutual information graphs. Consistent with the previous two plots, $R2$ and $R5$ generate graphs with higher disparity (indicated by the color of the node). (Compare to Figure \ref{fig:MI_network_late_time_avg} for $C=2$).

In the distributions of $\tau_z$ the random rule on $C=4$ has a lower standard deviation and fewer instances of non-(C)P propagator maps compared to $C=2$. In keeping with the results in Figure \ref{fig:C4_seed_45} above, only $R2$ and $R5$ show large ensemble-averaged $\sigma(\tau_z)$ and high probability for non-(C)P dynamics when $C=4$. For $C=2$, $R2-R5$ all had an order of magnitude higher $\sigma(\tau_z)$ compared to $R1$. For the other mixed central states, {\bf CS2} and {\bf CS3}, only $R2$ and $R3$, the global maximization rules, maintain high standard deviation and high probability for non-(C)P dynamics.

\begin{figure}
    \centering
\includegraphics[width=\textwidth,height=0.5\textheight]{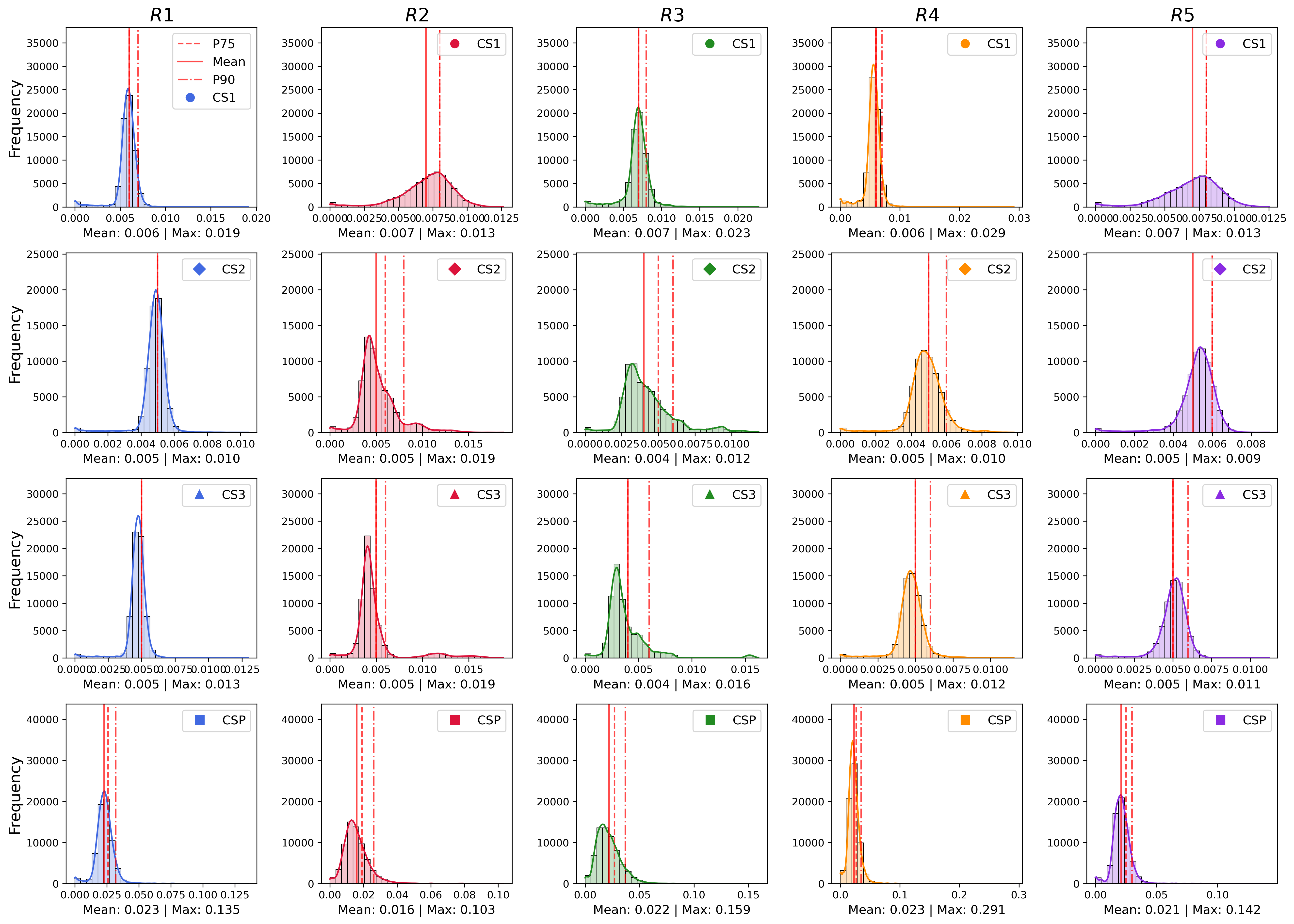}
    \caption{The distribution of the correlation magnitudes, $\frac{1}{4}|C^{xx}_{ab}|$, between all possible pairs of qubits ($Q_a$, $Q_b$) on 12-qubit networks with $C=4$ connectivity. The distributions include correlations at all circuit layers and for all members of the central state ensembles. The mean and maximum amplitudes are noted under each plot, and the solid line shows the mean. The dashed and dot-dashed vertical lines show the values below which $75\%$, and $90\%$ of values occur, respectively. \label{fig:tz_corr_histogram_c4}}
\end{figure}

As discussed in the main body of the paper, much of the behavior of the $\tau_z$ distributions can be traced back to the two-qubit correlations. If $\tau_z$ has high variance, large correlations must develop on the network. However, significantly different distributions in the correlations and mutual information can all generate large $\sigma(\tau_z)$. Figure \ref{fig:tz_corr_histogram_c4} shows the distribution of the magnitudes of correlations for different central states and different update rules, on the connectivity $C=4$ networks. (The same information for $C=2$ was shown in Figure \ref{fig:Corr_12_Q}.) For both connectivities, the networks that have inhomogeneous subsystem dynamics have correlation distributions skewed to the left. In those cases, a few qubit pairs share higher magnitude of correlation at the expense of most pairs having lower correlation.

\begin{figure*}
    \centering
    \includegraphics[width=\columnwidth]{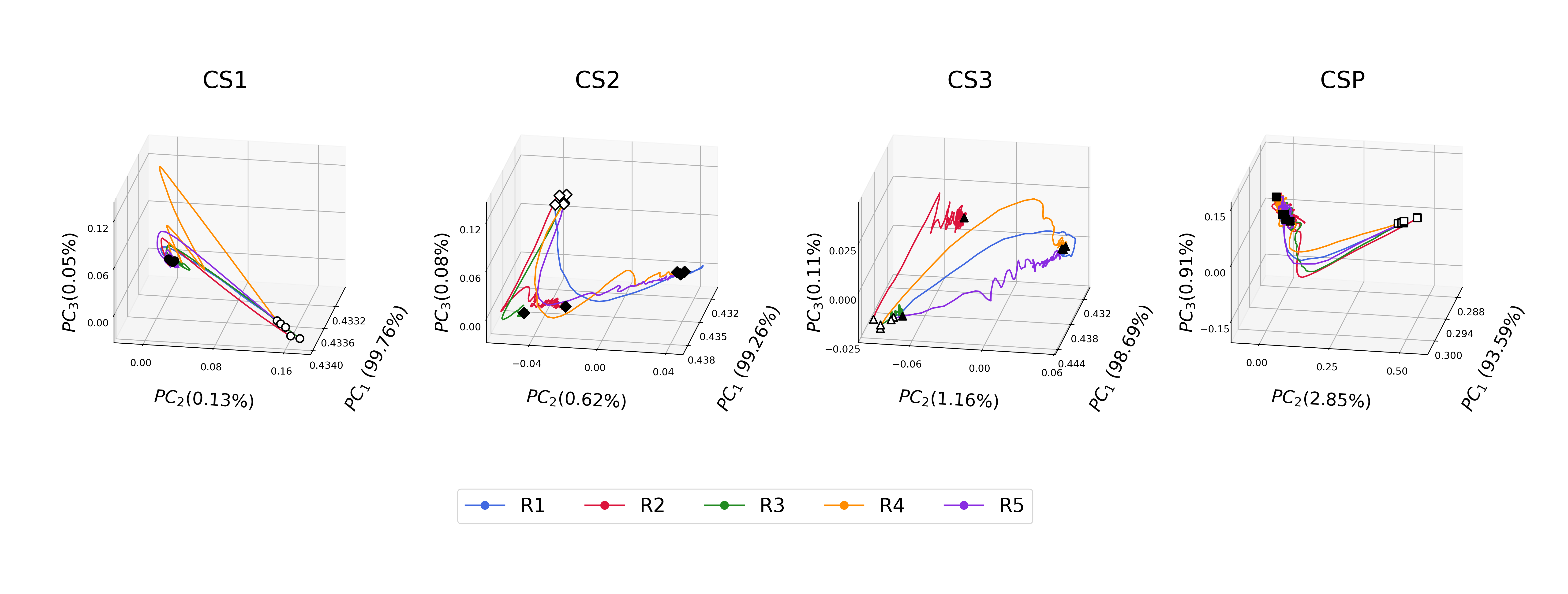}
    \caption{PCA plots for the $\langle\sigma_z\rangle$ values for all the central states and rules for 12-qubit networks with connectivity $C=4$. The PCA was performed on the dataset containing all the rules for all the members of a central state. The unfilled symbols indicate the projected state at layer $\ell =10$ and filled symbols are the last layer.     \label{fig:PCA_C4}}
\end{figure*}

The PCA state space analysis shown in Figure \ref{fig:PCA_C4} shows a very different behavior at $C=4$, without the steady-state cycles seen in $C=2$ (Figure \ref{fig:PCA_across_ICs}). For most central states, most rules take the networks to the same part of the single-qubit state space (convergence of filled symbols). The exceptions are $R2$ and $R3$, for {\bf CS2} and {\bf CS3}, which remain in a distinct part of the space.

\begin{figure}
\centering
\begin{minipage}[b]{0.48\textwidth}
\centering
\includegraphics[width=0.85\textwidth, height=0.3\textheight]{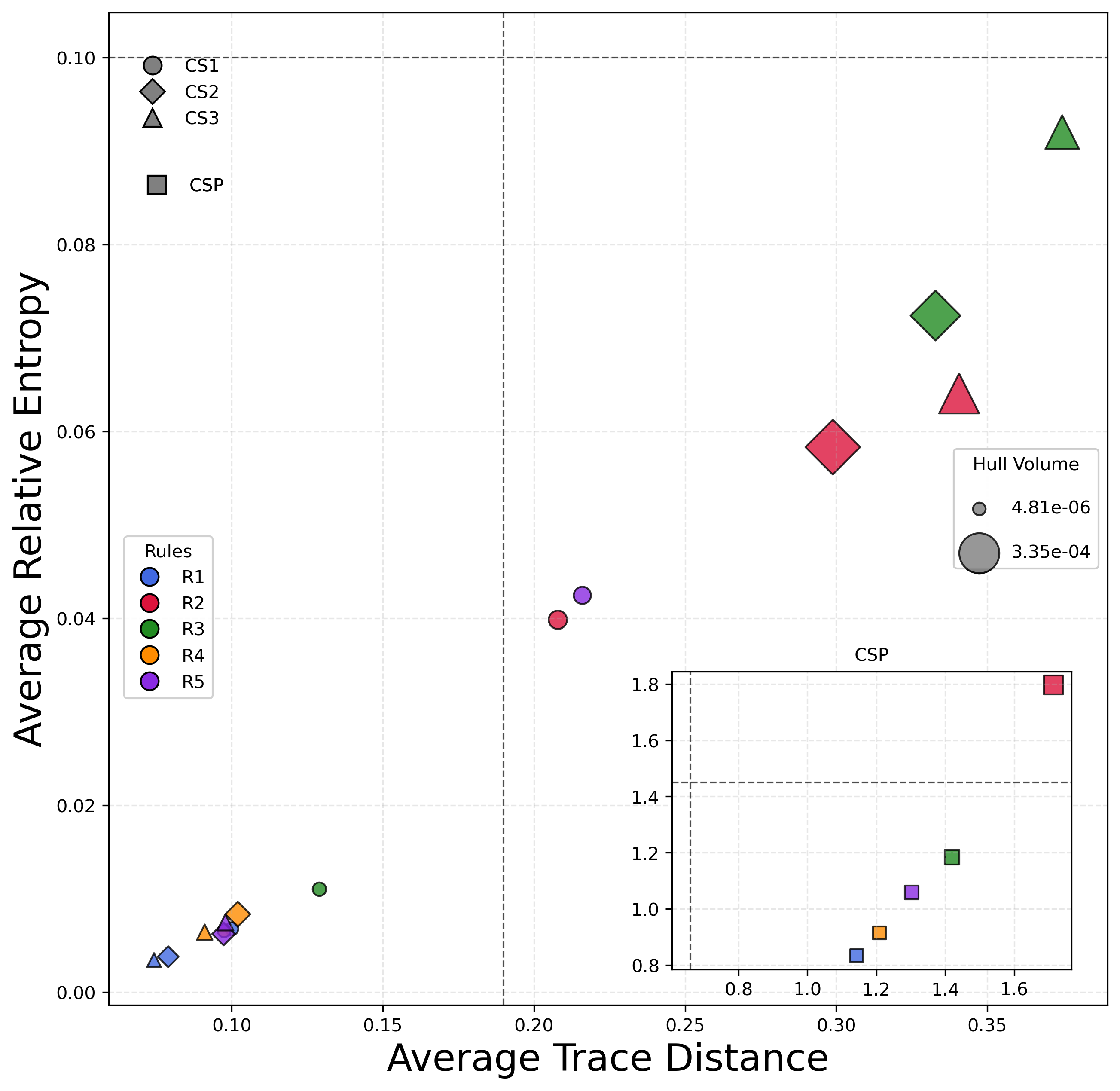}
\end{minipage}
\hfill
\begin{minipage}[b]{0.48\textwidth}    
\centering
\includegraphics[width=0.85\textwidth,height=0.3 \textheight]{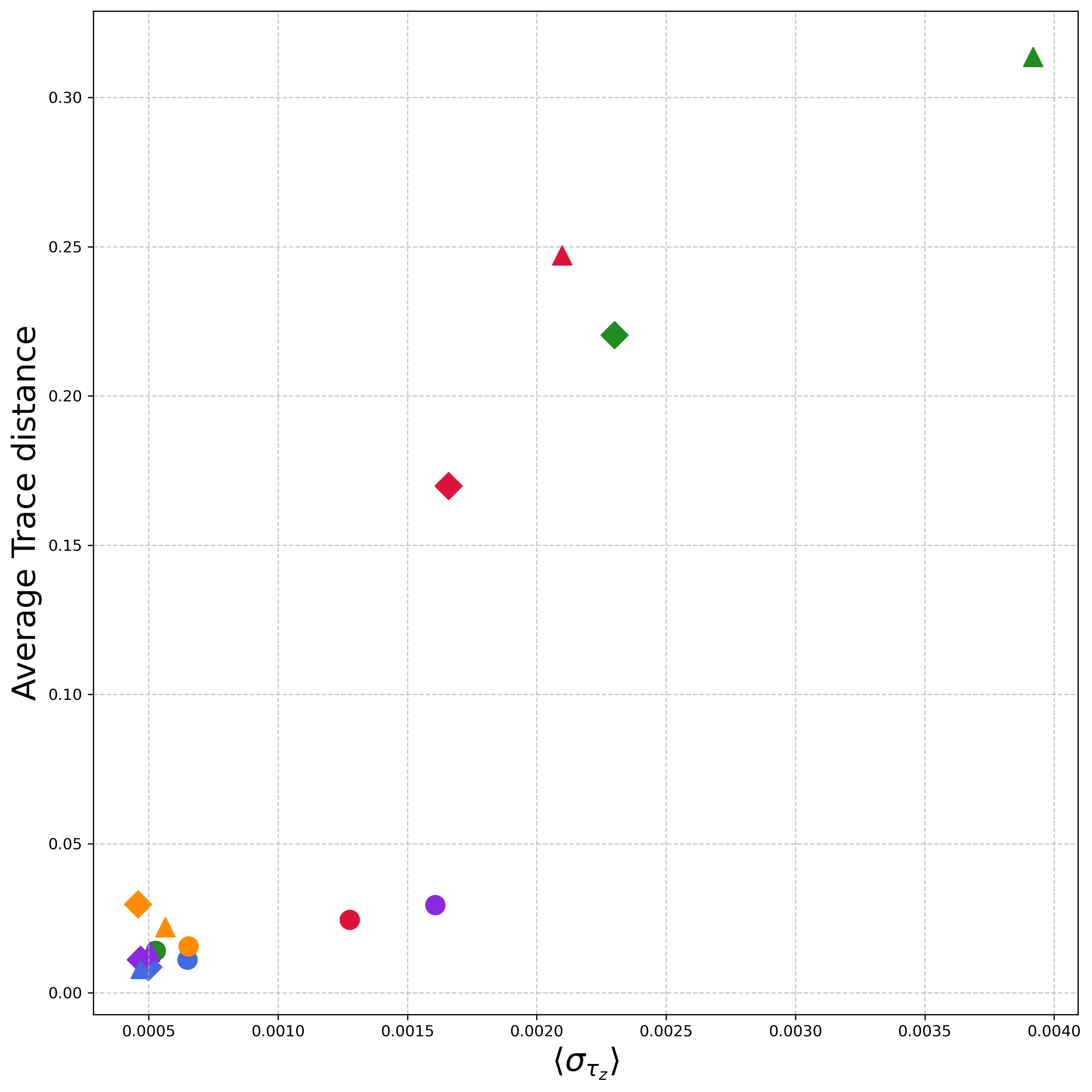}
\end{minipage}
\caption{\textbf{Left:} Ensemble-averaged, late-time total relative entropy $\langle \langle \mathcal{D}(\{\rho_q\},\bar{\rho})\rangle\rangle_{\rm ens}$, total trace distance, $\langle \langle {Tr(\rho^{(N)},\bar{\rho}^{(N)}}\rangle\rangle_{\rm ens}$, and volume of the convex hull $V_{CH}$ for 12-qubit networks with connectivity $C=4$. The symbol size scales with the convex hull volume, with the relationship shown for two examples in the inset legend. The vertical dashed line is the average trace distance of the initial state distributions about the three mixed central states. The horizontal dashed line is the mean relative entropy of the three central states after the 10 layers. \textbf{Right:} The average trace distance vs $\langle \langle\sigma(\tau_z)\rangle\rangle_{\rm ens}$. \label{fig:KL_vol_TD_C4}}
\end{figure}

Finally, Figure \ref{fig:KL_vol_TD_C4} shows the relationship between several state space measures for all rules, all central states on the $C=4$ networks. (Compare Figure \ref{fig:KL_vol_TD} for $C=2$.) These results show that for {\bf CS1} (circles) $R2$ and $R5$ do stay away from the thermal state better than the other rules. However, the more significant outliers are $R2$ and $R3$, for {\bf CS2} and {\bf CS3}.

\subsection{Approach to the steady state}
\label{sec:timeEvol}
Many of the qubit networks with connectivity $C=2$ reach a steady state. This section shows several examples of the evolution across many layers to show the approach to the steady state, and how it compares between central state distributions and update rules. 

\begin{figure}[t]
    \includegraphics[width=\columnwidth]{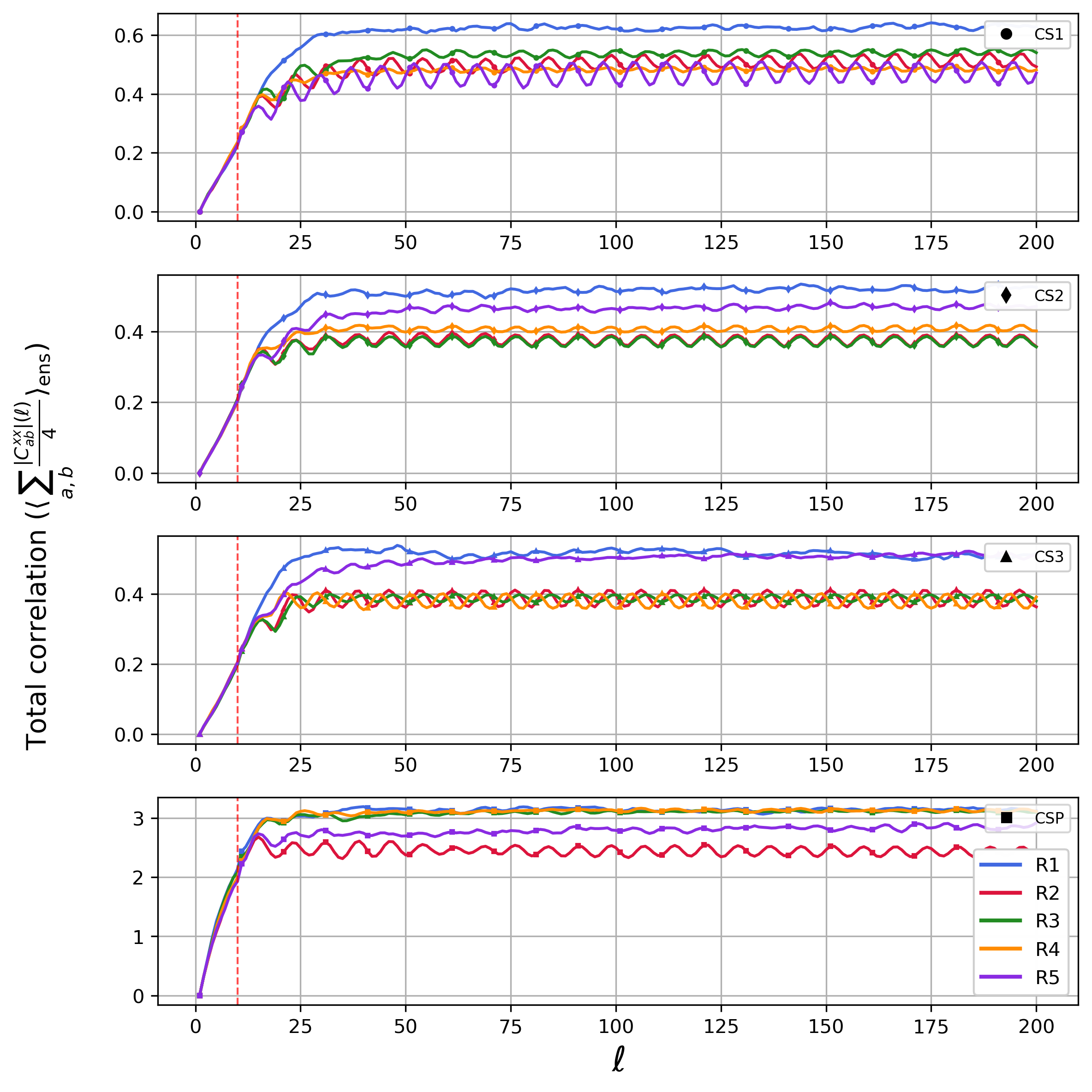}
    \caption{The evolution of total correlations in two-qubit density matrices on 12-qubit networks with connectivity $C=2$, averaged over all members of each central state distribution. The vertical dashed red line shows circuit layer 10, where the transition from random gates on a fully connected ($C=N-1)$ network to evolution with $C=2$ and the update rules $R1$-$R5$ occurs. \label{fig:Corr_growth}}
\end{figure}

Figure \ref{fig:Corr_growth} shows the time evolution of the total magnitude of correlations, Eq.(\ref{eq:CorrTotal}) for all central states (one per panel) and all update rules (colored trajectories). In all cases, the total two-qubit correlation in the network reaches a stable value, although the steady-state value differs for different rules. Random dynamics ($R1$, the blue lines), generates the largest amplitude of correlations for all mixed central states (top three panels). This is consistent with the expectation that thermalizing dynamics must move more information into correlations. The relative behavior of the non-thermalizing dynamics, $R2-R5$ in the top three panels, depends on the central state. This indicates that for dynamics that retains a memory of the initial state, the non-equilibrium steady state is a function of both the configuration and the dynamics. Finally, notice that for the pure central state (bottom panel), there is not a sharp distinction between $R1$ and the non-equilibrating rules. For {\bf CSP} and system size $N=12$, $R1$ does not generate a steady state that is particularly close to thermalized.
It is particularly interesting to compare $R2$ and $R5$ (see also Figure \ref{fig:KL_vol_TD} and the discussion on this point). While $R5$ has the lowest total correlation for {\bf CS1}, it behaves much more like the random rule for the other mixed central states. In contrast, $R2$ performs similarly to $R3$ and $R4$ for {\bf CS2}, {\bf CS3}, but for $CSP$ it generates the lowest total correlation.

\begin{figure}
    \includegraphics[width=\columnwidth]{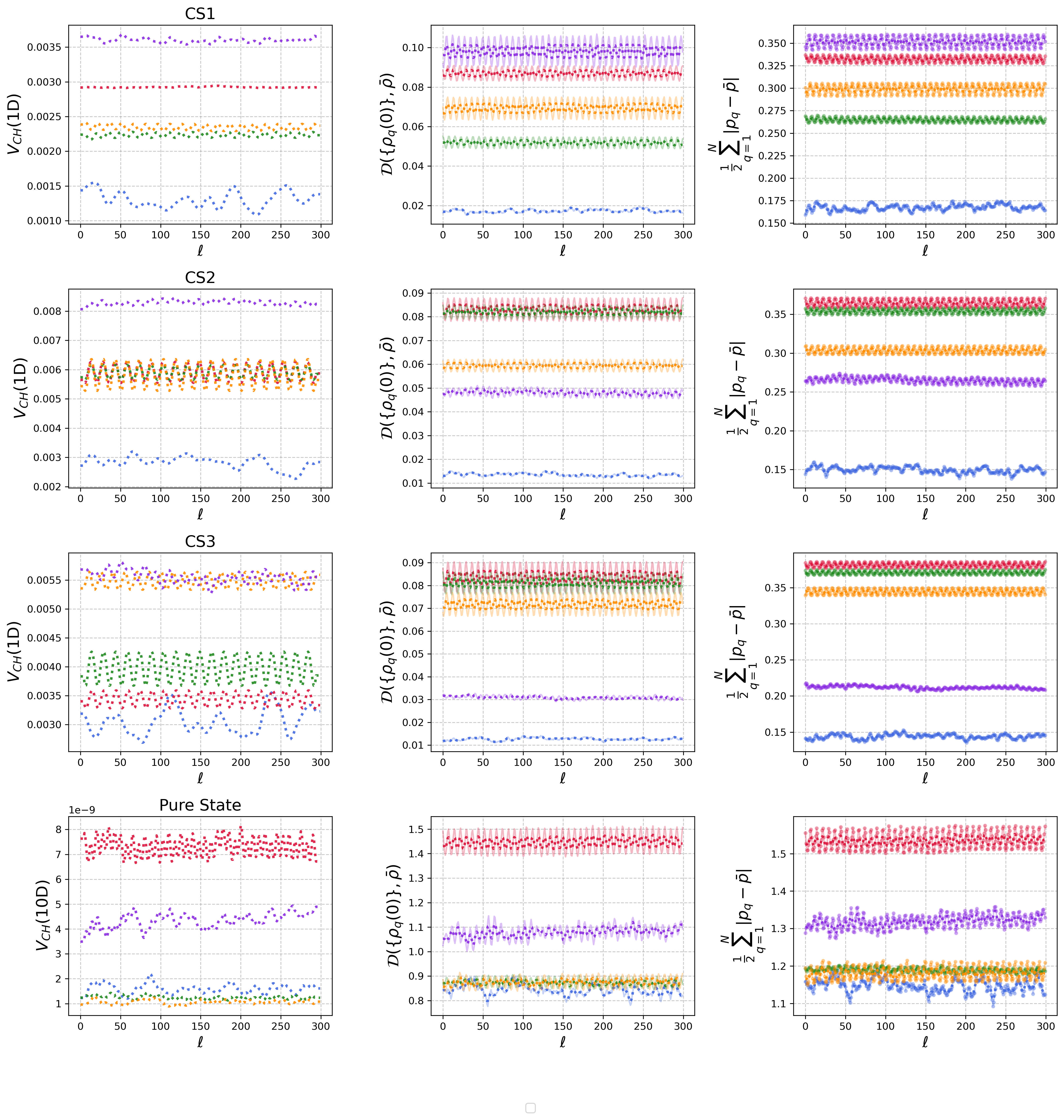}
    \caption{The ensemble-averaged convex hull volume, the average relative entropy and trace distance as a function of circuit layer for 12-qubit networks with connectivity $C=2$.  
    \label{fig:StateSpaceLayers}}
\end{figure}

Figure \ref{fig:StateSpaceLayers} shows how the relative entropy, trace distance, and convex hull volume evolve as a function of layer. The non-random rules frequently display a periodicity related to the fixed angle in $U^{*}$ together with the development of locked interaction neighborhoods, while the random rule, $R1$, shows a less regular structure.

\subsection{Non-(C)P maps and system size effects}
\label{sec:NCPscaling}
\begin{figure*}
\includegraphics[width=0.5\linewidth]{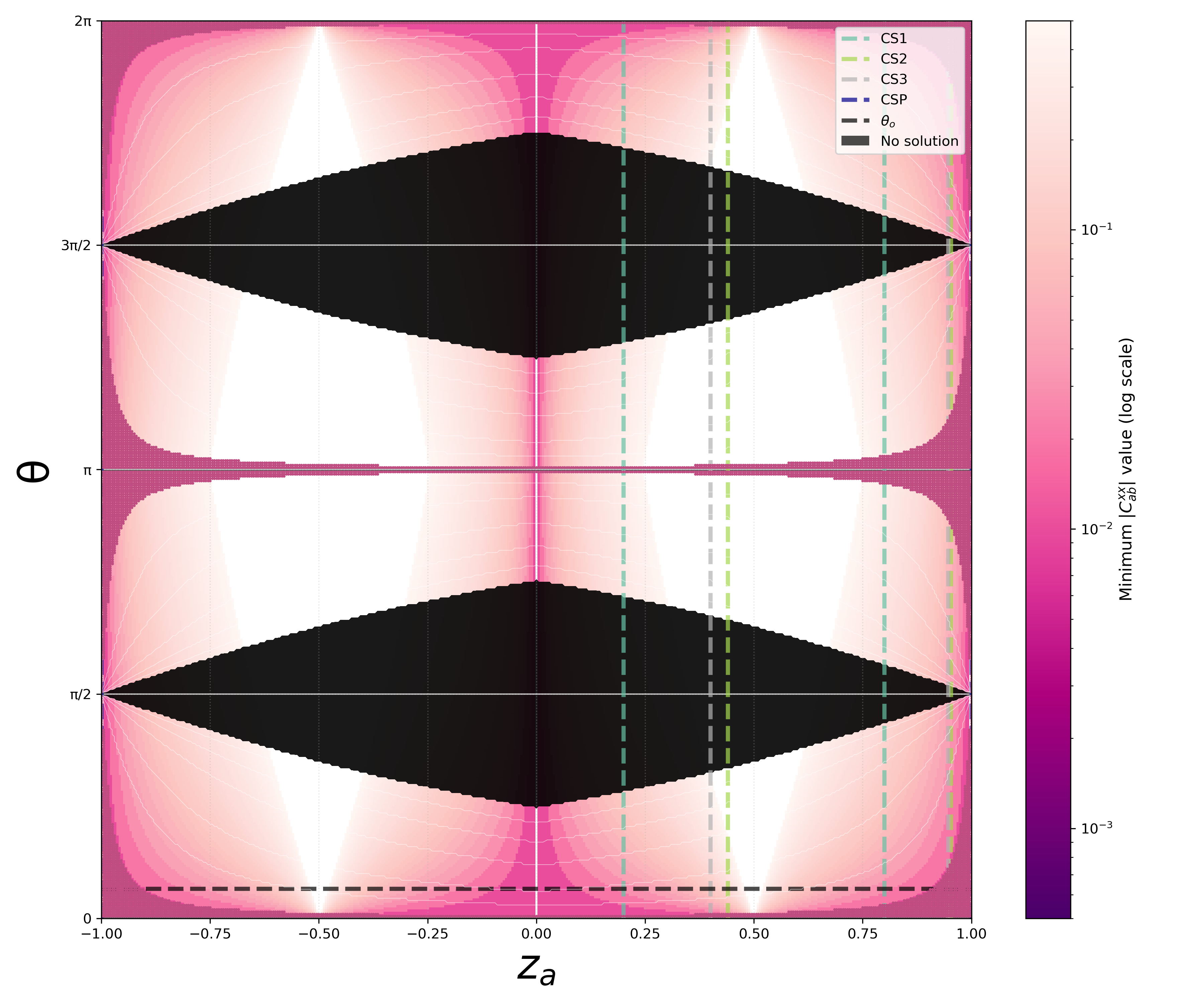}
    \caption{The minimum magnitude of two-qubit correlation, $C^{xx}$, needed to break the (complete) positivity condition, Eq.(\ref{eqn::CPcond}), as a function of the $z$-component of the Bloch vector of the partner qubit and the rotation angle $\theta$ in the unitary. The black region corresponds to values of $\theta$ for which there is no admissible value of the correlation that can generate NCP dynamics. The color scheme used has a log scale to emphasize the regions of the parameter space where low magnitude of correlations can support NCP maps. The horizontal black dashed line corresponds to $\theta$ used in all examples presented in this paper. The vertical lines show the bounds on the $z$-component of the Bloch vector imposed by the different central states presented in the paper. \label{fig:nonCPconditions}}
\end{figure*}

The condition for propagator maps to be (completely) positive was given in Eq.(\ref{eq:oneParamCP}). Figure \ref{fig:nonCPconditions} shows the minimum magnitude of two-qubit correlation, $|C^{xx}|$, needed to break this condition for any single-parameter two-qubit unitary and any mixed initial state of the type considered in this paper.

\begin{figure}
\begin{minipage}[b]{0.48\textwidth}
\includegraphics[width=\textwidth, height=0.6\textheight]{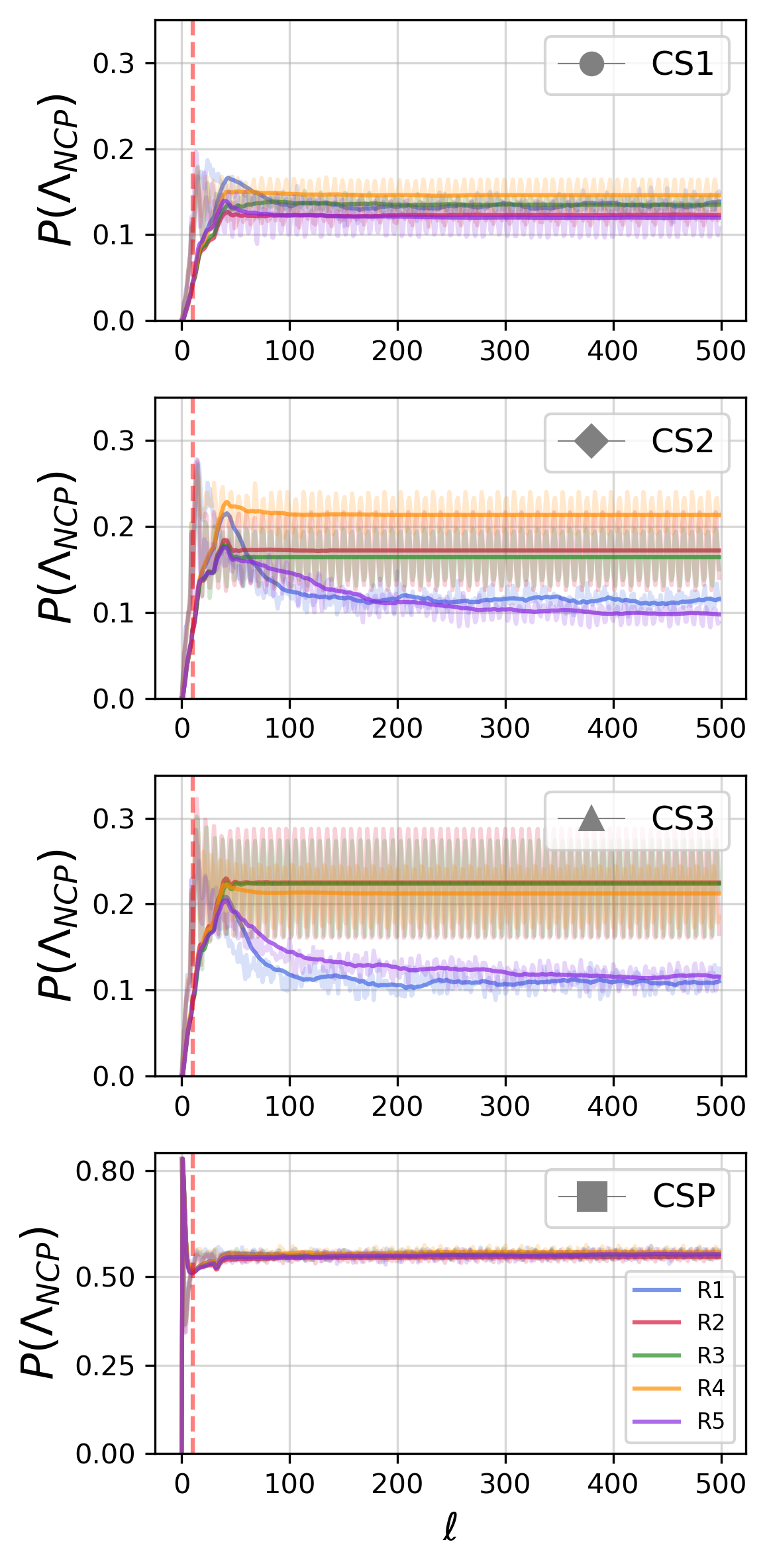}
\end{minipage}
\begin{minipage}[b]{0.48\textwidth}
\includegraphics[width=\textwidth, height=0.6\textheight]{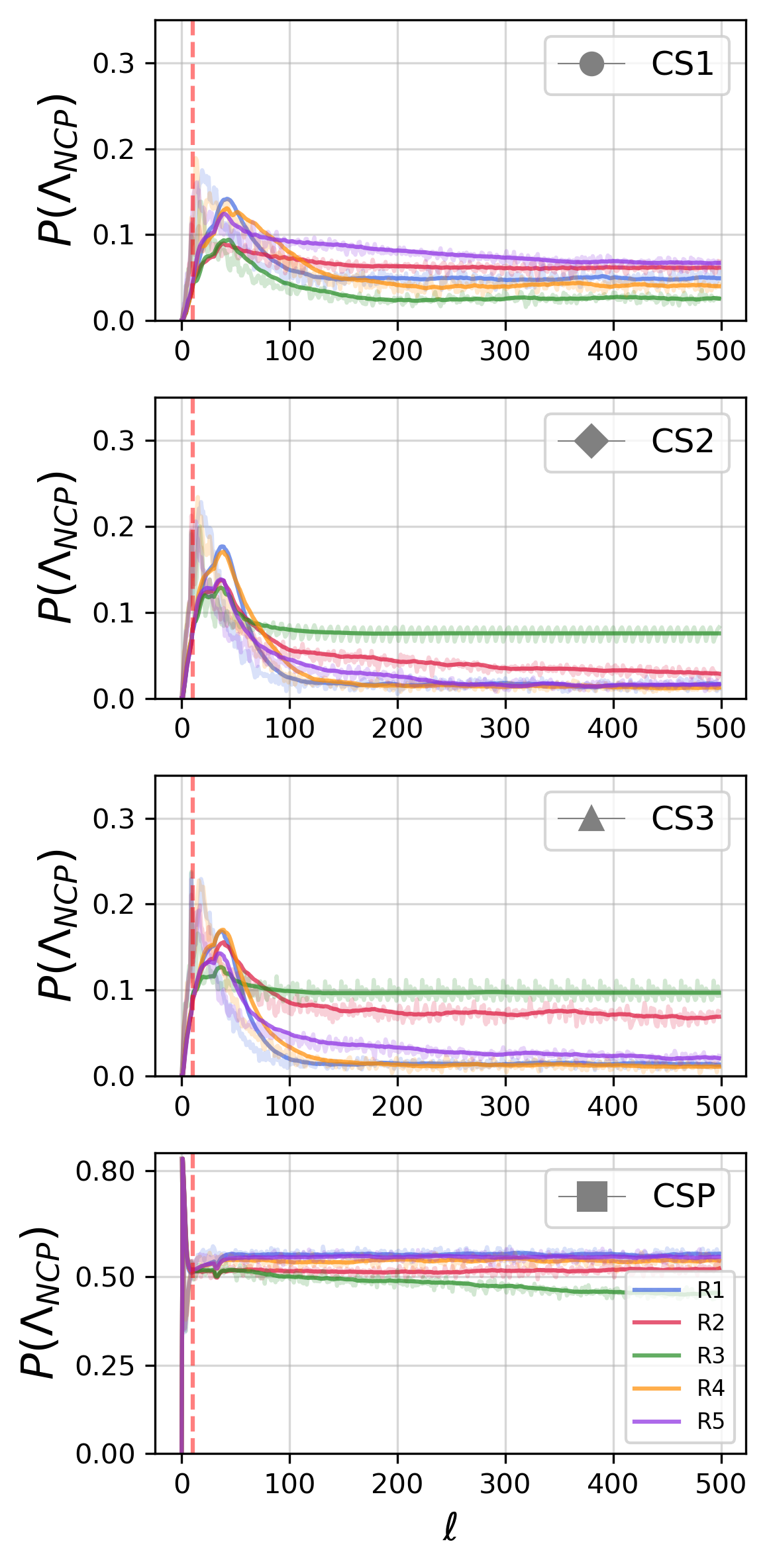}
\end{minipage}
    \caption{Probability of non-(C)P dynamics in 12-qubit networks for connectivity $C=2$ (left) and $C=4$ (right). These are the original dynamics, not noise-reduced. \label{fig:NCP_prob}}
\end{figure}

The histograms for the propagator map parameter $\tau_z$ (Figure \ref{fig:TauzDistributions}) reported the percentage of layers with propagator maps that violated the (C)P condition. To provide a bit more detail about differences between central states and rules, and to connect with the time-evolution show in the previous section, Figure \ref{fig:NCP_prob} shows the probability as a function of circuit layer. 

We expect that the non-(C)P behavior seen in the random rule is a feature of the finite system size. In a larger network with homogenizing dynamics, all maps should tend rapidly toward the map whose fixed point is the maximum entropy single-qubit state, $\rho^*$, Eq.(\ref{eq:mapInvariantState}). This map must be C(P), which we can explicitly confirm for the dynamics used in this paper as follows.
The symmetric dynamics imposed here restrict the Bloch vector for the single-qubit state $\rho^*$ to have only one non-trivial component, $z^*$. If the rest of the qubits act as a perfect bath for each individual qubit at late times, then all qubits share the same value of $z^*$ for $\ell \geq \ell^*$, and all qubits evolve by the dynamical map whose fixed point state has a Bloch vector described by $z^*$. Using Eq.(\ref{eq:zEvolve}) and Eq.(\ref{eq:lambdaTau}), that means 
\begin{align}
    z_{q}(\ell)  &= \cos^2 (\theta)  z_q(\ell-1) + \tau_{z,q}(\ell -1) \\
    z ^* &= \cos^2 (\theta)  z ^* + \tau_{z,q}(\ell -1) \\
    \implies \tau_{z,q} &= z^* \sin^2 (\theta)
\end{align}
for $\ell \geq \ell^*$ and for all the qubits on the network. It follows from this that for a thermalizing network, the (C)P condition at late times must be obeyed by all dynamical maps since $\cos^2 (\theta) + |z^* \sin^2(\theta)|\leq1$ is trivially satisfied $\forall\, |z^*|\leq 1$. 
However, $\tau_{z,q} = z^* \sin^2 (\theta)$ can only be true as an exact statement when two-qubit correlations are zero. That is, the map with $\tau_z (\ell) = z^* \sin^2 (\theta)$ exactly corresponds to the dynamics produced in a collisional model where the system qubit always interacts with an uncorrelated bath qubit with the $z$-component of its Bloch vector $z^*$. 

\begin{figure}
    \subfloat[]{%
  \includegraphics[width=0.48\columnwidth]{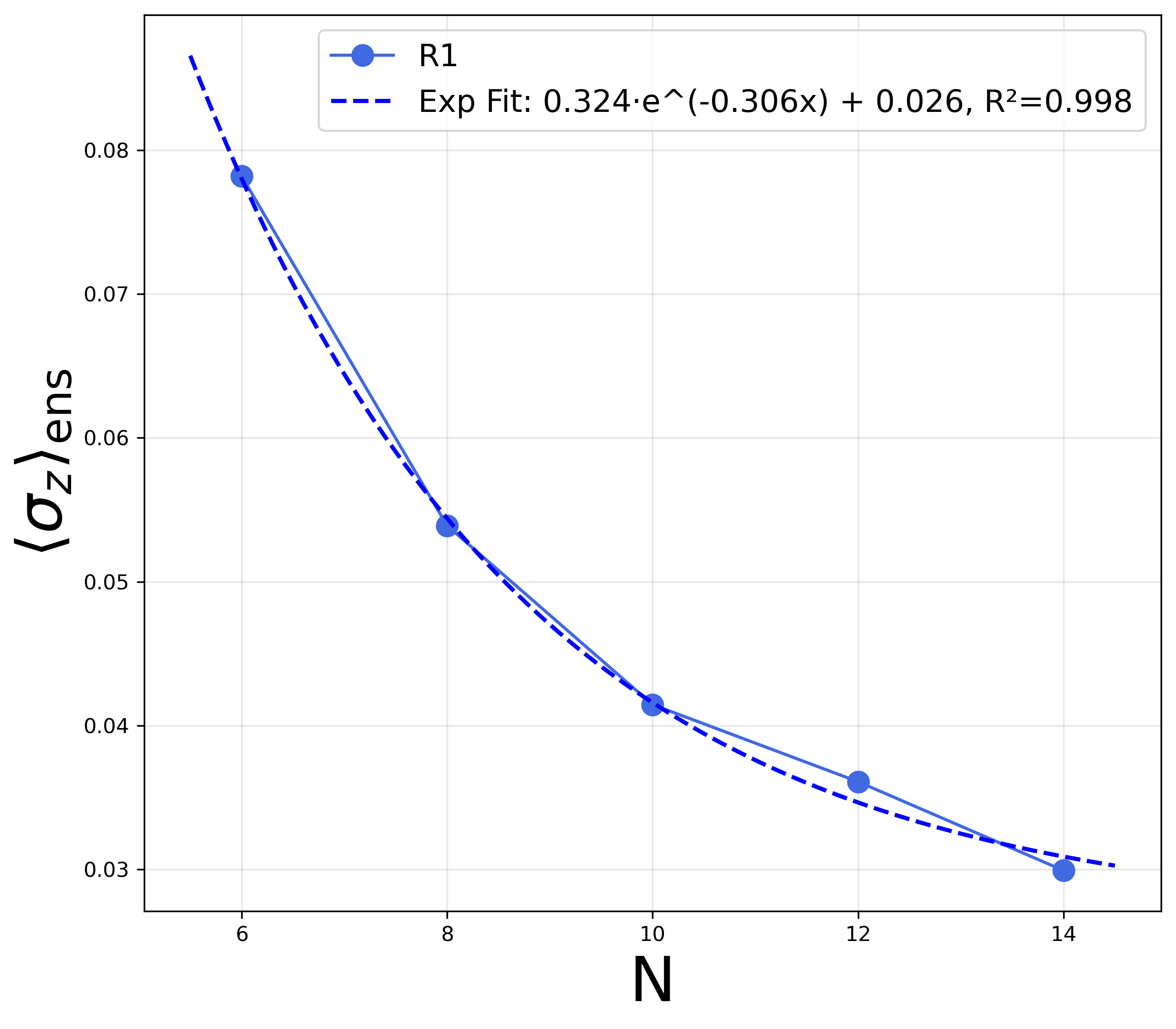}%
}
    \hfill
    \subfloat[]{%
\includegraphics[width=0.48\columnwidth]{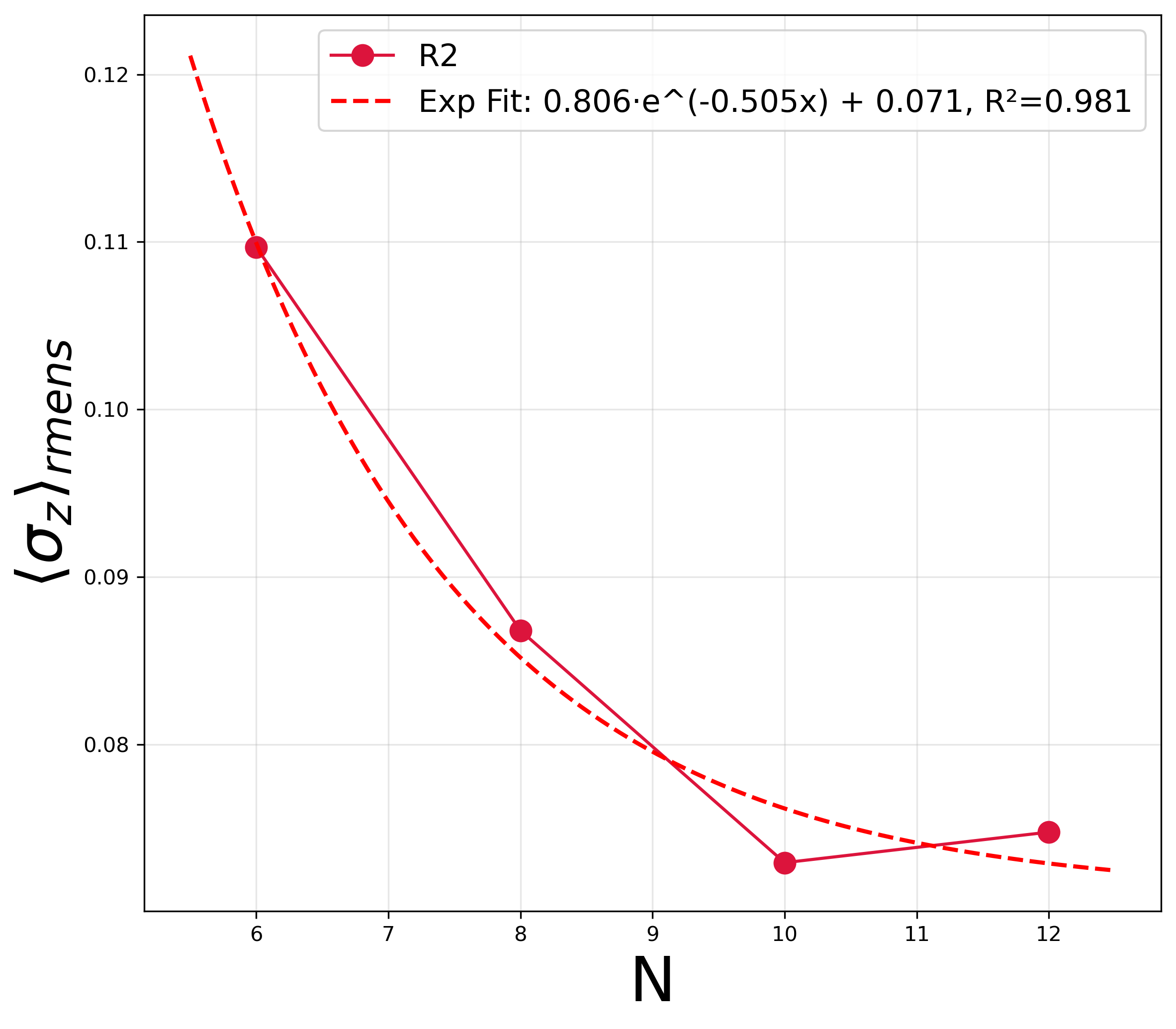}%
}
    \caption{The standard deviation of $\langle \sigma_{z}\rangle_{\rm ens}$ on $C=2$ networks as a function of system size $N$, averaged over all qubits, circuit layers $\ell=300-500$, and each of the 100 members of the {\bf CS1}. At each system size random evolution, $R1$, produces $\langle \sigma_z\rangle$ significantly lower than that generated under $R_2$. The large-$N$ limit of the exponential fit is an order of magnitude higher value in $R2$ compared to $R1$. }
    \label{fig:fss-pops-std}
\end{figure}

For finite $N$ and $\ell$, we expect that approximately thermalizing networks will have $|z_q(\ell) - z^*|\leq \epsilon $ for some finite $\epsilon>0$ that depends on the size of the system and tends to zero. Figure \ref{fig:fss-pops-std} confirms that with the random rule, $R1$, the typical variance in $\langle\sigma_z\rangle$ decreases rapidly with increasing system size. The non-random rule $R2$, shows a significantly higher variance at low $N$, and is an order-of-magnitude larger in the the large-$N$ limit of the fit.

\begin{figure}
\subfloat[]{%
\includegraphics[width=0.48\columnwidth]{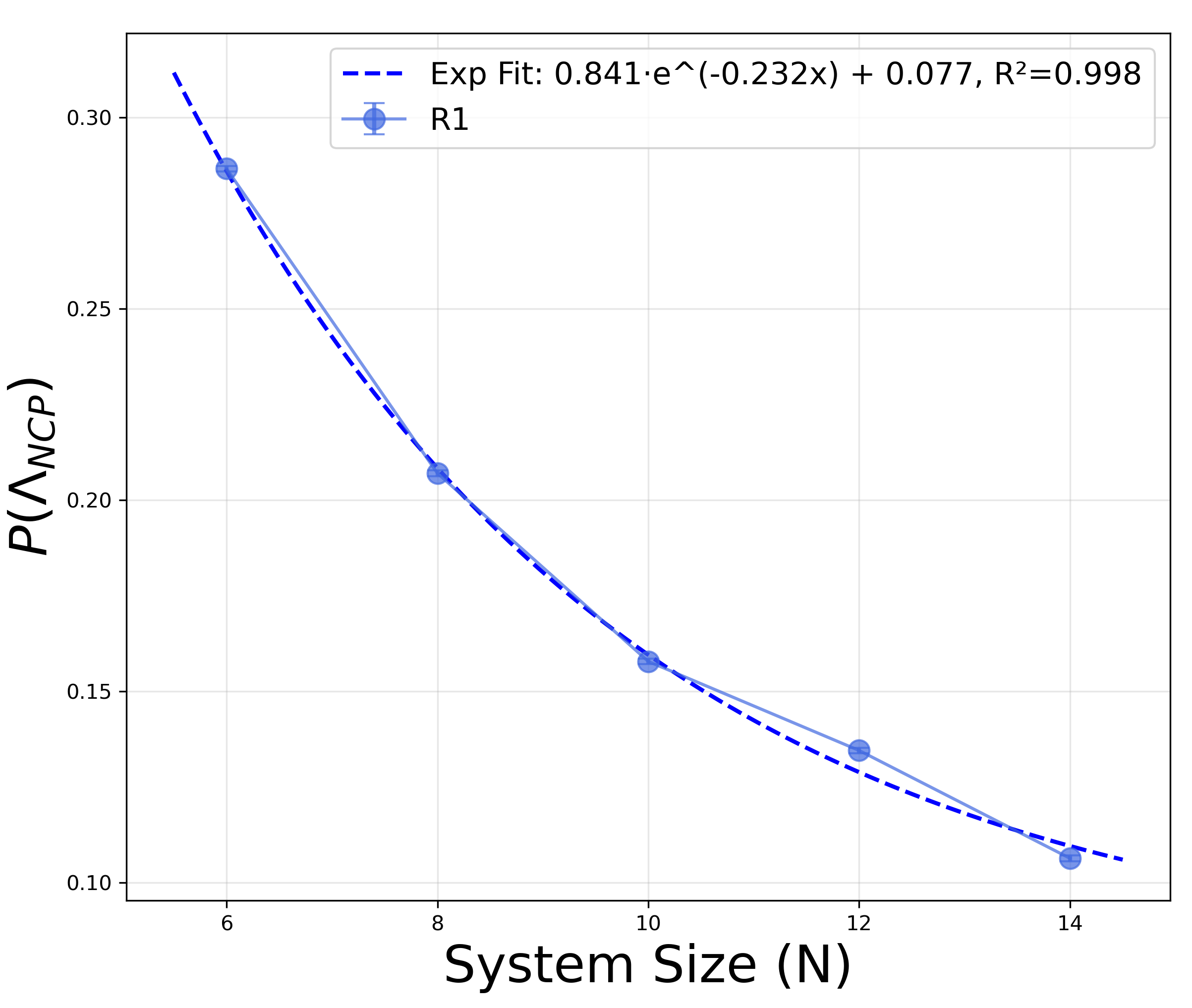}%
}
    \hfill
    \subfloat[]{%
\includegraphics[width=0.48\columnwidth]{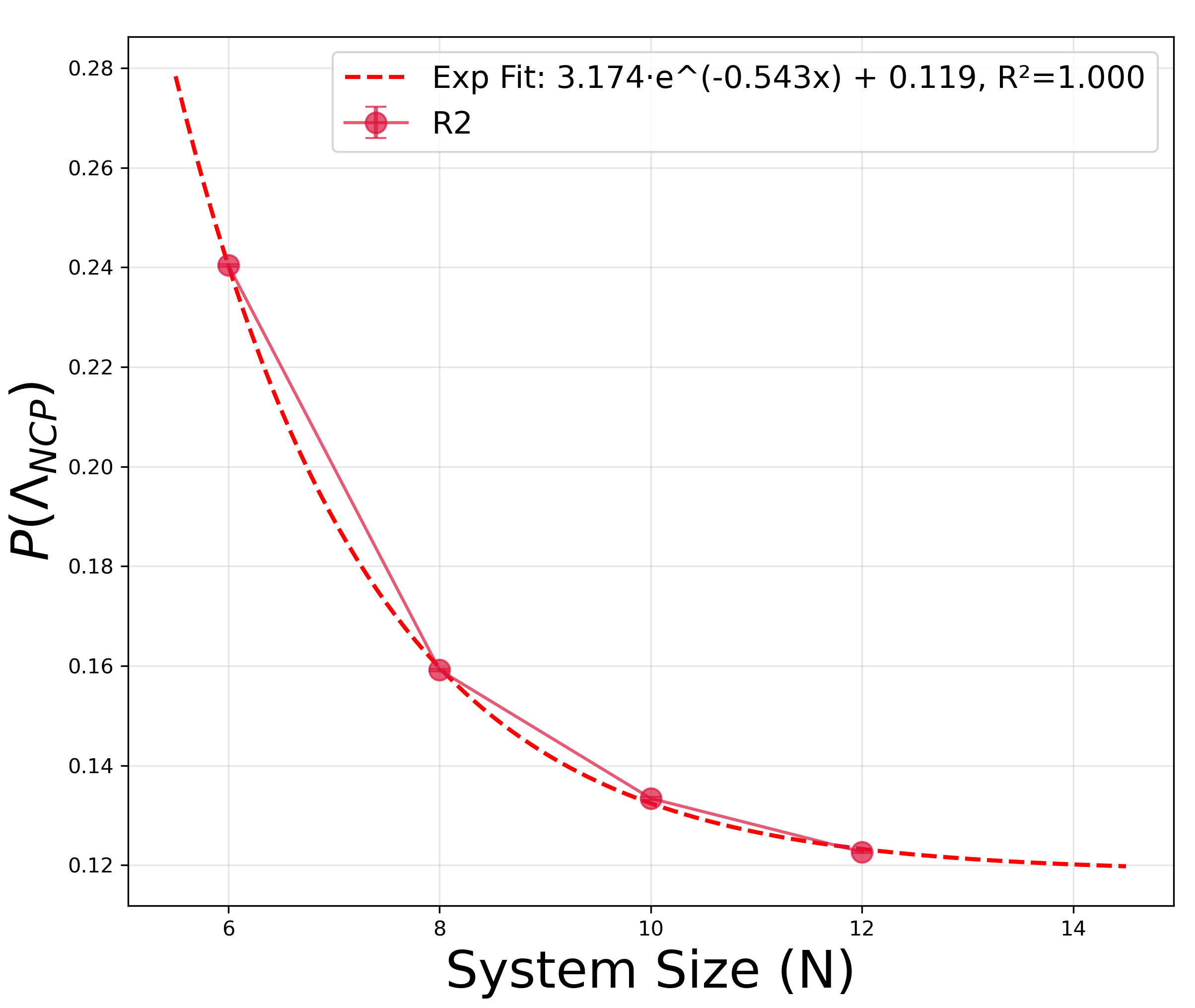}%
}
    \caption{The fraction of propagator maps that break the (complete) positivity condition, among layers $\ell=300-500$, over all 100 members of the {\bf CS1} ensemble with $C=2$. The presence of non-(C)P dynamics imply the presence of correlations that prevent the rest system from acting as a perfect thermal bath for any individual qubit.  \label{fig:fss-NCP}}
\end{figure}

Figure \ref{fig:fss-NCP} shows the corresponding scaling of the fraction of non-(C)P dynamics in the high-$\ell$ propagator maps. Again, $R2$ has a different scaling with $N$, suggestive that the dynamics will maintain non-(C)P dynamics even for very large systems.
For these reasons, we expect that the non-(C)P dynamics in the networks evolved with the random rule is largely a finite-system effect. This is the motivation for examining the noise-reduced maps instead in Section \ref{sec:BernouliResults}. 

 \section{Principal Component Analysis}
\label{sec:AppPCA}
Principal component analysis is a dimensional reduction technique that enables analysis of a high dimensional data set along a smaller number of modes with the most variance. We use PCA on the space of values of $\langle\sigma_z\rangle$ (which is linearly related to the excited state population $p_q = \frac{1-<\sigma_{z,q}>}{2}$). The full set of data contains values of $\langle\sigma_z\rangle$ labeled by qubit number ($q$), circuit layer ($\ell$), and trial number ($T$) labeling the position in the central state ensemble. That is, each entry is $\langle\sigma_z(q,\ell,T)\rangle$. For each network of size $N\in \{6,8,10,12,14\}$ and each central state $CSj\in \{CSP_1, CSP_2,CSP_3,CSP\}$, evolved with a rule $R_i\in \{R_1,R_2,R_3,R_4,R_5\}$, we store the data for each qubit, across 500 circuit layers, for each of the 100 networks in the central state ensemble. This gives 20 different $N\times 500\times100$ arrays $\{X_{i,j}\}$ where $i$ labels the rule and $j$ labels the central state. Although PCA is often performed by centering the data around the mean value of the dataset, here the $\langle\sigma_z\rangle$ data for each qubit generally do not stabilize to a fixed value at late times. We perform uncentered PCA to capture variations in $\langle\sigma_z\rangle$ as a function of the depth of the circuit. 

To display trajectories in the PCA space with less noise, we use ensemble-averaged data 
\begin{equation}
    \{\bar{X}_{i,j}\}=\{\langle\langle\sigma_z(q,\ell,T)\rangle\rangle_T\}
\end{equation}
These arrays are $N\times500$. The ensemble-averaged data is used in Figures \ref{fig:PCA_across_ICs} and \ref{fig:PCA_across_rules}. The convex hull volume, however, is the space spanned by the members of the ensemble and so uses the full $\{X_{i,j}\}$ (Figures \ref{fig:first_10_steps_IC}, \ref{fig:KL_vol_TD} and \ref{fig:KL_vol_TD_C4}). In addition, one can select for just some circuit layers. Any data sets can be combined, for example,  to analyze the $\langle\sigma_z\rangle$ common to a fixed rule across central states. Because these data sets sometimes have overlapping values in the first ten steps, we add a small Gaussian noise with mean 0 and standard deviation $10^{-6}$ for numerical stability. 

For any data set, we first flatten the array so that it is a list of $N$-tuples ($N$ is the number of qubits). For this data set $X$, we compute the covariance matrix $C$ of dimension $N\times N$
\begin{equation}
C = \frac{1}{N-1} X^TX\,.
\end{equation}
Then the algorithm computes the eigenvectors and eigenvalues of the covariance matrix and sorts them in descending order based on the eigenvalues. The eigenvectors corresponding to the largest eigenvalues are the principal components that capture the most variance in the data.
The eigen-decomposition gives us a set of eigenvectors $v_1,v_2,v_3 \ldots v_N$ each of length $N$ and corresponding eigenvalues $\lambda_1,\lambda_2 \ldots \lambda_N$. The top $m$ eigenvectors can be used for the $m-$dimensional PCA decomposition. The projection of the data onto the eigenvectors is
\begin{equation}
T = X M\,
\end{equation}
where $M$ is a matrix generated from concatenating the $m$ eigenvectors $v_1,v_2\ldots v_m$.

As the number of eigenvectors $m$, is increased, the PCA captures more of the variance of the data. To compute the convex hull volume, we used only enough components to capture variance greater than or equal to 0.85. The convex-hull volume is computed using SciPy's \href{https://docs. scipy.org/doc/scipy/reference/generated/scipy.spatial.ConvexHull.html}{ConvexHull} function. The code finds the smallest convex set that contains all the projected points. The volume is then calculated by decomposing the full hull into triangle simplices. The sum of the volumes of each simplex then gives the estimate for the full volume. 

PCA can be sensitive to outliers, which can significantly distort the principal components and lead to inaccurate representations of the data. So, it is important that any physical conclusions drawn from the PCA results are confirmed by also examining other quantities.
\endwidetext 
\bibliography{BigLandscape}

\end{document}